\documentclass[11pt]{article}

\usepackage[margin=1in]{geometry}
\usepackage{amsmath,amssymb,amsfonts,amsthm}
\usepackage[numbers]{natbib}
\usepackage[colorlinks,citecolor=blue,urlcolor=blue]{hyperref}
\usepackage{graphicx}

\theoremstyle{plain}
\newtheorem{axiom}{Axiom}
\newtheorem{Pro}[axiom]{Proposition}
\newtheorem{Thm}{Theorem}[section]
\newtheorem{Lemma}[Thm]{Lemma}
\newtheorem{Coro}[Thm]{Corollary}

\theoremstyle{definition}
\newtheorem{Def}[Thm]{Definition}

\theoremstyle{remark}

\newcommand{\ee}{\mathrm{e}}
\newcommand{\PP}{\mathbb{P}}
\newcommand{\EE}{\mathbb{E}}
\newcommand{\RR}{\mathbb{R}}

\title{Group Competition Test: A P-value-free False Discovery Rate Control Framework for Grouped Hypotheses}
\author{Mingzhou Deng\thanks{State Key Laboratory of Mathematical Sciences, Academy of Mathematics and Systems Science, Chinese Academy of Sciences. E-mail: dengmingzhou@amss.ac.cn} \and Yan Fu\thanks{State Key Laboratory of Mathematical Sciences, Academy of Mathematics and Systems Science, Chinese Academy of Sciences. E-mail: yfu@amss.ac.cn}}
\date{}

\begin{document}
\maketitle

\begin{abstract}
Grouped multiple testing arises in many high-dimensional applications, where hypotheses may differ across groups and valid p-values are unavailable. Applying competition-based procedures separately within groups can inflate the global false discovery rate (FDR), while conservative corrections may substantially reduce power. We extend the competition test framework to the grouped setting and propose two procedures. The group competition (GC) test provides finite-sample FDR bounds under arbitrary inter-group dependence. For homogeneous groups, the required correction has logarithmic order in the number of groups, and this order is unavoidable for the GC test. The consistent group competition (CGC) filter instead couples group-specific thresholds through a single global normalization. Under the stated basic-independence conditions, it controls the FDR in finite samples without a correction that grows with the number of groups; under arbitrary inter-group dependence, the same guarantee is retained with a per-group correction that depends on the symmetry parameter of each group. We further develop a parameterized-threshold analysis to establish asymptotic guarantees and to characterize when the per-group correction can be dropped, so that a simple filter still controls the FDR approximately. Simulation studies show that the GC test and the CGC filter can improve power over pooled competition testing in heterogeneous settings, whereas the pooled test may remain more powerful under homogeneity. A peptide-spectrum matching benchmark involving post-translational modifications illustrates the practical relevance of the proposed methods.
\end{abstract}


\section{Introduction}
\label{sec:intro}

Multiple hypothesis testing is fundamental in statistics, bioinformatics and machine learning, where many inferences are performed simultaneously. The false discovery rate (FDR), proposed by \cite{benjamini1995}, is the expected proportion of false rejections among all rejections, defined as
\[
\mathrm{FDR} = \EE\left[\frac{|\mathcal{H}_0\cap\mathcal{R}|}{1\vee|\mathcal{R}|}\right],
\]
where $\mathcal{H}$ is the set of hypotheses, $\mathcal{H}_0$ the true nulls and $\mathcal{R}$ the rejection set. Due to the favorable balance between error control and power, FDR has become a widely used criterion, with the Benjamini-Hochberg (BH) procedure being a standard approach for controlling the FDR.

While widely used, the BH procedure and its variants fundamentally rely on valid p-values from known or asymptotically justified null distributions. In many contemporary settings---high-dimensional regression, nonparametric inference or complex dependent data---p-values may be unavailable, computationally infeasible or invalid. These challenges have motivated p-value-free methods that control the FDR without explicit null distributions.

Several influential works exemplify this direction, and they fall into two regimes of symmetry. The first regime constructs pseudo variables that compete with the original ones, and it satisfies an exact conditional symmetry. The Target-Decoy approach \cite{elias2007,he2015} validates peptide-spectrum matches by competing them against decoy sequences, and it extends to case-control studies \cite{he2022}; it requires independence, which is sufficient for the conditional symmetry above. The knockoff framework \cite{barber2015,candes2018} constructs synthetic predictors that mimic the original dependence structure but are independent of the response, and it selects variables by comparing their importance with that of the knockoffs; it has been extended to multi-layer testing \cite{eugene2019,eugene2020}, change detection \cite{liu2022}, directional FDR control \cite{cao2024} and machine learning \cite{lu2018,romano2019,zhu2021}. The second regime is related but weaker. Data splitting \cite{dai2023} (DS) and Symmetrized Data Aggregation \cite{du2023} (SDA) construct symmetry statistics through sample splitting or distributional symmetry, and the Gaussian Mirror \cite{xing2023} generates a pseudo null sample for each hypothesis by perturbing the observed data. These methods satisfy only marginal or approximate symmetry instead of the exact conditional symmetry, and their FDR control is asymptotic or approximate with additional regularity conditions.

Despite these differences, these methods share a competition mechanism. They construct pseudo variables that compete with the originals and use the resulting win-loss labels to determine a rejection threshold, so that the FDR can be controlled without p-values. A unified treatment of this mechanism, and of its limits in complex data settings, is still lacking. We therefore formalize the competition test framework, which also provides the basis for the grouped extensions developed below.

While the competition test avoids the need for p-values, it loses several advantages of p-values. A p-value is a standardized measure of evidence on a fixed scale, so p-value-based methods can split, merge or update hypotheses sequentially while preserving FDR control \cite{li2017,javanmard2018}. Competition scores have no such scale. In multi-center studies, multi-omics integration or proteomics with multiple modifications, the hypotheses fall into groups that differ in size, in the distribution of the competition scores and in the symmetry parameters. The question we address is how to control the FDR over all individual hypotheses when the groups are heterogeneous, without an unnecessary loss of power. This error rate is defined at the level of individual hypotheses and aggregated over all groups, and it is not an error rate over groups; methods that aggregate related hypotheses into a new group-level testing problem control the latter \cite{eugene2019}. Two natural approaches fail. Applying one common threshold to all hypotheses cannot adapt to the differences among the groups and may cause a large loss of power. Running a standard competition test within each group and taking the union of the rejections can push the global FDR above the target level even for two independent groups, as Proposition~\ref{pro:loss_control} shows. Local FDR control therefore does not combine into global FDR control.

Group-walk \cite{freestone2022} is the closest method to ours. It also handles grouped hypotheses in the competition setting, and it provides FDR control under its stated assumptions. It raises the thresholds of all groups on a common rank scale and stops on a single pooled estimate of the false discovery proportion. It is motivated by the analysis of proteomics data. We provide the GC test and the CGC test, which are different in two main respects. Group-walk requires the probability that a true null wins the competition to be the same in every group, whereas we allow group-specific symmetry parameters. It also sets no error level for an individual group, so its rejections may concentrate in a few groups. Table~\ref{tab:compare} summarizes the differences.

\begin{table}[t]
\centering
\small
\renewcommand{\arraystretch}{1.25}
\hyphenpenalty=10000
\exhyphenpenalty=10000
\caption{Comparison of the proposed methods with Group-walk.}
\label{tab:compare}
\begin{tabular}{@{}p{0.17\textwidth}p{0.24\textwidth}p{0.21\textwidth}p{0.26\textwidth}@{}}
\hline
 & \textbf{Group-walk} & \textbf{GC test} & \textbf{CGC test} \\
\hline
{\raggedright\textbf{Symmetry parameters}} & {\raggedright Common value across all groups} & {\raggedright Group-specific $r_i$} & {\raggedright Group-specific $r_i$} \\
\hline
{\raggedright\textbf{Threshold rule}} & {\raggedright All groups advanced on a common rank scale} & {\raggedright Group-specific level $q_i$} & {\raggedright One global normalization} \\
\hline
{\raggedright\textbf{Error level per group}} & {\raggedright None; the stop depends on a pooled estimate} & {\raggedright Yes, $q_i$} & {\raggedright Yes, the share $q\,m_i/m$} \\
\hline
{\raggedright\textbf{Inter-group dependence}} & {\raggedright Independence} & {\raggedright Arbitrary} & {\raggedright Basic independence; \\arbitrary with a correction} \\
\hline
{\raggedright\textbf{FDR guarantee}} & {\raggedright Under TD assumptions} & {\raggedright Finite sample} & {\raggedright Finite sample} \\
\hline
\end{tabular}
\end{table}

Our contributions are threefold. First, we formulate a competition test framework that clarifies when competition-based procedures control the FDR, and we separate the exact conditional symmetry regime from the approximate symmetry regime. Second, we develop the GC test and derive finite-sample FDR bounds under arbitrary inter-group dependence. For homogeneous groups, we obtain a correction of logarithmic order in the number of groups, and we show that this order is necessary for the GC filter. The proof rests on a conditional bound on the false discovery proportion, built from a truncated geometric distribution, rather than the martingale argument of \cite{barber2015} or the flip-one (FO) technique of \cite{dai2023} used in earlier competition proofs. Third, we propose the CGC filter, which coordinates group-specific thresholds through a single global normalization. Under basic independence the CGC filter controls the FDR in finite samples without a correction that grows with the number of groups. Under arbitrary inter-group dependence, finite-sample control is retained through a per-group correction $c(r_i)$ that depends on the symmetry parameter of each group and not on the number of groups. Throughout, ``arbitrary dependence'' refers to the dependence between groups; the conditional symmetry within each group is assumed in all cases. In addition, we develop a parameterized-threshold analysis, which yields asymptotic guarantees and characterizes when the per-group correction can be dropped, so that a simple filter still controls the FDR approximately.

Finally, we evaluate the proposed methods through extensive simulations and an application to peptide-spectrum matching with post-translational modifications. The simulations show that the GC and CGC tests improve power over pooled competition testing in heterogeneous settings, while the pooled test may remain more powerful under homogeneous groups. The application shows the practical relevance of grouping in proteomics.

The paper is organized as follows. Section~\ref{sec:CS} formalizes the competition test and discusses its practical challenges. Section~\ref{sec:GCS} introduces the GC test with FDR control under arbitrary dependence. Section~\ref{sec:AGCS} develops asymptotic theory under inter-group independence with parameterization and proposes the CGC test. Sections~\ref{sec:simu} and~\ref{sec:psm} present simulations and the proteomics application. All proofs are deferred to the supplementary material.

\section{Competition Test: Framework and Theory}
\label{sec:CS}

This section abstracts the common mechanism behind the p-value-free procedures surveyed above. Our aim here is consolidation. We collect their shared mechanism into a single framework, the competition test, and record the properties that follow. The definitions and theorems below thus restate existing ideas in a common language. We regard this consolidation as a contribution, but not the main one; the main contributions are the grouped extensions developed in Sections~\ref{sec:GCS} and~\ref{sec:AGCS}.

\subsection{Notation}

For $n\in\mathbb{Z}_{+}$, let $[n]=\{1,\dots,n\}$. Let $\mathbb{R}_{+}$ denote the positive reals and $\mathbb{R}_{*}=\mathbb{R}_{+}\cup\{0\}$ the non-negative reals. For a vector $\mathbf{X}=(X_1,\dots,X_m)$ and $j\in[m]$, let $\mathbf{X}_{-j}$ be the subvector omitting the $j$th component; for $S\subseteq[m]$, let $\mathbf{X}_S=(X_i)_{i\in S}$ and $\mathbf{X}_{-S}=(X_i)_{i\notin S}$. Unless stated otherwise, $\mathbf{X}_{(\cdot)}$ and $\mathbf{X}_{(\cdot:S)}$ denote the order statistics of $\mathbf{X}$ and $\mathbf{X}_S$ in non-decreasing order. Let $\mathcal{H}=[m]$ be the set of hypotheses, with true nulls $\mathcal{H}_0\subseteq\mathcal{H}$ and alternatives $\mathcal{H}_1=\mathcal{H}\setminus\mathcal{H}_0$. For $d$ groups, let $\mathcal{H}^{i}$, $i\in[d]$, denote the $i$-th group, with nulls $\mathcal{H}_0^i$ and alternatives $\mathcal{H}_1^i$. For $a,b\in\mathbb{R}$, $a\vee b=\max\{a,b\}$, $a\wedge b=\min\{a,b\}$. For a set $A$, $|A|$ denotes its cardinality.

\subsection{Competition statistics and filter}

We first introduce the definition of the competition filter. The rejection threshold, which is uniquely determined by the samples, is constructed by solving an optimization problem. Based on this threshold and the labels of hypotheses, the final rejection set is then obtained.

\begin{Def}
    \label{def:ct}
    For the statistics $(W_j, L_j) \in \mathbb{R} \times \{0,1\}$, $j = 1, 2, \dots, m$ and given $q\in(0,1)$, consider an optimization problem,
    \begin{align*}
        &T= \arg\max_t R_+(t)\\
        \mathrm{s.t.}&\quad\frac{1+R_-(t)}{1\vee R_+(t)}\leq q\quad\text{or}\quad t=+\infty,\quad t\in\{W_j\}\cup\{+\infty\},
    \end{align*}
    where $R_+(t)=\sum_{j=1}^m\mathbf{1}\{W_j\geq t,L_j=1\},R_-(t)=\sum_{j=1}^m\mathbf{1}\{W_j\geq t,L_j=0\}$. Construct the rejection set
    \[
    \mathcal{R}_{q}(\mathbf{W}, \mathbf{L}) = \{ j \in [m] : W_j \geq T, L_j = 1 \},
    \]
    then $\mathcal{R}_q$ is a competition filter with parameter $q$. The feasible set is never empty, because $t=+\infty$ is always allowed. If no score satisfies the inequality, the only feasible choice is $t=+\infty$, so $T=+\infty$ and hence $\mathcal{R}_q=\emptyset$.
\end{Def}

Note that the maximizer in Definition~\ref{def:ct} need not be unique, because different candidates can give the same rejection counts and hence the same value of the objective. From a theoretical point of view, we take the smallest admissible maximizer, componentwise when the threshold is a vector, and we use the same convention for the thresholds defined below (Definitions~\ref{def:gct}, \ref{def:pt} and \ref{def:ggt}). In practice, the computation is simplified by restricting the candidates to $t\in\{W_j:L_j=1\}\cup\{+\infty\}$, because $R_+$ changes only at the scores of the hypotheses with $L_j=1$. This restriction does not change the rejection set, and it is the version we use in the computation.

We now introduce the definition of competition statistics. A competition statistic is a bivariate statistic consisting of a competition score, which possesses a total order structure, and a competition label taking one of two values, $0$ or $1$. Unless stated otherwise, a higher competition score indicates stronger evidence against the null hypothesis.

\begin{Def}
    \label{def:cs}
    The statistics $(\mathbf{W},\mathbf{L})\in\mathbb{R}^m\times\{0,1\}^m$ are competition statistics with symmetry parameter $r$ with respect to $\mathcal{N}\subseteq[m]$ if for any $j\in\mathcal{N}$
    \[
    \PP\{L_j=1 \mid \mathbf{W},\mathbf{L}_{-j}\}=r\,\PP\{L_j=0 \mid \mathbf{W},\mathbf{L}_{-j}\}.
    \]
\end{Def}

This conditional symmetry holds exactly for the knockoff and Target-Decoy methods. However, note that the DS and SDA methods \cite{dai2023,du2023} do not construct pseudo variables that can eliminate the dependence structure of the original variables, and that the Gaussian Mirror \cite{xing2023} perturbs the observed data rather than comparing it with such a variable. Consequently, their statistics do not satisfy the formal definition of a competition statistic. Specifically, they lack the symmetry of conditional probability and satisfy only the marginal symmetry condition. Nevertheless, DS and SDA can still control the FDR in an asymptotic sense \cite{dai2023,du2023}. This is because the competition test itself is stable with respect to symmetry. We therefore introduce the definition of an approximate competition statistic.

\begin{Def}
    \label{def:acs}
    The statistics $(\mathbf{W},\mathbf{L})\in\mathbb{R}^m\times\{0,1\}^m$ are $\varepsilon$-$\delta$ approximate competition statistics with symmetry parameter $r$ with respect to $\mathcal{N}\subseteq[m]$ if
    \[
    \PP\left\{\sup_{j\in\mathcal{N}}\left|\PP\{L_j=1\mid\mathbf{W},\mathbf{L}_{-j}\}-\frac{r}{1+r}\right|>\varepsilon\right\}\leq\delta.
    \]
\end{Def}

\subsection{FDR control for competition test}

A competition test is formally defined as the rejection set $\mathcal{R}_q$ obtained by applying a competition filter to a collection of competition statistics. The following theorem establishes that the competition test can control the FDR.

\begin{Thm}
    \label{thm:CT}
    Suppose the competition statistics $(\mathbf{W},\mathbf{L})\in\mathbb{R}^m\times\{0,1\}^m$ with respect to $\mathcal{H}_0$ have symmetry parameter $r$. Then the competition filter $\mathcal{R}_{r^{-1}\alpha}(\mathbf{W},\mathbf{L})$ controls the FDR at level $\alpha$.
\end{Thm}

This result shows that the competition procedure with competition statistics achieves finite-sample FDR control, without reliance on p-values, known null distributions or dependence assumptions. Furthermore, we demonstrate that the competition filter with approximate competition statistics can approximately control the FDR. We have the following theorem.

\begin{Thm}
    \label{thm:ACT}
    Suppose the $\varepsilon$-$\delta$ approximate competition statistics $(\mathbf{W},\mathbf{L})\in\mathbb{R}^m\times\{0,1\}^m$ with respect to $\mathcal{H}_0$ have symmetry parameter $r$ and satisfy $\varepsilon<1/(1+r)$. Then the competition filter $\mathcal{R}_{r^{-1}\alpha}(\mathbf{W},\mathbf{L})$ approximately controls the FDR at level $\alpha$, such that
    \[
    \mathrm{FDR}\leq \alpha\frac{r+(1+r)\varepsilon}{r-r(1+r)\varepsilon}+\delta.
    \]
\end{Thm}
When $\varepsilon = \delta = 0$, this recovers exact FDR control. The condition $\varepsilon<1/(1+r)$ is what makes the bound informative. The factor $\frac{r+(1+r)\varepsilon}{r-r(1+r)\varepsilon}$ equals $\frac{1+2\varepsilon}{1-2\varepsilon}$ when $r=1$ and diverges as $\varepsilon\uparrow 1/(1+r)$, so an approximate competition statistic is required to be accurate relative to the symmetry parameter.

\subsection{Limitations and motivation for grouping}

While the competition test provides a unified and p-value-free framework for FDR control, it faces several structural limitations when applied to complex modern data settings.

First, competition statistics lack a standardized scale.
Unlike p-values, which are confined to $[0,1]$ and admit well-understood aggregation and calibration, competition scores are unrestricted and method-dependent. When the null or alternative distributions are multimodal or heavily skewed, a single global threshold may fail to capture heterogeneous signals, leading to reduced power or discovery instability.

Second, the conditional independence structure underlying standard competition tests is often violated in practice. The competition statistic implies a strong independence structure, which we refer to as three-level independence. (i) The labels within $\mathbf{L}_{\mathcal{H}_0}$ are mutually independent. (ii) $\mathbf{L}_{\mathcal{H}_0}$ and $\mathbf{L}_{\mathcal{H}_1}$ are independent. (iii) $\mathbf{L}_{\mathcal{H}_0}$ and $\mathbf{W}$ are independent. In realistic high-dimensional or dependent data, these assumptions may hold only locally, rather than globally across all hypotheses.

Third, a competition test is not decomposable or mergeable in a naive way. Because the rejection threshold is constructed from the global joint distribution of scores and labels, splitting the hypotheses into independent sub-problems and testing each sub-problem separately breaks FDR control. Likewise, the per-sub-problem results cannot be combined into a single global decision rule, because the threshold of each part depends on the entire collection of statistics. The procedure therefore does not support splitting, merging or sequential updating of the hypothesis set in a straightforward way. This rigidity limits its direct use in distributed experiments, multi-resolution inference or hierarchically structured hypotheses, and it motivates an effective correction or a new approach.

These limitations become acute when the hypotheses are partitioned into groups, as in multi-omics, multi-center trials or proteomic analyses with heterogeneous modifications. Ignoring group structure can lead to overly conservative control, severe power loss or FDR inflation, as we formally demonstrate in the next section. Addressing this requires a framework that retains the p-value-free nature of competition tests while enabling principled grouping, localized thresholding and guaranteed global FDR control.

\section{Group Competition Test}
\label{sec:GCS}

\subsection{Motivation for grouping and adjustments}

A direct grouping strategy---decomposing the global problem into sub-problems, testing locally within each group and taking the union of rejection sets---can lead to uncontrolled FDR inflation even under independence, because the FDR is the expectation of a ratio and does not satisfy additivity. The following example illustrates this issue.

\begin{Pro}
    \label{pro:loss_control}
    There exist two independent sets of competition statistics $(\mathbf{W}^{1},\mathbf{L}^{1})\in\mathbb{R}_{*}^{m_{1}}\times\{0,1\}^{m_{1}}$ and $(\mathbf{W}^{2},\mathbf{L}^{2})\in\mathbb{R}_{*}^{m_{2}}\times\{0,1\}^{m_{2}}$ with respect to $\mathcal{H}^{1}_0$ and $\mathcal{H}^{2}_0$, whose parameters are $r_{1}$, $r_{2}$ respectively, such that the competition procedures $\mathcal{R}_{\alpha/r_{1}}(\mathbf{W}^{1},\mathbf{L}^{1})$ and $\mathcal{R}_{\alpha/r_{2}}(\mathbf{W}^{2},\mathbf{L}^{2})$ satisfy
    \[
    \mathrm{FDR} = \EE\left[\frac{\left|(\mathcal{H}_{0}^{1}\cap\mathcal{R}_{\alpha/r_{1}}(\mathbf{W}^{1},\mathbf{L}^{1}))\cup(\mathcal{H}_{0}^{2}\cap\mathcal{R}_{\alpha/r_{2}}(\mathbf{W}^{2},\mathbf{L}^{2}))\right|}{1\vee|\mathcal{R}_{\alpha/r_{1}}(\mathbf{W}^{1},\mathbf{L}^{1})\cup\mathcal{R}_{\alpha/r_{2}}(\mathbf{W}^{2},\mathbf{L}^{2})|}\right] > \alpha.
    \]
\end{Pro}

A straightforward remedy is to apply a correction to the local thresholding scheme, making it more conservative. The Bonferroni correction is a common choice but is often overly conservative, leading to severe power loss. We propose a more refined correction that provably controls the FDR while retaining higher power. Accordingly, we introduce the GC test, which applies the correction derived from the FDR upper bound of the GC filter.

\subsection{Group competition filter and FDR upper bound}

For a general multiple testing problem with $d$ groups of hypotheses, we define the GC filter as follows.

\begin{Def}
    \label{def:gct}
    For the sets of statistics $\{(\mathbf{W}_i,\mathbf{L}_i)\}_{i=1}^d$, where $(\mathbf{W}_i,\mathbf{L}_i)\in \mathbb{R}^{m_i} \times \{0,1\}^{m_i}$, consider an optimization problem,
    \begin{align*}
        &\mathbf{T}= \arg\max_{\boldsymbol{t}} \sum_{i=1}^dR^i_+(t_i)\\
        \mathrm{s.t.}&\quad\frac{1+R^i_-(t_i)}{1\vee R^i_+(t_i)}\leq q_i\quad\text{or}\quad t_i=+\infty,\\
        &\quad t_i\in\{W_{ij}\}\cup\{+\infty\},\quad i=1,2,\dots,d.
    \end{align*}
    where $R^i_+(t)=\sum_{j=1}^{m_i}\mathbf{1}\{W_{ij}\geq t,L_{ij}=1\},R^i_-(t)=\sum_{j=1}^{m_i}\mathbf{1}\{W_{ij}\geq t,L_{ij}=0\},i=1,2,\dots,d$. Construct the rejection set
    \[
    \mathcal{R}_{\boldsymbol{q}}(\mathbf{W}, \mathbf{L}) = \{ (i,j) : W_{ij} \geq T_i,L_{ij} = 1\},
    \]
    then $\mathcal{R}_{\boldsymbol{q}}$ is a group competition filter with parameter vector $\boldsymbol{q}$. The feasible set is never empty, because $t_i=+\infty$ is always allowed. For a group $i$ in which no score satisfies the inequality, the only feasible choice is $t_i=+\infty$, so $T_i=+\infty$ and group $i$ contains no rejections.
\end{Def}

We now extend the definition of competition statistics to the group setting.

\begin{Def}
    \label{def:gcs}
    The sets of statistics $\{(\mathbf{W}_i,\mathbf{L}_i)\}_{i=1}^d$, where $(\mathbf{W}_i,\mathbf{L}_i)\in \mathbb{R}^{m_i} \times \{0,1\}^{m_i}$ are group competition statistics with symmetry parameter vector $\boldsymbol{r}$ with respect to $\{\mathcal{N}_i\}_{i=1}^d$, if for any $i\in[d]$ and $j\in\mathcal{N}_i$,
    \[
    \PP\{L_{ij}=1 \mid \mathbf{W}_i,\mathbf{L}_{i,-j}\}=r_i\,\PP\{L_{ij}=0 \mid \mathbf{W}_i,\mathbf{L}_{i,-j}\}.
    \]
\end{Def}

Group competition statistics possess two important properties. First, they allow different groups to have distinct numbers of hypotheses $m_i$ and symmetry parameters $r_i$, enhancing adaptability to model heterogeneity. Second, the conditional probability equality involves only scores and labels within the same group, implying that scores and labels across different groups may exhibit arbitrary dependence, thereby relaxing the requirement of global independence.

Based on group competition statistics, we now present an upper bound for the FDR of the GC filter. We first consider the model-homogeneous case where all groups share the same number of hypotheses and the same symmetry parameter, deriving the FDR upper bound and the required parameter $\boldsymbol{q}$ to control the FDR at a target significance level.

\begin{Thm}
    \label{thm:GCT}
    Suppose the group competition statistics $\{(\mathbf{W}_i,\mathbf{L}_i)\}_{i=1}^d$, where $(\mathbf{W}_i,\mathbf{L}_i)\in \mathbb{R}^{m_i} \times \{0,1\}^{m_i}$ with respect to $\{\mathcal{H}^i_0\}_{i=1}^d$ have symmetry parameter vector $\boldsymbol{r}$ and satisfy that $m_1=m_2=\dots=m_d=m,r_1=r_2=\dots=r_d=r$. Then the GC filter $\mathcal{R}_{\boldsymbol{q}}(\mathbf{W},\mathbf{L})$ controls the FDR at level $\alpha$, where $q_1=q_2=\dots=q_d=q$,
    \[
    q=q_{r,d,m,\alpha}=\sup_{f\in\mathcal{F}}\sup_{\lambda\in\Lambda(f)}\frac{\lambda\alpha}{f^{-1}\left(\sum_{k=0}^{m-1}\frac{dr^k}{(1+r)^{k+1}}f(\lambda k)+d\left(\frac{r}{1+r}\right)^mf(\lambda m)\right)},
    \]
    where $\mathcal{F}=\{f:\RR_*\to\RR_*\mid f'>0,f''\geq 0\}$ is a set of convex functions and
    \[
    \Lambda(f)=\left\{\lambda>0:\sum_{k=0}^{m-1}\frac{dr^k}{(1+r)^{k+1}}f(\lambda k)+d\left(\frac{r}{1+r}\right)^mf(\lambda m)\in f(\RR_*)\right\}
    \]
    which makes $f^{-1}$ valid.
\end{Thm}

The new ingredient in the proof of Theorem~\ref{thm:GCT} is a conditional bound on the false discovery proportion. We construct a truncated geometric distribution and use it to bound the conditional expectation of the local false discovery proportion, instead of the martingale argument of \cite{barber2015} or the flip-one (FO) technique of \cite{dai2023} that the earlier competition proofs rely on. This different route is what yields an explicit upper bound in the grouped setting. We return to the comparison in the proof sketch below.

Note that the adjusted parameter $q$ depends on the group size $m$, the symmetry parameter $r$, the number of groups $d$ and the target FDR level $\alpha$, where $r,d,\alpha$ are the key parameters and $m$ is a secondary parameter. The form of $q$ is complex, because it is obtained by two supremum operations, one of which is over a function space. However, we can obtain an explicit form of $q$ by choosing a specific function $f$. For example, if we choose $f(x)=\exp(x)$, then we have the following corollary.

\begin{Coro}
    \label{coro:GCT}
    With the assumptions of Theorem~\ref{thm:GCT}, let $f(x)=\exp(x)$, then
    \[
    q=\sup_{\lambda\in(0,-\log\frac{r}{1+r})}\frac{\lambda\alpha}{\log d-\log(1+r)-\log\left(1-\ee^{\lambda}\frac{r}{1+r}\right)}\vee\frac{\alpha}{rd}.
    \]
\end{Coro}

The bound in Corollary~\ref{coro:GCT} is more explicit and easier to compute than the original bound in Theorem~\ref{thm:GCT}. The order of the correction factor is $\mathcal{O}(\log d)$, which is much smaller than that of the Bonferroni correction and analogous to the classical Benjamini-Yekutieli (BY) adjustment $\mathcal{O}(1/\sum_{i=1}^{m}i^{-1})$ \cite{benjamini2001}. Given the known tightness of the classical BY bound, we expect the order of our bound to be sharp as well. The next proposition shows that this order cannot be improved. Its construction is given in the supplementary material.

\begin{Pro}
    \label{pro:necessity}
    The order $\mathcal{O}(\log d)$ of the correction factor is necessary for the GC filter. There exist group competition statistics for which no correction of smaller order controls the FDR at level $\alpha$.
\end{Pro}

The tightness of the order is of independent interest, because the logarithmic factor is not specific to the grouped competition test. The same order appears in the BY procedure \cite{benjamini2001}, which controls the FDR for p-values under arbitrary dependence, where the adjustment is of order $1/\log m$ in the number of hypotheses $m$. Our adjustment is of order $1/\log d$ in the number of groups $d$. The two adjustments thus share the same logarithmic order, with $d$ playing the role that the number of hypotheses plays in the BY procedure. The mechanisms differ, since BY adjusts p-values while ours adjusts competition statistics, so the correspondence is a formal one and is suggestive rather than exact.

For the general heterogeneous case, we have the following theorem regarding the FDR upper bound with a weight vector.

\begin{Thm}
    \label{thm:iGCT}
    Suppose the group competition statistics $\{(\mathbf{W}_i,\mathbf{L}_i)\}_{i=1}^d$, where $(\mathbf{W}_i,\mathbf{L}_i)\in \mathbb{R}^{m_i} \times \{0,1\}^{m_i}$ with respect to $\{\mathcal{H}^i_0\}_{i=1}^d$ have symmetry parameter vector $\boldsymbol{r}$. With any given weight vector $\boldsymbol{w}\in\RR_+^d$, the GC filter $\mathcal{R}_{\boldsymbol{q}}(\mathbf{W},\mathbf{L})$ controls the FDR such that
    \[
    \mathrm{FDR}\leq \inf_{f\in\mathcal{F}}\inf_{\lambda\in\Lambda(f)}\lambda^{-1}f^{-1}\left(\sum_{i=1}^d\EE [f(\lambda w_iY_i)]\right)\cdot\max_i\frac{q_i}{w_i},
    \]
    where $Y_i,i\in[d]$ are random variables following truncated geometric distributions with parameter $(m_i,r_i/(1+r_i))$.
\end{Thm}

Based on this upper bound, we propose a practical parameter selection that controls the FDR at the target level. Specifically, the local threshold $q_i$ for group $i$ is determined as a function of the group size $m_i$, the symmetry parameter $r_i$ and a group-specific weight $w_i$.

\begin{Coro}
    \label{coro:iGCT}
    With the assumptions of Theorem~\ref{thm:iGCT}, let
    \[
    q_i=\alpha w_i\left(\sum_{j=1}^dc_j(a_d)w_j^{a_d}\right)^{-\frac{1}{a_d}},\quad i=1,2,\dots,d,
    \]
    where
    \[
    c_j(a)=\sum_{k=1}^{m_j-1}\frac{r_j^k}{(1+r_j)^{k+1}}k^a+\left(\frac{r_j}{1+r_j}\right)^{m_j}m_j^a,\quad j=1,2,\dots,d,
    \]
    and
    \[
    a_d=\arg\sup_{a\in[1,d]}\left(\sum_{j=1}^dc_j(a)w_j^a\right)^{-\frac{1}{a}},
    \]
    we have $\mathrm{FDR}\leq\alpha$.
\end{Coro}

\section{Asymptotic Theory and the Consistent Filter}
\label{sec:AGCS}

In this section, we develop the theory of the group competition test under basic independence in three steps. First, we ask when a correction is unnecessary. Under inter-group independence the GC test controls the FDR asymptotically at the nominal level without any per-group correction, and the sufficient conditions are practically verifiable (Lemma~\ref{pro:ac} and Theorem~\ref{thm:medcondi}). Second, we introduce the tool that makes this analysis possible, a parameterized threshold that separates the scale, adjustment and level parameters and satisfies an FO stability property (Definition~\ref{def:pt} and Lemma~\ref{lem:pt}). Third, we address finite-sample control in two regimes. Under inter-group independence we first derive a fixed bound (Theorem~\ref{thm:ft_gct}) and then a consistent filter that replaces the per-group correction by a single global normalization (Theorem~\ref{thm:ggt}). Under arbitrary inter-group dependence, the same filter keeps a per-group correction $c(r_i)$ that depends only on the symmetry parameter, so the resulting bound is independent of the number of groups (Corollary~\ref{coro:ggt}).

\subsection{The fixed bound under independence}

Although the FDR upper bound in Theorem~\ref{thm:GCT} is finite-sample valid under arbitrary dependence, it is conservative in practice. Numerical evidence suggests that when group competition statistics are independent across groups, the FDR remains close to the target level even without aggressive correction. Thus we investigate the FDR under the assumption of inter-group independence and present the following upper bound theorem.

\begin{Thm}
    \label{thm:ft_gct}
    Suppose the group competition statistics $\{(\mathbf{W}_i,\mathbf{L}_i)\}_{i=1}^d$, where $(\mathbf{W}_i,\mathbf{L}_i)\in \mathbb{R}^{m_i} \times \{0,1\}^{m_i}$ with respect to $\{\mathcal{H}^i_0\}_{i=1}^d$ have symmetry parameter vector $\boldsymbol{r}$ with $r_1=\dots=r_d=r$ and satisfy
    \[
    \PP\{L_{ij}=1\mid\mathbf{W},\mathbf{L}_{-i},\mathbf{L}_{i,-j}\}=r\PP\{L_{ij}=0\mid\mathbf{W},\mathbf{L}_{-i},\mathbf{L}_{i,-j}\}.
    \]
    Then the GC filter $\mathcal{R}_{\boldsymbol{q}}(\mathbf{W},\mathbf{L})$ with $q_i=\alpha/c(r)$, $i\in[d]$, controls the FDR at level $\alpha$, where
    \[
    c(r)=\inf_{\varepsilon>0}(1+\varepsilon)r\vee1+\frac{r}{\log\left((1+\varepsilon)\vee\frac{1}{r}\right)}.
    \]
\end{Thm}

This bound is conservative. The technique behind this bound involves only the binomial distribution, so it is possible to obtain sharper bounds in specific cases by more tailored arguments. For example, when $r=1$, which implies that under the null the label equals $0$ or $1$ with equal probability, setting $q_i=0.518\alpha$ can also control the FDR, where $c(1)\approx 3.4>q_i^{-1}$. More details can be found in the supplementary material.

We refer to group competition statistics that satisfy the conditions in the theorem as having basic independence. The theorem shows that under the condition of independence across groups, the FDR upper bound is independent of the number of groups $d$. It is worth noting that this upper bound closely resembles the bounds established in \cite{eugene2019} for multi-layer testing and in \cite{barber2025} for the BH procedure with independent compound p-values. Moreover, the conservative correction of this bound was shown to be necessary in those works.

\subsection{Asymptotic FDR control under basic independence}

Establishing asymptotic FDR control for GC tests is technically challenging because the threshold $\mathbf{T}$ depends on all the data through optimization. To address this, we adopt a parameterized threshold representation (Definition~\ref{def:pt}) and analyze its stability properties (Lemma~\ref{lem:pt}). This allows us to derive an explicit FDR upper bound (Theorem~\ref{thm:aGCT}) and obtain interpretable sufficient conditions for asymptotic control (Lemma~\ref{pro:ac}). Finally, we provide a practically verifiable condition based on empirical process theory (Theorem~\ref{thm:medcondi}).

First, we define the parameterized rejection threshold.

\begin{Def}
    \label{def:pt}
    For the statistics $(\mathbf{W},\mathbf{L})$, the parameterized threshold is defined as
    \[
    T^{a,b}(\beta;\mathbf{W},\mathbf{L})=\inf\left\{{t\in\{W_j\}}:\frac{a\sum_{j=1}^m\mathbf{1}\{W_j\geq t,L_j=0\}+b}{1\vee\sum_{j=1}^m\mathbf{1}\{W_j\geq t,L_j=1\}}\leq\beta\right\},
    \]
    and the FO parameterized threshold is defined as
    \[
    T^{a,b}_i(\beta;\mathbf{W},\mathbf{L})=\inf\left\{{t\in\{W'_j\}}:\frac{a\sum_{j=1}^m\mathbf{1}\{W'_j\geq t,L'_j=0\}+b}{1\vee\sum_{j=1}^m\mathbf{1}\{W'_j\geq t,L'_j=1\}}\leq\beta\right\},
    \]
    where
    \[
    W'_j=W_j,\quad L'_j=\left\{\begin{array}{ll}L_j,&\quad j\neq i,\\1,&\quad j=i.\end{array}\right.
    \]
\end{Def}

Let $a$, $b$ and $\beta$ denote the scale parameter, adjustment parameter and nominal level parameter, respectively. We parameterize the threshold to better understand its properties. Since the threshold itself generally lacks an explicit expression, clarifying the relationships among these parameters enables a more thorough analysis of FDR control. We present the following lemma, which states that the adjustment parameter, the level parameter and the FO adjustment can be mutually transformed into each other with the scale parameter.

\begin{Lemma}
    \label{lem:pt}
    The parameters in the parameterized threshold can be transformed into each other.
    \begin{enumerate}
        \item The adjustment parameter and the level parameter can be transformed into each other.
        If $b \ge \beta$ and $c \ge 0$, then
        \begin{align*}
            T^{a,b}(\beta;\mathbf{W},\mathbf{L}) = T^{a,b-c}\left(\beta - \frac{c}{1 \vee R_+\bigl(T^{a,b}(\beta;\mathbf{W},\mathbf{L})\bigr)};\,\mathbf{W},\mathbf{L}\right), \\
            T^{a,b-c}(\beta;\mathbf{W},\mathbf{L}) \ge T^{a,b}\left(\beta + \frac{c}{1 \vee R_+\bigl(T^{a,b}(\beta;\mathbf{W},\mathbf{L})\bigr)};\,\mathbf{W},\mathbf{L}\right).
        \end{align*}
        \item The adjustment parameter and the FO operation can be transformed into each other.
        If $b \ge 2\beta$ and $W_i < T^{a,b}_i(\beta;\mathbf{W},\mathbf{L})$, then
        \[
            T^{a,b}_i(\beta;\mathbf{W},\mathbf{L}) = T^{a,b}(\beta;\mathbf{W},\mathbf{L}),
        \]
        and if $b \ge 2\beta$ and $W_i \ge T^{a,b}_i(\beta;\mathbf{W},\mathbf{L})$, then
        \[
            T^{a,b}_i(\beta;\mathbf{W},\mathbf{L}) = T^{a,b-a-\beta}(\beta;\mathbf{W},\mathbf{L}).
        \]
    \end{enumerate}
\end{Lemma}

In this lemma, we remove the group-indexing superscript $i$ from $R_+, R_-$ and instead treat the statistics $\mathbf{W},\mathbf{L}$ as observed values. We define
\[
V_+(t;\mathbf{W},\mathbf{L},\mathcal{H}_0)=\sum_{j\in\mathcal{H}_0}\mathbf{1}\{W_j\geq t,L_j=1\},
\]
\[V_-(t;\mathbf{W},\mathbf{L},\mathcal{H}_0)=\sum_{j\in\mathcal{H}_0}\mathbf{1}\{W_j\geq t,L_j=0\}.
\]

Lemma~\ref{lem:pt} shows that the parameterized threshold is stable under small perturbations of the adjustment parameter $b$ and the FO operation. This stability is the key to establishing asymptotic FDR control without requiring explicit closed-form expressions for the threshold. We establish an upper bound on the FDR for the parameterized threshold.

\begin{Thm}
    \label{thm:aGCT}
    Suppose the group competition statistics $\{(\mathbf{W}_i,\mathbf{L}_i)\}_{i=1}^d$, where $(\mathbf{W}_i,\mathbf{L}_i)\in \mathbb{R}^{m_i} \times \{0,1\}^{m_i}$ with respect to $\{\mathcal{H}^i_0\}_{i=1}^d$ have basic independence and a symmetry parameter vector $\boldsymbol{r}$, where $\min r_i\geq 2\alpha$. Then the GC filter $\mathcal{R}_{\boldsymbol{q}}(\mathbf{W},\mathbf{L})$, where $q_i=\alpha/r_i,i\in[d]$, controls the FDR in the sense that for $0<\varepsilon<\min_i r_i$,
    \[
    \mathrm{FDR}\leq \alpha\sum_{i=1}^d\EE\left[\frac{r_iV_-\left(T^{r_i,-\alpha}(\alpha);\mathbf{W}_i,\mathbf{L}_i,\mathcal{H}_0^i\right)}{\varepsilon\vee\sum_{k=1}^dr_kR_-\left(T^{r_k,r_k-(r_k+\alpha)\mathbf{1}\{k=i\}}(\alpha);\mathbf{W}_k,\mathbf{L}_k\right)}\right].
    \]
\end{Thm}

From the above theorem, we can obtain sufficient conditions for asymptotic FDR control.

\begin{Lemma}
    \label{pro:ac}
    The following conditions are sufficient for the asymptotic control
    \[
    \limsup_{\min_i m_i \to \infty} \mathrm{FDR} \le \alpha.
    \]
    \begin{enumerate}
        \item For every $i$, as $m_i \to \infty$,
        \[
        \EE\left[ \frac{r_i V_{-}\bigl(T^{r_i,-\alpha}(\alpha);\mathbf{W}_i,\mathbf{L}_i,\mathcal{H}_0^i\bigr) - r_i V_{-}\bigl(T^{r_i,r_i}(\alpha);\mathbf{W}_i,\mathbf{L}_i,\mathcal{H}_0^i\bigr)} {\varepsilon \vee \sum_{k=1}^d r_k R_{-}\bigl(T^{r_k,\,r_k-(r_k+\alpha)\mathbf{1}\{k=i\}}(\alpha);\mathbf{W}_k,\mathbf{L}_k\bigr)} \right] \to 0.
        \]
        \item For every $i$, as $m_i \to \infty$,
        \[
        \frac{r_i V_{-}\bigl(T^{r_i,-\alpha}(\alpha);\mathbf{W}_i,\mathbf{L}_i,\mathcal{H}_0^i\bigr) - r_i V_{-}\bigl(T^{r_i,r_i}(\alpha);\mathbf{W}_i,\mathbf{L}_i,\mathcal{H}_0^i\bigr)} {\varepsilon \vee \sum_{k=1}^d r_k R_{-}\bigl(T^{r_k,\,r_k-(r_k+\alpha)\mathbf{1}\{k=i\}}(\alpha);\mathbf{W}_k,\mathbf{L}_k\bigr)} \overset{P}{\to} 0.
        \]
    \end{enumerate}
\end{Lemma}

The sufficient conditions decompose the complex global expectation into constraints on individual groups. Specifically, they require that for any group, at least one of the following two conditions holds. (i) The set of hypotheses with label $0$ that lie above the threshold is stable with respect to changes in the adjustment parameter of the threshold, that is, the number of such hypotheses changes relatively little while the parameter changes. (ii) The number of rejections in that group is negligible compared with the other groups. To make the conditions practically verifiable, we present the following theorem as an intermediate condition.

\begin{Thm}
    \label{thm:medcondi}
    The following conditions are sufficient for Lemma~\ref{pro:ac}.
    For each $i \in [d]$, there exist continuous positive functions $G_+^i$, $G_-^i$ and a nondecreasing sequence $\{b_{i,n}\}$ with $b_{i,n} \to +\infty$ as $m_i \to +\infty$ such that
    \begin{align*}
        &\sup_{t \in (t_i^L, t_i^U)} \Bigl| b_{i,m_i}^{-1} R_+(t;\mathbf{W}_i,\mathbf{L}_i) - G_+^i(t) \Bigr| \overset{P}{\longrightarrow} 0, \\
        &\sup_{t \in (t_i^L, t_i^U)} \Bigl| b_{i,m_i}^{-1} R_-(t;\mathbf{W}_i,\mathbf{L}_i) - G_-^i(t) \Bigr| \overset{P}{\longrightarrow} 0,
    \end{align*}
    where $(t_i^L, t_i^U)$ is a prescribed interval. Define $G^i(t) = G_-^i(t)/G_+^i(t)$. Assume that there exist constants $d_i^L, d_i^U$ satisfying
    \[
    -d_i^U \le \inf_{t \in (t_i^L, t_i^U)} \frac{\partial G^i}{\partial t}
          \le \sup_{t \in (t_i^L, t_i^U)} \frac{\partial G^i}{\partial t}
          \le -d_i^L < 0,
    \]
    and that $t_{i,\alpha} \in (t_i^L, t_i^U)$ is the unique largest root of $G^i(t)=\alpha$.  Furthermore, for each $i$ there exists $\delta_i > 0$ such that
    \[
    \lim_{m_i \to +\infty}
    \PP\Biggl\{\,
        \inf_{t \le t_i^L}
        \frac{R_-(t;\mathbf{W}_i,\mathbf{L}_i)}{1 \vee R_+(t;\mathbf{W}_i,\mathbf{L}_i)} - \alpha
        > \delta_i
    \Biggr\} = 1.
    \]
\end{Thm}

This condition holds under most asymptotic frameworks, e.g., in DS \cite{dai2023} and SDA \cite{du2023} settings under regularity conditions; for general settings, see \cite{fan2025}. The condition comprises two parts---convergence of a statistic resembling the empirical distribution, and characterization of the limiting function---but only the first typically requires explicit verification, via the weak Glivenko-Cantelli theorem for three cases: independent but not identically distributed \cite{wellner1981}, weakly dependent identically distributed data \cite{wellner2006} and more general settings \cite{wellner2023}.

In addition, the parameterization can extend the competition statistics to a more general form, where the symmetry parameter $r$ can be replaced by a function of the score. We show that FDR control still holds under this generalization and some regularity conditions in the supplementary material.

\subsection{Consistent group competition filter}
\label{sec:cgc}

For group competition statistics with basic independence, we construct a novel and more powerful rejection threshold and its filter. The idea is to replace group-specific denominators with a global one.

\begin{Def}
    \label{def:ggt}
    For the family of statistics $\{(\mathbf{W}_i,\mathbf{L}_i)\}_{i=1}^d$, where $(\mathbf{W}_i,\mathbf{L}_i)\in\mathbb{R}^{m_i}\times\{0,1\}^{m_i}$, consider the consistent threshold vector
    \[
    \mathbf{T}^{\mathbf{r}}(q;\mathbf{W},\mathbf{L})=\inf\left\{\boldsymbol{t}\in\mathcal{A}_{\mathbf{W},\mathbf{L}}:\sup_{\{i\in[d]:R_+^i(t_i)>0\}}\frac{r_i+r_iR_-^i(t_i)}{m_i}\bigg/\frac{1\vee\sum_{k=1}^dR^k_+(t_k)}{\sum_{k=1}^dm_k}\leq q\right\},
    \]
    where the $\inf$ is taken component-wise and
    \[
    \mathcal{A}_{\mathbf{W},\mathbf{L}}=\{\boldsymbol{t}:t_i\in\{W_{ij}\}\cup\{+\infty\},i=1,\dots,d\}.
    \]
    Construct the rejection set
    \[
    \mathcal{R}_{q,\boldsymbol{r}}(\mathbf{W},\mathbf{L})=\{(i,j):W_{ij}\geq T^{\boldsymbol{r}}_i(q),L_{ij}=1\},
    \]
    then $\mathcal{R}_{q,\boldsymbol{r}}$ is a CGC filter with parameters $q,\boldsymbol{r}$.
\end{Def}

This construction replaces the group-specific rejection counts in Definition~\ref{def:gct} with a global aggregation. Specifically, the denominator is the total number of rejections across all groups, and each group contributes to the constraint proportionally to its size. Equivalently, for every group with $R_+^i(t_i)>0$ the constraint reads $r_i+r_iR_-^i(t_i)\le q\,(m_i/m)\,(1\vee\sum_{k=1}^dR_+^k(t_k))$, where $m=\sum_{k=1}^d m_k$ is the total number of hypotheses. The left side is the error budget allowed in group $i$, and the right side is proportional to the size $m_i$ of that group. The global level is thus divided among the groups in proportion to the number of hypotheses they contain, and group $i$ receives the share $q\,m_i/m$. This allocation justifies the term ``consistent'', because all groups are held to one common level through a single normalization and no group requires a separately tuned correction. A critical requirement for this construction is to demonstrate that the threshold is well-defined, which amounts to the existence and uniqueness of the threshold vector $\mathbf{T}^{\boldsymbol{r}}(q;\mathbf{W},\mathbf{L})$. We therefore present the following theorem.

\begin{Thm}
    \label{thm:valid_ggt}
    The consistent vector threshold $\mathbf{T}^{\boldsymbol{r}}(q;\mathbf{W},\mathbf{L})$ is well-defined: for any given statistics $\mathbf{W},\mathbf{L}$, the consistent vector threshold $\mathbf{T}^{\boldsymbol{r}}(q;\mathbf{W},\mathbf{L})$ exists and is uniquely determined.
\end{Thm}

When using group competition statistics that are independent across groups, the CGC filter controls the FDR.

\begin{Thm}
    \label{thm:ggt}
    Suppose the group competition statistics $\{(\mathbf{W}_i,\mathbf{L}_i)\}_{i=1}^d$, where $(\mathbf{W}_i,\mathbf{L}_i)\in\RR^{m_i}\times\{0,1\}^{m_i}$ with respect to $\{\mathcal{H}_0^i\}_{i=1}^d$ have basic independence and a symmetry parameter vector $\boldsymbol{r}$. Then the CGC filter $\mathcal{R}_{\alpha,\boldsymbol{r}}$ controls the FDR such that $\mathrm{FDR}\leq\alpha$.
\end{Thm}

Although Theorem~\ref{thm:ggt} assumes group independence, the consistent filter can be adapted to arbitrary dependence structures and still controls the FDR with a correction. The following corollary is derived by combining Theorem~\ref{thm:ggt} with the correction $c(r_i)$ of Theorem~\ref{thm:ft_gct}.

\begin{Coro}
    \label{coro:ggt}
    Suppose the group competition statistics $\{(\mathbf{W}_i,\mathbf{L}_i)\}_{i=1}^d$, where $(\mathbf{W}_i,\mathbf{L}_i)\in\RR^{m_i}\times\{0,1\}^{m_i}$ with respect to $\{\mathcal{H}_0^i\}_{i=1}^d$ have a symmetry parameter vector $\boldsymbol{r}$. Then the CGC filter $\mathcal{R}_{\alpha,\boldsymbol{r'}}$ controls the FDR such that $\mathrm{FDR}\leq\alpha$, where $r'_i=c(r_i)r_i$.
\end{Coro}

Unlike the GC filter, the CGC filter does not incur a logarithmic factor in the number of groups. Proposition~\ref{pro:necessity} shows that a correction of order $\log d$ is necessary for the GC filter under arbitrary dependence, whereas the global normalization of the CGC filter removes this factor, and the bound in Corollary~\ref{coro:ggt} is independent of the number of groups, with the per-group correction $c(r_i)$ not growing with $d$.

The consistent GC filter is computationally scalable. Recall that $m$ is the total number of hypotheses. The threshold $\mathbf{T}^{\boldsymbol{r}}$ can be computed by first sorting the scores within each group ($\mathcal{O}(\sum m_i \log m_i)$) and then performing a global scan over the sorted candidates ($\mathcal{O}(m \log m)$). Hence, the overall time complexity is
\[
\mathcal{O}\!\left(\sum_{i=1}^d m_i \log m_i \;+\; m \log m\right),
\]
which is near-linear in the total hypothesis size. This is comparable to running a standard BH procedure on the pooled data.

\section{Simulation}
\label{sec:simu}

In this section, we evaluate the finite-sample performance of the GC filter and the CGC filter via two simulation studies. Since the competition statistics are generated directly, the p-value-free methods that share the competition mechanism are unified by the competition test, so comparing against the competition test also covers these existing methods.

\subsection{Groups of multiple hypothesis testing}

In the first simulation, we directly generate grouped multiple testing problems. Each experiment is repeated $300$ times; the average false discovery proportion (FDP) and true positive proportion (TPP) estimate the FDR and power. ``CT'' denotes the competition test, ``GCT'' the GC filter without local level adjustment and ``CGC'' the CGC filter under independence. For each hypothesis we generate two random variables, the original variable and the pseudo variable, with the following distributions:
\[
X^{\text{pse}}_{ij}\sim F_{ij,0},\quad
\begin{cases}
X^{\text{org}}_{ij}\sim F_{ij,1}, & j\in\mathcal{H}^i_1\\
X^{\text{org}}_{ij}\sim F_{ij,0}, & j\in\mathcal{H}^i_0
\end{cases}
\]
Subsequently, we use the original variable and the pseudo variable to construct the statistics
\[
W_{ij}=f(X_{ij}^{\text{org}},X_{ij}^{\text{pse}}),\quad
L_{ij}=\mathbf{1}\{|X^{\text{org}}_{ij}|>|X^{\text{pse}}_{ij}|\}+\varepsilon_{ij}\mathbf{1}\{|X^{\text{org}}_{ij}|=|X^{\text{pse}}_{ij}|\},
\]
where $f$ is a symmetric function; here we take $f(x,y)=|x|\vee|y|$, and $\varepsilon_{ij}$ are independent Bernoulli random variables with success probability $0.5$. The resulting statistics are competition statistics with symmetry parameter $r=1$.

\begin{figure}[t]
    \centering
    \includegraphics[width=0.85\linewidth]{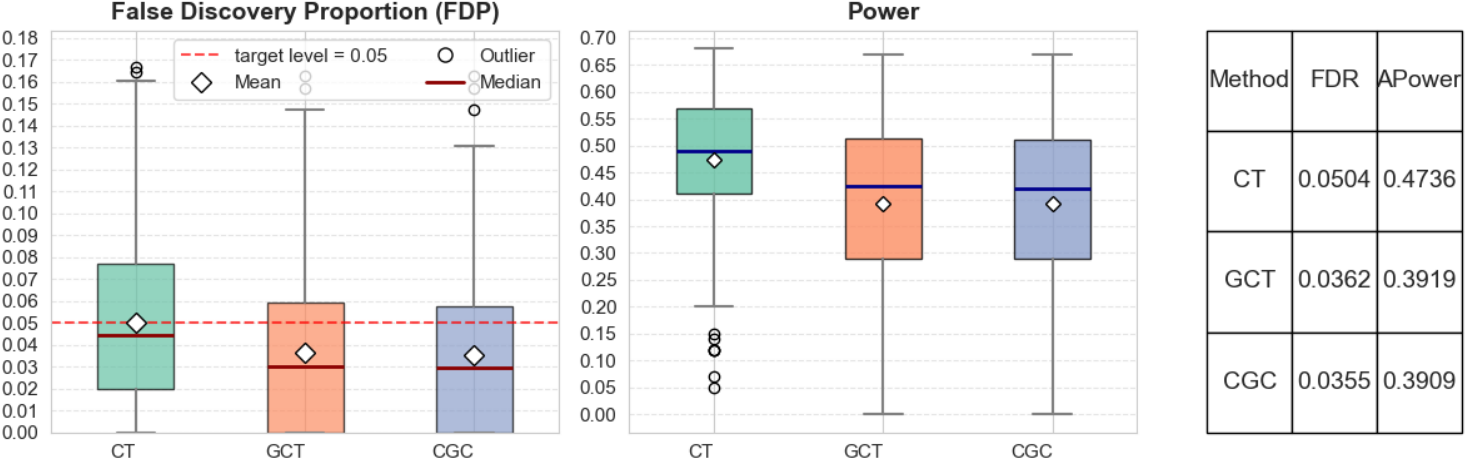}
    \caption{The single-model experiment, where parameters are consistent across groups.}
    \label{fig:exp_unimodel_cons_para}
\end{figure}

First, we conduct experiments under consistent parameters and model settings. Consider
\[
F_{ij,0}\sim \mathrm{N}(0,1),\quad F_{ij,1}\sim\mathrm{N}(\beta,1)
\]
where $\beta=2.7$, $m_i=|\mathcal{H}^i|=500, d=20$, and $\pi_{i0}=0.8$ denotes the proportion of true null hypotheses in group $i$, i.e., $|\mathcal{H}^i_0|=\pi_{i0}m_i$. The experimental results are shown in Figure~\ref{fig:exp_unimodel_cons_para}.

The experimental results show that when the parameters and the model are inherently consistent, there is no need to split the hypotheses into groups. The problem can be effectively handled as a single, unified one. The competition test achieves the highest power because pooling all hypotheses yields a larger number of tests. Moreover, the GC test without local parameter adjustment still maintains FDR control. As anticipated, these observations are consistent with our theoretical analysis.

\begin{figure}[t]
    \centering
    \includegraphics[width=0.85\linewidth]{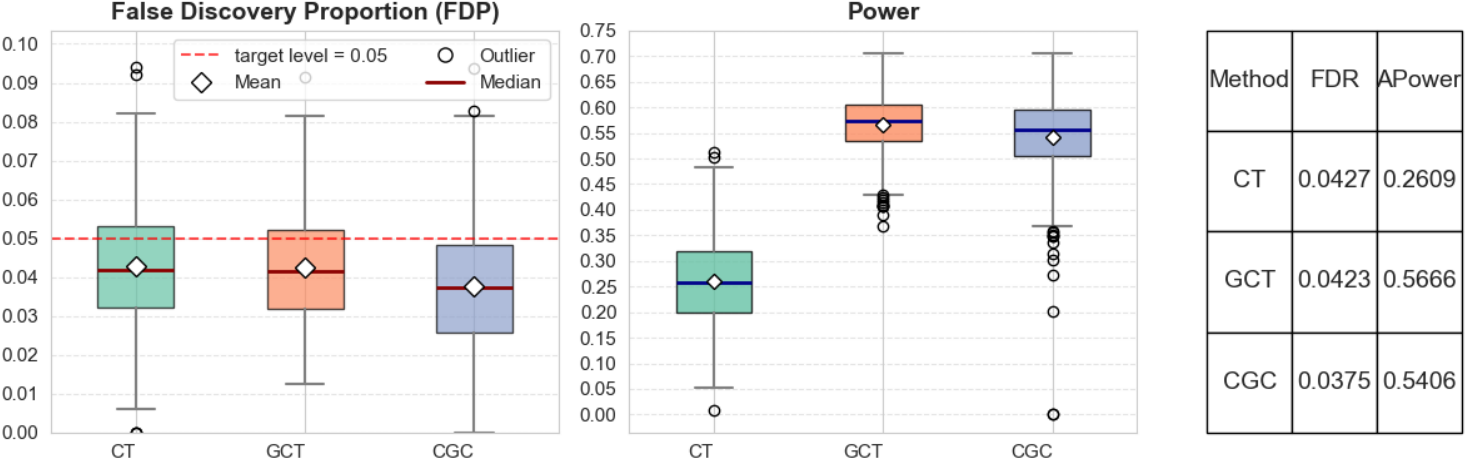}
    \caption{The single-model experiment with different parameters across groups.}
    \label{fig:exp_unimodel_diff_para}
\end{figure}

The signal strength is held constant at $\beta_i = \beta = 2.7$, while $m_i$ and $\pi_{i0}$ vary across groups. Furthermore, we generate scale parameters $c^{\text{sca}}_i$ and location parameters $c^{\text{loc}}_i$ for each group, and let $W_{ij} = c^{\text{loc}}_i + c^{\text{sca}}_i \, f(X_{ij}^{\text{org}}, X_{ij}^{\text{pse}})$. This generates a model in which the parameter settings vary across groups. Detailed settings are provided in the supplementary material. The experimental results are shown in Figure~\ref{fig:exp_unimodel_diff_para}.

Finally, we consider groups following different distribution types (e.g., Gaussian and uniform). Detailed settings are in the supplementary material. Results are shown in Figure~\ref{fig:exp_multimodel}.

\begin{figure}[t]
    \centering
    \includegraphics[width=0.85\linewidth]{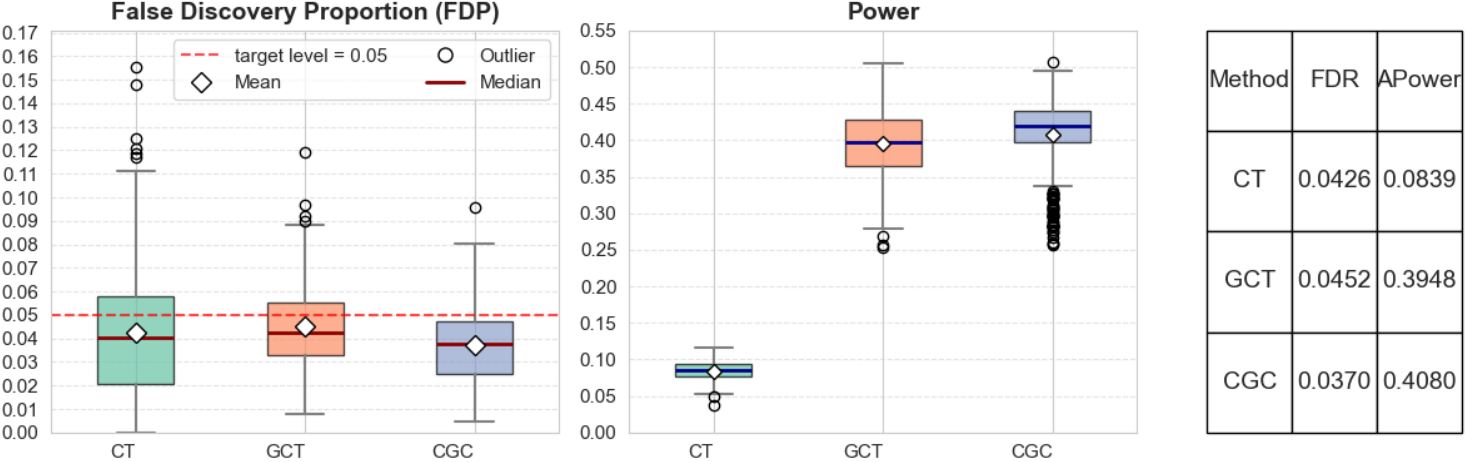}
    \caption{The multi-model experiment.}
    \label{fig:exp_multimodel}
\end{figure}

\subsection{Edge inference in Gaussian graphical models}
\label{sec:ggm}

In this set of simulations, we consider the problem of edge inference in a Gaussian graphical model.
\[
H_{ij}:\Omega_{ij}=0\leftrightarrow\Omega_{ij}\neq0,\quad j>i,
\]
where $\Omega$ is the precision matrix of the Gaussian graphical model.

\begin{figure}[t]
    \centering
    \includegraphics[width=0.85\linewidth]{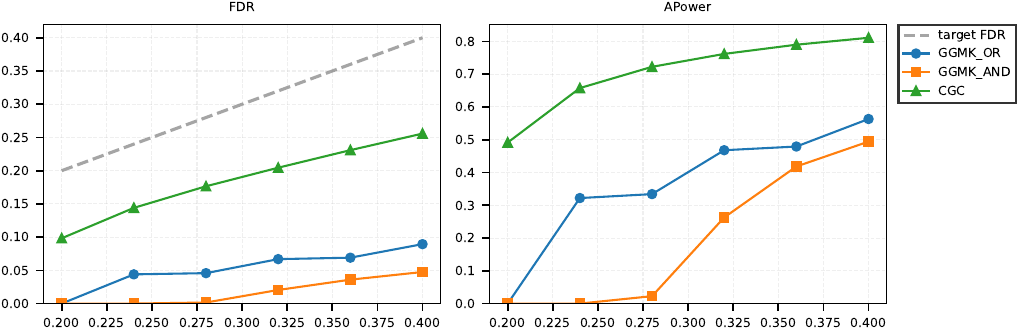}
    \caption{Results of edge inference in GGMs across target FDR control levels, with a sample size of $2000$.}
    \label{fig:q_list_exp}
\end{figure}

\begin{figure}[t]
    \centering
    \includegraphics[width=0.85\linewidth]{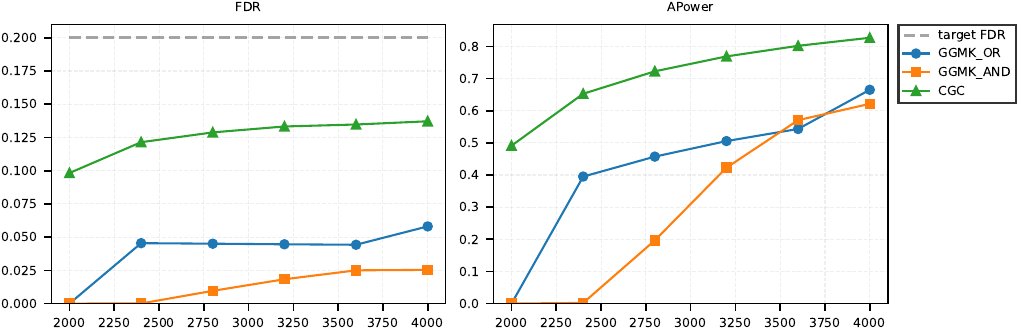}
    \caption{Results of edge inference in GGMs across sample sizes, with a target FDR control level of $0.2$.}
    \label{fig:n_list_exp}
\end{figure}

We compare three methods: the CGC test; the Gaussian graphical model knockoff filter \cite{li2021} with the rule $\mathfrak{R}=\mathfrak{R}_{OR}$, denoted as ``GGMK OR''; the Gaussian graphical model knockoff filter with the rule $\mathfrak{R}=\mathfrak{R}_{AND}$, denoted as ``GGMK AND''. The Gaussian graphical model is generated using the Erd\H{o}s-R\'{e}nyi graph. Specifically, the precision matrix is constructed as $\Omega=\Omega^0+(|\min \lambda(\Omega^0)|+0.5)\mathbf{I}_p\in\mathbb{R}^{p\times p}$, where $\lambda(\Omega)$ denotes the spectrum of the precision matrix, $p=250$, and $\Omega^0$ is given by $\Omega^0_{ii}=1$ and $\Omega^0_{ji}=\Omega^0_{ij}=\omega_{ij}\psi_{ij}$ for $j>i$. Here each $\omega_{ij}$ is drawn uniformly from $[-0.6,0.2]\cup[0.2,0.6]$, and the $\psi_{ij}$ are independent Bernoulli$(0.2)$ variables, drawn independently of the $\omega_{ij}$. We conduct simulations under different target FDR control levels and sample sizes, and obtain the following two sets of results shown in Figures~\ref{fig:q_list_exp} and~\ref{fig:n_list_exp}, respectively.

\section{Group Competition Test for Peptide-Spectrum Matching}
\label{sec:psm}

To further assess the proposed methods, we apply the GC filter and the CGC filter to peptide-spectrum matches (PSMs) with post-translational modifications. This application is critical because it represents the exact scenario where the current standard---the competition test based on the Target-Decoy approach---faces challenges due to group heterogeneity.

We analyze a benchmark dataset of tandem mass spectra \cite{fu2014transferred} simulated from known peptide sequences with $10$ predefined modification types \cite{an2019ptminer}. The dataset contains approximately $95,\!000$ spectra, including background spectra. Each spectrum is searched using the pFind engine \cite{chi2018comprehensive} with the standard Target-Decoy competition setup \cite{elias2007,he2015}. Spectra without a valid match are discarded. More details about the dataset and the search process are provided in the supplementary material.

We group hypotheses (PSMs) by modification identification type, resulting in $d = 13$ groups (including the no-modification group and the multi-modification group), which we treat as mutually independent. For each spectrum, we use two scores exported by pFind. (i) The raw PSM score. (ii) The processed PSM score, a score derived from the raw score through an empirical transformation. We evaluate three methods. (i) CT (standard), the standard competition test pooling all modification types \cite{elias2007}. (ii) GCT, the GC test with conservative correction for independent groups (Definition~\ref{def:gct} and Theorem~\ref{thm:ft_gct}). (iii) CGC, the CGC filter (Definition~\ref{def:ggt}). All methods target an FDR of $0.01$, and Table~\ref{tab:psm_results} reports the results.

\begin{table}[t]
\centering
\caption{Performance on the PSM dataset ($\alpha = 0.01$). CT refers to the standard Target-Decoy competition test.}
\label{tab:psm_results}
\begin{tabular}{lcccc}
\hline
Method & Score Type & Observed FDP & Power & \# Discoveries \\
\hline
CT (Standard) & Raw     & 0.0093 & 0.602 & 18,\!400 \\
GCT           & Raw     & 0.0056 & 0.803 & 24,\!457 \\
CGC           & Raw     & 0.0099 & 0.802 & 24,\!519 \\
\hline
CT (Standard) & Processed   & 0.0162 & 0.991 & 30,\!504 \\
GCT           & Processed   & 0.0077 & 0.985 & 30,\!063 \\
CGC           & Processed   & 0.0135 & 0.991 & 30,\!418 \\
\hline
\end{tabular}
\end{table}

These results validate our theoretical analysis in a real-world scientific setting. Under the raw score, all three methods control the FDR. However, the GC filter and the CGC filter achieve substantially higher power (approximately $33\%$ higher than CT). Under the processed score---where empirical calibration typically promotes PSMs labeled as ``Target''---all methods exhibit near-perfect power (close to $1.0$). In this regime, the critical difference lies in FDR stability. The standard CT exhibits noticeable FDR inflation. In contrast, the proposed GCT and CGC filters are more robust to heterogeneity. In summary, while the standard CT remains viable for raw scores, our proposed filters provide a more robust and powerful solution for real-world proteomics pipelines, particularly when dealing with processed scores that induce complex group heterogeneity.

\section{Conclusion and Discussion}
\label{sec:cad}

This paper addresses multiple hypothesis testing without valid p-values, a challenge in high-dimensional and nonparametric inference. We develop a unified competition test framework that synthesizes methods such as knockoff and the Target-Decoy approach under a single competition mechanism. Extending this framework to grouped hypotheses through the GC and CGC tests resolves a key limitation of existing competition tests in handling complex data structures. We introduce a parameterized threshold representation $T^{a,b}(\beta;\mathbf{W},\mathbf{L})$ that unifies the scale, adjustment and level parameters, and show via the FO technique that these parameters are mutually transformable. This parameterization provides a tractable tool for analyzing competition test thresholds.

Simulations and a peptide-spectrum match application demonstrate that: (i) under homogeneous settings, the standard competition test may achieve the highest power; (ii) under inter-group heterogeneity, GCT and CGC enhance power while maintaining FDR control.

Notably, e-values, introduced by \cite{wang2022,ramdas2024}, offer a unified metric for p-value-free multiple testing methods. A theorem in \cite{ignatiadis2023} shows that any multiple testing procedure can be transformed into the e-BH procedure, which has been used in the knockoff framework \cite{ren2024}. This raises the question of whether a generic e-value transformation could replace the CGC filter for the grouped problem considered here. We argue that it does not do so automatically. For a multiple testing procedure that controls a given level, the e-values induced by the generic transformation are nearly equal to the threshold of the procedure itself, so they carry essentially no information beyond that threshold. Using them to accomplish the same task therefore leaves only two options. Either the level of the original procedure is lowered, which greatly reduces power, or the e-values are passed to e-BH as they are, in which case almost no rejection is produced, because the generated e-values are either $0$ or not large enough to exceed the data-dependent e-BH threshold. How to choose such an underlying control level is itself a major problem that remains unsolved, and there is no guarantee that any such choice is effective. The CGC filter avoids this difficulty, hence providing a better solution for handling heterogeneous data. A numerical study of this limitation is provided in the supplementary material.

Several avenues warrant further investigation. First, deriving sharper FDR bounds for GCT under arbitrary dependence is essential. Our current bounds and the proof techniques, while finite-sample valid, may be conservative. We believe exploring connections to the BY procedure could yield insights into dependency adjustment and optimal bound derivation. Second, for the GCT and CGC, developing self-adaptive grouping strategies and weight schemes would enhance their applicability to complex real-world datasets. Finally, extending the competition test framework to online settings or causal inference represents a promising direction for future statistical methodology research.

\bibliographystyle{plainnat}
\bibliography{references}

@article{an2019ptminer,
  title     = {{PTMiner}: localization and quality control of protein modifications detected
               in an open search and its application to comprehensive post-translational
               modification characterization in human proteome},
  author    = {An, Zhiwu and Zhai, Linhui and Ying, Wantao and Qian, Xiaohong
               and Gong, Fuzhou and Tan, Minjia and Fu, Yan},
  journal   = {Molecular \& Cellular Proteomics},
  volume    = {18},
  number    = {2},
  pages     = {391--405},
  year      = {2019},
  publisher = {Elsevier}
}

@article{barber2015,
  title     = {Controlling the false discovery rate via knockoffs},
  author    = {Barber, Rina Foygel and Cand{\`e}s, Emmanuel J},
  journal   = {The Annals of Statistics},
  volume    = {43},
  number    = {5},
  pages     = {2055--2085},
  year      = {2015},
  publisher = {JSTOR}
}

@article{barber2025,
  title        = {False discovery rate control with compound p-values},
  author       = {{Foygel Barber}, Rina and Samworth, Richard J},
  journal      = {Journal of the Royal Statistical Society Series B: Statistical Methodology},
  year         = {2026},
  note         = {Advance access},
  doi          = {10.1093/jrsssb/qkag116},
  eprint       = {2507.21465},
  archivePrefix= {arXiv},
  primaryClass = {math.ST}
}

@article{benjamini1995,
  title     = {Controlling the false discovery rate: a practical and powerful approach
               to multiple testing},
  author    = {Benjamini, Yoav and Hochberg, Yosef},
  journal   = {Journal of the Royal Statistical Society: Series B (Methodological)},
  volume    = {57},
  number    = {1},
  pages     = {289--300},
  year      = {1995},
  publisher = {Wiley}
}

@article{benjamini2001,
  title     = {The control of the false discovery rate in multiple testing under dependency},
  author    = {Benjamini, Yoav and Yekutieli, Daniel},
  journal   = {The Annals of Statistics},
  volume    = {29},
  number    = {4},
  pages     = {1165--1188},
  year      = {2001},
  publisher = {JSTOR}
}

@article{candes2018,
  title     = {Panning for gold: {`Model-X'} knockoffs for high dimensional controlled
               variable selection},
  author    = {Cand{\`e}s, Emmanuel and Fan, Yingying and Janson, Lucas and Lv, Jinchi},
  journal   = {Journal of the Royal Statistical Society Series B: Statistical Methodology},
  volume    = {80},
  number    = {3},
  pages     = {551--577},
  year      = {2018},
  publisher = {Oxford University Press}
}

@article{cao2024,
  title     = {Split knockoffs for multiple comparisons: controlling the directional
               false discovery rate},
  author    = {Cao, Yang and Sun, Xinwei and Yao, Yuan},
  journal   = {Journal of the American Statistical Association},
  volume    = {119},
  number    = {548},
  pages     = {2822--2832},
  year      = {2024},
  publisher = {Taylor \& Francis},
  doi       = {10.1080/01621459.2023.2279292}
}

@article{chi2018comprehensive,
  title     = {Comprehensive identification of peptides in tandem mass spectra using
               an efficient open search engine},
  author    = {Chi, Hao and Liu, Chao and Yang, Hao and Zeng, Wen-Feng and Wu, Long
               and Zhou, Wen-Jing and Wang, Rui-Min and Niu, Xiu-Nan and Ding, Yue-He
               and Zhang, Yao and others},
  journal   = {Nature Biotechnology},
  volume    = {36},
  number    = {11},
  pages     = {1059--1061},
  year      = {2018},
  publisher = {Nature Publishing Group}
}

@article{dai2023,
  title     = {False discovery rate control via data splitting},
  author    = {Dai, Chenguang and Lin, Buyu and Xing, Xin and Liu, Jun S},
  journal   = {Journal of the American Statistical Association},
  volume    = {118},
  number    = {544},
  pages     = {2503--2520},
  year      = {2023},
  publisher = {Taylor \& Francis}
}

@article{du2023,
  title     = {False discovery rate control under general dependence by symmetrized
               data aggregation},
  author    = {Du, Lilun and Guo, Xu and Sun, Wenguang and Zou, Changliang},
  journal   = {Journal of the American Statistical Association},
  volume    = {118},
  number    = {541},
  pages     = {607--621},
  year      = {2023},
  publisher = {Taylor \& Francis}
}

@article{elias2007,
  title     = {Target-decoy search strategy for increased confidence in large-scale
               protein identifications by mass spectrometry},
  author    = {Elias, Joshua E and Gygi, Steven P},
  journal   = {Nature Methods},
  volume    = {4},
  number    = {3},
  pages     = {207--214},
  year      = {2007},
  publisher = {Nature Publishing Group}
}

@article{fan2025,
  title     = {{ARK}: robust knockoffs inference with coupling},
  author    = {Fan, Yingying and Gao, Lan and Lv, Jinchi},
  journal   = {The Annals of Statistics},
  volume    = {53},
  number    = {2},
  pages     = {749--773},
  year      = {2025},
  publisher = {Institute of Mathematical Statistics},
  doi       = {10.1214/24-AOS2480}
}

@article{fu2014transferred,
  title     = {Transferred subgroup false discovery rate for rare post-translational
               modifications detected by mass spectrometry},
  author    = {Fu, Yan and Qian, Xiaohong},
  journal   = {Molecular \& Cellular Proteomics},
  volume    = {13},
  number    = {5},
  pages     = {1359--1368},
  year      = {2014},
  publisher = {Elsevier}
}

@article{freestone2022,
  title     = {Group-walk: a rigorous approach to group-wise false discovery rate
               analysis by target-decoy competition},
  author    = {Freestone, Jack and Short, Temana and Noble, William Stafford and
               Keich, Uri},
  journal   = {Bioinformatics},
  volume    = {38},
  number    = {Supplement\_2},
  pages     = {ii82--ii88},
  year      = {2022},
  publisher = {Oxford University Press},
  doi       = {10.1093/bioinformatics/btac471}
}

@misc{he2015,
  title        = {A theoretical foundation of the target-decoy search strategy for
                  false discovery rate control in proteomics},
  author       = {He, Kun and Fu, Yan and Zeng, Wen-Feng and Luo, Lan and Chi, Hao
                  and Liu, Chao and Qing, Lai-Yun and Sun, Rui-Xiang and He, Si-Min},
  year         = {2015},
  howpublished = {arXiv preprint arXiv:1501.00537},
  eprint       = {1501.00537},
  archivePrefix= {arXiv},
  primaryClass = {stat.AP}
}

@article{he2022,
  title     = {Null-free false discovery rate control using decoy permutations},
  author    = {He, Kun and Li, Meng-jie and Fu, Yan and Gong, Fu-zhou and Sun, Xiao-ming},
  journal   = {Acta Mathematicae Applicatae Sinica, English Series},
  volume    = {38},
  number    = {2},
  pages     = {235--253},
  year      = {2022},
  publisher = {Springer}
}

@article{ignatiadis2023,
  title   = {E-values as unnormalized weights in multiple testing},
  author  = {Ignatiadis, Nikolaos and Wang, Ruodu and Ramdas, Aaditya},
  journal = {Biometrika},
  volume  = {111},
  number  = {2},
  pages   = {417--439},
  year    = {2024}
}

@article{javanmard2018,
  title     = {Online rules for control of false discovery rate and false discovery
               exceedance},
  author    = {Javanmard, Adel and Montanari, Andrea},
  journal   = {The Annals of Statistics},
  volume    = {46},
  number    = {2},
  pages     = {526--554},
  year      = {2018},
  publisher = {Institute of Mathematical Statistics}
}

@article{eugene2019,
  title     = {Multilayer knockoff filter: controlled variable selection at multiple
               resolutions},
  author    = {Katsevich, Eugene and Sabatti, Chiara},
  journal   = {The Annals of Applied Statistics},
  volume    = {13},
  number    = {1},
  pages     = {1--33},
  year      = {2019},
  publisher = {Institute of Mathematical Statistics},
  doi       = {10.1214/18-AOAS1185}
}

@article{eugene2020,
  title     = {Simultaneous high-probability bounds on the false discovery proportion
               in structured, regression and online settings},
  author    = {Katsevich, Eugene and Ramdas, Aaditya},
  journal   = {The Annals of Statistics},
  volume    = {48},
  number    = {6},
  pages     = {3465--3487},
  year      = {2020},
  publisher = {Institute of Mathematical Statistics}
}

@article{li2017,
  title     = {Accumulation tests for {FDR} control in ordered hypothesis testing},
  author    = {Li, Ang and {Foygel Barber}, Rina},
  journal   = {Journal of the American Statistical Association},
  volume    = {112},
  number    = {518},
  pages     = {837--849},
  year      = {2017},
  publisher = {Taylor \& Francis},
  doi       = {10.1080/01621459.2016.1180989}
}

@article{li2021,
  title     = {{GGM} knockoff filter: false discovery rate control for Gaussian
               graphical models},
  author    = {Li, Jinzhou and Maathuis, Marloes H.},
  journal   = {Journal of the Royal Statistical Society Series B: Statistical Methodology},
  volume    = {83},
  number    = {3},
  pages     = {534--558},
  year      = {2021},
  publisher = {Oxford University Press},
  doi       = {10.1111/rssb.12430}
}

@article{liu2022,
  title     = {A generalized knockoff procedure for {FDR} control in structural
               change detection},
  author    = {Liu, Jingyuan and Sun, Ao and Ke, Yuan},
  journal   = {Journal of Econometrics},
  volume    = {239},
  number    = {2},
  pages     = {105331},
  year      = {2024},
  publisher = {Elsevier},
  doi       = {10.1016/j.jeconom.2022.07.008}
}

@inproceedings{lu2018,
  title        = {{DeepPINK}: reproducible feature selection in deep neural networks},
  author       = {Lu, Yang Young and Fan, Yingying and Lv, Jinchi and Noble, William Stafford},
  booktitle    = {Advances in Neural Information Processing Systems 31},
  pages        = {8676--8686},
  year         = {2018},
  publisher    = {Curran Associates, Inc.},
  eprint       = {1809.01185},
  archivePrefix= {arXiv},
  primaryClass = {stat.ML}
}

@misc{ramdas2024,
  title        = {Hypothesis testing with e-values},
  author       = {Ramdas, Aaditya and Wang, Ruodu},
  year         = {2024},
  howpublished = {arXiv preprint arXiv:2410.23614},
  eprint       = {2410.23614},
  archivePrefix= {arXiv},
  primaryClass = {math.ST}
}

@article{ren2024,
  title     = {Derandomised knockoffs: leveraging e-values for false discovery rate control},
  author    = {Ren, Zhimei and {Foygel Barber}, Rina},
  journal   = {Journal of the Royal Statistical Society Series B: Statistical Methodology},
  volume    = {86},
  number    = {1},
  pages     = {122--154},
  year      = {2024},
  publisher = {Oxford University Press}
}

@article{romano2019,
  title     = {Deep knockoffs},
  author    = {Romano, Yaniv and Sesia, Matteo and Cand{\`e}s, Emmanuel},
  journal   = {Journal of the American Statistical Association},
  volume    = {115},
  number    = {532},
  pages     = {1861--1872},
  year      = {2020},
  publisher = {Taylor \& Francis},
  doi       = {10.1080/01621459.2019.1660174}
}

@article{wang2022,
  title   = {False discovery rate control with e-values},
  author  = {Wang, Ruodu and Ramdas, Aaditya},
  journal = {Journal of the Royal Statistical Society Series B: Statistical Methodology},
  volume  = {84},
  number  = {3},
  pages   = {822--852},
  year    = {2022},
  doi     = {10.1111/rssb.12489}
}

@article{wellner1981,
  title   = {A {Glivenko-Cantelli} theorem for empirical measures of independent but
             non-identically distributed random variables},
  author  = {Wellner, Jon A.},
  journal = {Stochastic Processes and their Applications},
  volume  = {11},
  number  = {3},
  pages   = {309--312},
  year    = {1981},
  doi     = {10.1016/0304-4149(81)90033-8}
}

@incollection{wellner2006,
  title     = {Empirical processes},
  author    = {Wellner, Jon A.},
  booktitle = {Encyclopedia of Statistical Sciences},
  publisher = {John Wiley \& Sons},
  year      = {2006},
  doi       = {10.1002/9781118445112.stat02986}
}

@book{wellner2023,
  title  = {Weak convergence and empirical processes},
  author = {van der Vaart, A. W. and Wellner, Jon A.},
  year   = {2023},
  doi    = {10.1007/978-3-031-29040-4}
}

@article{xing2023,
  title     = {Controlling false discovery rate using Gaussian mirrors},
  author    = {Xing, Xin and Zhao, Zhigen and Liu, Jun S},
  journal   = {Journal of the American Statistical Association},
  volume    = {118},
  number    = {541},
  pages     = {222--241},
  year      = {2023},
  publisher = {Taylor \& Francis}
}

@article{zhu2021,
  title   = {{DeepLINK}: deep learning inference using knockoffs with applications
             to genomics},
  author  = {Zhu, Zifan and Fan, Yingying and Kong, Yinfei and Lv, Jinchi and Sun, Fengzhu},
  journal = {Proceedings of the National Academy of Sciences},
  volume  = {118},
  number  = {36},
  pages   = {e2104683118},
  year    = {2021},
  doi     = {10.1073/pnas.2104683118}
}

\end{document}


\maketitle

\appendix

\section{FDR Control of Competition Test}
\label{ssec:ctfdr}

\begin{proof}[Proof of Theorem~\ref{thm:CT}]

We begin by presenting the following derivation. Regardless of whether $(\mathbf{W},\mathbf{L})$ are competition statistics, with the definition of $T$, we always have the following inequality.
\begin{align*}
    &\textrm{FDR}=\EE\left[\frac{\sum_{j\in\mathcal{H}_0}\mathbf{1}\left\{W_j\geq T,L_j=1\right\}}{\left(\sum_{j\in\mathcal{H}}\mathbf{1}\left\{W_j\geq T,L_j=1\right\}\right)\vee1}\right]\\
    &=\EE\left[\frac{r\sum_{j\in\mathcal{H}_0}\mathbf{1}\left\{W_j\geq T,L_j=0\right\}+r}{\left(\sum_{j\in\mathcal{H}}\mathbf{1}\left\{W_j\geq T,L_j=1\right\}\right)\vee1}\cdot\frac{\sum_{j\in\mathcal{H}_0}\mathbf{1}\left\{W_j\geq T,L_j=1\right\}}{r\sum_{j\in\mathcal{H}}\mathbf{1}\left\{W_j\geq T,L_j=0\right\}+r}\right]\\
    &\leq\alpha\EE\left[\frac{\sum_{j\in\mathcal{H}_0}\mathbf{1}\left\{W_j\geq T,L_j=1\right\}}{r\sum_{j\in\mathcal{H}}\mathbf{1}\left\{W_j\geq T,L_j=0\right\}+r}\right]
\end{align*}

Thus, we obtain the following sufficient condition to control the FDR.

\[
\EE\left[\frac{\sum_{j\in\mathcal{H}_0}\mathbf{1}\left\{W_j\geq T,L_j=1\right\}}{r\sum_{j\in\mathcal{H}}\mathbf{1}\left\{W_j\geq T,L_j=0\right\}+r}\right]\leq1\Rightarrow \textrm{FDR}\leq\alpha.
\]

Let $W_{(0)}=+\infty$, and define $V_+(t)=\sum_{j\in\mathcal{H}_0}\mathbf{1}\{W_j\geq t,L_j=1\},V_-(t)=\sum_{j\in\mathcal{H}_0}\mathbf{1}\{W_j\geq t,L_j=0\}$. Then we have
\[
M_t=\frac{V_+(t)}{1+V_-(t)},\qquad\widetilde{M}_k=M_{W_{(k)}}.
\]

Let $\mathcal{H}_0(t)=\{j\in\mathcal{H}_0:W_j\geq t\}$, $k_t=\arg\min_{j}\{W_j:j\in\mathcal{H}_0(t)\}$. If $V_+(t)>0$, we have
\begin{align*}
    \PP\left(L_{k_t}=1\big| V_+(t),V_-(t),\mathcal{H}_0(t)\right)=\frac{V_+(t)}{V_+(t)+V_-(t)}
\end{align*}
and then for any $k$,
\begin{align*}
    &\EE\left[\widetilde{M}_{k}-\widetilde{M}_{k+1}\Big|\sigma\{\widetilde{M}_{k}\}\right]=\EE\left[\frac{V_+(W_{(k)})}{1+V_-(W_{(k)})}-\frac{V_+(W_{(k+1)})}{1+V_-(W_{(k+1)})}\Big|\sigma\{\widetilde{M}_{k}\}\right]=\\
    &\EE\left[\EE\left[\frac{V_+(W_{(k)})}{1+V_-(W_{(k)})}-\frac{V_+(W_{(k+1)})}{1+V_-(W_{(k+1)})}\Big|V_+(W_{(k)}),V_-(W_{(k)}),\mathcal{H}_0(W_{(k)})\right]\Big|\sigma\left\{\widetilde{M}_{k}\right\}\right]\\
    &\leq 0.
\end{align*}
So $\{\widetilde{M}_k\}$ is a reversed supermartingale, and by Doob's theorem, we have
\begin{align*}
    &\EE\left[\frac{\sum_{j\in\mathcal{H}_0}\mathbf{1}\left\{W_j\geq T,L_j=1\right\}}{r\sum_{j\in\mathcal{H}}\mathbf{1}\left\{W_j\geq T,L_j=0\right\}+r}\right]\leq\EE\left[\frac{\sum_{j\in\mathcal{H}_0}\mathbf{1}\left\{W_j\geq 0,L_j=1\right\}}{r\sum_{j\in\mathcal{H}_0}\mathbf{1}\left\{W_j\geq 0,L_j=0\right\}+r}\right]\\
    =&\sum_{k=0}^{|\mathcal{H}_0|}\frac{1}{r}\frac{k}{1+|\mathcal{H}_0|-k}\binom{|\mathcal{H}_0|}{k}\left(\frac{r}{1+r}\right)^{k}\left(\frac{1}{1+r}\right)^{|\mathcal{H}_0|-k}<1.
\end{align*}

Thus, the FDR is controlled.

\end{proof}

\begin{proof}[Proof of Theorem~\ref{thm:ACT}]

To simplify notation, let $\Delta$ denote the event that
\[
\Delta=\left\{\left|p_{j,\mathbf{W},\mathbf{L}_{-j}}-\frac{r}{1+r}\right|\leq\varepsilon,\forall j\in\mathcal{H}_0\right\},\quad p_{j,\mathbf{W},\mathbf{L}_{-j}}=\PP\{L_j=1\mid\mathbf{W},\mathbf{L}_{-j}\},
\]
then
\[
\mathrm{FDR}\leq\EE\left[\frac{V_+(T)}{1\vee R_+(T)}\mid\Delta\right]+\PP\{\Delta^c\}
\]
We first consider only the case where event $\Delta$ holds. Without loss of generality, we assume $W_1<W_2<W_3\cdots<W_m$. Construct a sequence of random variables $Y_1,Y_2,\cdots,Y_m$ one by one, such that their distributions satisfy
\begin{align*}
&\PP\{Y_j=1\mid \mathbf{L}_{\{i:i\leq j\}},\mathbf{Y}_{\{i:i<j\}},\mathbf{L}_{\mathcal{H}_1},\mathbf{W}\}\\
&=1-L_j-(1-L_j)\left(\frac{r}{1+r}+\varepsilon\right)(1-p_j(\mathbf{L}_{\{i:i< j\}}))^{-1},\quad\forall j\in\mathcal{H}_0
\end{align*}
and $Y_j=0,\forall j\in\mathcal{H}_1$, where
\[
p_j(\mathbf{L}_{\{i:i< j\}})=\PP\{L_j=1\mid \mathbf{L}_{\{i:i<j\}},\mathbf{L}_{\mathcal{H}_1},\mathbf{W}\}.
\]
At this point, we have $L_j\perp \mathbf{Y}_{\{i:i<j\}}\mid\mathbf{L}_{\{i:i<j\}},\forall j\in\mathcal{H}_0$, and
\[
\PP\left\{L_j+Y_j=1\mid\mathbf{Y}_{\{i:i<j\}},\mathbf{L}_{\{i:i<j\}},\mathbf{L}_{\mathcal{H}_1},\mathbf{W}\right\}=\frac{r}{1+r}+\varepsilon,\quad \forall j\in\mathcal{H}_0.
\]
Furthermore, for any $k>j,k\in\mathcal{H}_0$,
\begin{align*}
&\PP\left\{L_{k}+Y_{k}=1\mid\mathbf{Y}_{\{i:i<j\}},\mathbf{L}_{\{i:i<j\}},\mathbf{L}_{\mathcal{H}_1},\mathbf{W}\right\}\\
=&\EE\left[\PP\left\{L_{k}+Y_{k}=1\mid\mathbf{Y}_{\{i:i<k\}},\mathbf{L}_{\{i:i<k\}},\mathbf{L}_{\mathcal{H}_1},\mathbf{W}\right\}\mid\mathbf{Y}_{\{i:i<j\}},\mathbf{L}_{\{i:i<j\}},\mathbf{L}_{\mathcal{H}_1},\mathbf{W}\right]\\
=&\EE\left[\frac{r}{1+r}+\varepsilon\mid\mathbf{Y}_{\{i:i<j\}},\mathbf{L}_{\{i:i<j\}},\mathbf{L}_{\mathcal{H}_1},\mathbf{W}\right]=\frac{r}{1+r}+\varepsilon.
\end{align*}
Let $Z_j=L_j+Y_j$. Then we have
\[
\PP\{Z_j=1\mid \mathbf{Z}_{\{i:i<j\}},\mathbf{L}_{\mathcal{H}_1},\mathbf{W}\}=\frac{r}{1+r}+\varepsilon,\quad\forall j\in\mathcal{H}_0,
\]
and for any $k>j,k\in\mathcal{H}_0$,
\begin{align*}
    &\PP\{Z_j=1\mid \mathbf{Z}_{\{i:i<j\}},Z_{j+1},\mathbf{L}_{\mathcal{H}_1},\mathbf{W}\}\\
    =&\frac{\PP\{Z_{j+1}\mid \mathbf{Z}_{\{i:i\leq j\}},\mathbf{L}_{\mathcal{H}_1},\mathbf{W}\}\PP\{Z_{j}\mid \mathbf{Z}_{\{i:i<j\}},\mathbf{L}_{\mathcal{H}_1},\mathbf{W}\}}{\PP\{Z_{j+1}\mid \mathbf{Z}_{\{i:i<j\}},\mathbf{L}_{\mathcal{H}_1},\mathbf{W}\}}\\
    =&\PP\{Z_{j}\mid \mathbf{Z}_{\{i:i<j\}},\mathbf{L}_{\mathcal{H}_1},\mathbf{W}\}=\frac{r}{1+r}+\varepsilon.
\end{align*}

Let
\[
M^Z_t=\frac{\sum_{j\in\mathcal{H}_0}\mathbf{1}\{Z_j=1,W_j\geq t\}}{1+\sum_{j\in\mathcal{H}_0}\mathbf{1}\{Z_j=0,W_j\geq t\}},\quad \widetilde{M}^Z_k=M^Z_{W_{m-k+1}}
\]
and consider the filtration $\{\sigma_{t}\}$
\begin{align*}
\sigma_t=&\sigma\left\{\sum_{j\in\mathcal{H}_0(u)}\mathbf{1}\{Y_j=1\},\sum_{j\in\mathcal{H}_0(u)}\mathbf{1}\{Y_j=0\},\sum_{j\in\mathcal{H}_0(u)}\mathbf{1}\{L_j=1\},\right.\\
&\left.\sum_{j\in\mathcal{H}_0(u)}\mathbf{1}\{L_j=0\},\mathbf{L}_{\mathcal{H}_1},\mathbf{W},\forall u\leq t\right\},
\end{align*}
Then $T$ is a stopping time and $\{\widetilde{M}^Z_k\}$ is a reversed supermartingale. Because
\[
\EE\left[\frac{\sum_{j\in\mathcal{H}_0}\mathbf{1}\{Z_j=1\}}{r+r\sum_{j\in\mathcal{H}_0}\mathbf{1}\{Z_j=0\}}\right]\leq \frac{r+(1+r)\varepsilon}{r-r(1+r)\varepsilon}
\]
we have
\[
\mathrm{FDR}\leq \alpha\frac{r+(1+r)\varepsilon}{r-r(1+r)\varepsilon}+\delta.
\]

\end{proof}

\section{FDR Control of Group Competition Test}
\label{ssec:gctfdr}

\subsection{Proofs of Proposition~\ref{pro:loss_control}, Theorem~\ref{thm:GCT} and Theorem~\ref{thm:iGCT}}

\begin{proof}[Proof of Proposition~\ref{pro:loss_control}]

We construct an example. Without loss of generality, we consider two groups of competition statistics with parameter $r_i=1$. Each group involves $11$ hypotheses, of which the first $9$ are alternatives and the last $2$ are nulls. When the alternative hypothesis holds, the label is $1$ and the score is always higher than when the null hypothesis holds. For the target FDR control level $\alpha=0.1$, the constraint inequality of the filter is satisfied in three cases: first, both null hypotheses have label $1$; second, only the null hypothesis with higher score has label $1$; third, the null hypothesis with higher score has label $0$. Because $r=1$, the corresponding probabilities for these three cases are $0.25$, $0.25$ and $0.5$, and the numbers of rejections and false rejections are $(11,2)$, $(10,1)$ and $(0,0)$, respectively. Hence we have
\begin{align*}
    \mathrm{FDR}=&\EE\left[\frac{|(\mathcal{H}_0^1\cap\mathcal{R}_{0.1}(\mathbf{W}^1,\mathbf{L}^1))\cup(\mathcal{H}_0^2\cap\mathcal{R}_{0.1}(\mathbf{W}^2,\mathbf{L}^2))|}{1\vee|\mathcal{R}_{0.1}(\mathbf{W}^1,\mathbf{L}^1)\cup\mathcal{R}_{0.1}(\mathbf{W}^2,\mathbf{L}^2)|}\right]\\
    =&\frac{1}{4}\times\frac{1}{4}\times\frac{4}{22}+\frac{1}{4}\times\frac{1}{4}\times\frac{2}{20}+2\times\frac{1}{4}\times\frac{1}{4}\times\frac{3}{21}\\
    &\quad+2\times\frac{1}{4}\times\frac{1}{2}\times\frac{2}{11}+2\times\frac{1}{4}\times\frac{1}{2}\times\frac{1}{10}\\
    =&\frac{261}{2464}>0.1=\alpha.
\end{align*}

We briefly analyze why the FDR may lose control. When grouping, although the FDR can be controlled locally, the local FDP is not necessarily controlled. The global FDP is essentially a weighted average of the local FDPs of each group, with the weights being the number of rejections in the corresponding groups. Therefore, if groups with larger local FDPs tend to have more rejections, the global FDR may lose control due to the rearrangement inequality. In general, when both the scores and the labels are well designed---as in the example above, where the scores of the alternatives are sufficiently large and almost all labels are $1$---having more null hypotheses with label $1$ increases the number of rejections while also enlarging the local FDP, which leads to the loss of FDR control described above. This situation is more likely to occur and harder to analyze when the local groups are not independent.

\end{proof}

\begin{proof}[Proof of Theorem~\ref{thm:GCT}]

Let $V^i_+(t)=\sum_{j\in\mathcal{H}_0^i}\mathbf{1}\{W_{ij}\geq t,L_{ij}=1\},V^i_-(t)=\sum_{j\in\mathcal{H}_0^i}\mathbf{1}\{W_{ij}\geq t,L_{ij}=0\}$, and construct a set
\[
A=A(\mathbf{W},\mathbf{L})=\left\{i\in[d]:R^i_+(T_i)>0\right\}
\], which is the set of groups that have at least one rejection. Then, by analogy with the competition test, we have
\begin{align*}
    \frac{\sum_{i=1}^dV^i_+(t_i)}{1\vee\sum_{i=1}^dR^i_+(t_i)}
    =&\left(\sum_{i\in A}\frac{1+R^i_-(t_i)}{R^i_+(t_i)}\cdot\frac{R^i_+(t_i)}{\sum_{i\in A}R^i_+(t_i)}\right)\cdot\frac{\sum_{i\in A}V^i_+(t_i)}{\sum_{i\in A}(1+R^i_-(t_i))}\\
    \leq&\frac{\sum_{i\in A}q_iR^i_+(t_i)}{\sum_{i\in A}R^i_+(t_i)}\cdot\frac{\sum_{i\in A}V^i_+(t_i)}{\sum_{i\in A}(1+R^i_-(t_i))}=q\frac{\sum_{i\in A}V^i_+(t_i)}{\sum_{i\in A}(1+R^i_-(t_i))}.
\end{align*}

Construct a sequence of parameterized random variables $\{\delta_i(t_i)\}$, where
\[
\delta_i(t_i)=\max_{t\geq t_i}\left\{V_+^i(t):V_-^i(t)=0\right\}.
\]
The following lemma characterizes $\delta_i$.

\begin{Lemma}
\label{slem:deltacontrol}
$\delta_i(T_i)$ follows a conditional distribution,
\[
\PP\left\{\delta_i(T_i)=k\mid V_+^i(T_i),V_-^i(T_i)\right\}=f\left(k;V_+^i(T_i),V_-^i(T_i)\right)
\]
which does not depend on the value of $T_i$ once $V_+^i(T_i)$ and $V_-^i(T_i)$ are conditioned on. Then the conditional expectation satisfies
\[
\EE\left[\delta_i(T_i)\mid V_+^i(T_i),V_-^i(T_i)\right]=\frac{V_+^i(T_i)}{1+V_-^i(T_i)}
\]
\end{Lemma}
Then we construct a set of random variables $\{Y_j\}_{j=1}^d$ such that, for $j=1,\dots,d$,
\[
\PP\{Y_j=k\mid V^i_+(T_i),V^i_-(T_i),i=1,\dots,d\}=\PP\{\delta_j(T_j)=k\mid V^j_+(T_j),V^j_-(T_j)\}.
\]
$Y_j$ and $\delta_j(T_j)$ share the same marginal and conditional distributions given $V_+^j(T_j),V_-^j(T_j)$; moreover, the $Y_j$ are conditionally independent given $\left\{V_+^i(T_i),V_-^i(T_i)\right\}_{i=1}^d$, and for every $j$,
\[
Y_j\perp\!\!\!\perp\left\{V_+^i(T_i),V_-^i(T_i),i\neq j\right\}\mid V_+^j(T_j),V_-^j(T_j).
\]

So there is the same conditional expectation
\[
\EE\left[Y_i\mid V_+^i(T_i),V_-^i(T_i)\right]=\frac{V_+^i(T_i)}{1+V_-^i(T_i)}.
\]
With such construction and
\[
\frac{\sum_{i\in A}V_+^i(T_i)}{\sum_{i\in A}(1+R_-^i(T_i))}\leq\max_{i\in A}\frac{V_+^i(T_i)}{1+V_-^i(T_i)}\leq\max_{1\leq i\leq d}\frac{V_+^i(T_i)}{1+V_-^i(T_i)}
\]
we have
\begin{align*}
    &\EE\left[\frac{\sum_{i\in A}V_+^i(T_i)}{\sum_{i\in A}(1+R_-^i(T_i))}\right]\leq\EE\left[\max_{1\leq i\leq d}\frac{V_+^i(T_i)}{1+V_-^i(T_i)}\right]\\
   =&\EE\left[\max_{1\leq i\leq d}\EE\left[Y_i\mid V_+^i(T_i),V_-^i(T_i)\right]\right]\\
   =&\EE\left[\max_{1\leq i\leq d}\EE\left[Y_i\mid V_+^1(T_1),V_-^1(T_1),V_+^2(T_2),V_-^2(T_2),\cdots,V_+^d(T_d),V_-^d(T_d)\right]\right]\\
   \leq&\EE\left[\EE\left[\max_{1\leq i\leq d}Y_i\mid V_+^1(T_1),V_-^1(T_1),V_+^2(T_2),V_-^2(T_2),\cdots,V_+^d(T_d),V_-^d(T_d)\right]\right]\\
   =&\EE\left[\max_{1\leq i\leq d}Y_i\right]\\
\end{align*}

Regarding the upper bound of this expectation, we present the following lemma.
\begin{Lemma}
\label{slem:cher}
    For a collection of variables $\{X_i\}_{i=1}^m$, and any function $f:\mathbb{R}\to\mathbb{R}$ satisfying $\partial f/\partial x>0,\partial^2 f/\partial x^2\geq0,\forall x\in \RR_+$, and $\PP\left(f(\lambda X_i)<0\right)=0,\forall \lambda>0$, we have
    \[
    \EE\left[\max_{1\leq i\leq D}X_i\right]\leq\lambda^{-1}f^{-1}\left(\EE\left[f(\lambda\max_{1\leq i\leq D}X_i)\right]\right)\leq\lambda^{-1}f^{-1}\left(\sum_{i=1}^D\EE[f(\lambda X_i)]\right).
    \]
\end{Lemma}
Then with Lemma~\ref{slem:cher}, we can obtain that for any function $f$ that satisfies the regularity conditions,
\[
\EE\left[\max_{1\leq i\leq d}Y_i\right]\leq\inf_{\lambda}\lambda^{-1}f^{-1}\left(\sum_{i=1}^d\EE[f(\lambda Y_i)]\right).
\]
Because $Y_i$ and $\delta_i(T_i)$ have identical marginal distribution, we can construct a random variable $Z_i$ such that
\[
\PP\{Z_i=k\mid Y_i\}=\PP\{\delta_i(0)=k\mid\delta_i(T_i)\},
\]
which implies that $Z_j\geq Y_j$ and
\[
    \PP\{Z_i=k\}=\left\{
    \begin{array}{ccc}
    r_i^k(1+r_i)^{-(k+1)},&k=0,1,\cdots,m_i-|\mathcal{H}^i_1|\\
    \left(\frac{r_i}{1+r_i}\right)^{m_i-|\mathcal{H}^i_1|},&k=m_i-|\mathcal{H}^i_1|\\
    0,&else
    \end{array}
    \right.
\]
Then we have
\begin{align*}
\inf_{\lambda}\lambda^{-1}f^{-1}\left(\sum_{i=1}^d\EE[f(\lambda Y_i)]\right)&\leq\inf_{\lambda}\lambda^{-1}f^{-1}\left(\sum_{i=1}^d\EE[f(\lambda Z_i)]\right)\\
&=\inf_{\lambda}\lambda^{-1}f^{-1}\left(\sum_{i=1}^d\sum_{k=0}^{m_i}\PP(Z_i=k)f(\lambda k)\right).
\end{align*}
If $q\leq \alpha(\EE[\max_{1\leq i\leq d}Y_i])^{-1}$, the FDR is controlled. So
\[
q=q_{r,d,m,\alpha}=\sup_{f\in\mathcal{F}}\sup_{\lambda\in\Lambda(f)}\frac{\lambda\alpha}{f^{-1}\left(\sum_{k=0}^{m-1}\frac{dr^k}{(1+r)^{k+1}}f(\lambda k)+d\left(\frac{r}{1+r}\right)^mf(\lambda m)\right)},
\]
where $\mathcal{F}=\{f:\RR_*\to\RR_*\mid f'>0,f''\geq 0\}$ is a set of convex functions and
\[
\Lambda(f)=\left\{\lambda>0:\sum_{k=0}^{m-1}\frac{dr^k}{(1+r)^{k+1}}f(\lambda k)+d\left(\frac{r}{1+r}\right)^mf(\lambda m)\in f(\RR_*)\right\}
\]
which makes $f^{-1}$ valid.

\end{proof}

\begin{proof}[Proof of Theorem~\ref{thm:iGCT}]

Following the same argument as in the proof of Theorem~\ref{thm:GCT}, we have
\begin{align*}
    &\frac{\sum_{i=1}^dV_+^i(T_i)}{1\vee\sum_{i=1}^dR_+^i(T_i)}=\sum_{i\in A}\frac{V_+^i(T_i)}{\sum_{i\in A}R_+^i(T_i)}\\
    =&\sum_{i\in A}\frac{w_iV_+^i(T_i)}{1+R_-^i(T_i)}\cdot\frac{1+R_-^i(T_i)}{w_iR_+^i(T_i)}\cdot\frac{R_+^i(T_i)}{\sum_{i\in A}R_+^i(T_i)}\\
    \leq&\sum_{i\in A}\frac{w_iV_+^i(T_i)}{1+R_-^i(T_i)}\cdot\left(\frac{q_i}{w_i}\cdot\frac{R_+^i(T_i)}{\sum_{i\in A}R_+^i(T_i)}\right)\\
    \leq&\max_{i\in[d]}\left\{\frac{w_iV_+^i(T_i)}{1+R_-^i(T_i)}\right\}\cdot\max_{i\in[d]}\left\{\frac{q_i}{w_i}\right\}\cdot\sum_{i\in A}\frac{R_+^i(T_i)}{\sum_{i\in A}R_+^i(T_i)}\\
    =&\max_{i\in[d]}\left\{w_i\frac{V_+^i(T_i)}{1+R_-^i(T_i)}\right\}\cdot\max_{i\in[d]}\left\{\frac{\alpha'_i}{w_i}\right\}
\end{align*}
and then
\[
\mathrm{FDR}\leq \inf_{f\in\mathcal{F}}\inf_{\lambda\in\Lambda(f)}\lambda^{-1}f^{-1}\left(\sum_{i=1}^d\EE [f(\lambda w_iY_i)]\right)\cdot\max_i\frac{q_i}{w_i},
\]
where $Y_i,i\in[d]$ are random variables following truncated geometric distributions with parameter $(m_i,r_i/(1+r_i))$.

Corollary~\ref{coro:GCT} and Corollary~\ref{coro:iGCT} are obtained by substituting $f$ and computing.

\end{proof}

\subsection{The Upper Bound of FDR for GCT}
\label{ssec:tightness}

This section proves Proposition~\ref{pro:necessity} of the main text and completes the discussion of the bound in Corollary~\ref{coro:GCT}. We first construct a simple example to show that the order $\mathcal{O}(\log d)$ is necessary. For any target FDR control level $\alpha$, consider a multiple testing problem with multiple groups $\{\mathcal{H}^i\}_{i=1}^d$, where $|\mathcal{H}^i_0|=|\mathcal{H}^i|=\lceil 1/\alpha\rceil+1$ for $i=1,\dots,d$, and $d=2^m$. Since we assume that all hypotheses are null, the false discovery proportion (FDP) can only be either $0$ or $1$, and the FDR is essentially the probability of making at least one rejection. Under arbitrary dependence, let $\{\mathbf{L}_i\}_{i=1}^d$ be a complete enumeration of $\{0,1\}^m$. This ensures that the statistics remain group competition statistics, and simultaneously guarantees that there is always exactly one group in which a rejection occurs. In this case, $\mathrm{FDR}=1$, i.e.,
\[
\mathrm{FDR}\approx\alpha\frac{\log d}{\log2},
\]
which implies that a correction factor of order $\mathcal{O}(\log d)$ is necessary. Further, we present a more complex but tighter bound. Take $f(x)=\exp(a_d\log x)$. Then we have
\[
q=\alpha\left(\sum_{k=0}^{m-1}\frac{dr^k}{(1+r)^{k+1}}k^{a_d}+d\left(\frac{r}{1+r}\right)^mm^{a_d}\right)^{-\frac{1}{a_d}},
\]
where
\[
a_d=\arg\sup_{a\in[1,d]}\left(\sum_{k=0}^{m-1}\frac{dr^k}{(1+r)^{k+1}}k^{a}+d\left(\frac{r}{1+r}\right)^mm^{a}\right)^{-\frac{1}{a}}.
\]
Or, as a simplification, we take $a_d = \log d$.

\begin{figure}[t]
    \centering
    \includegraphics[width=0.9\linewidth]{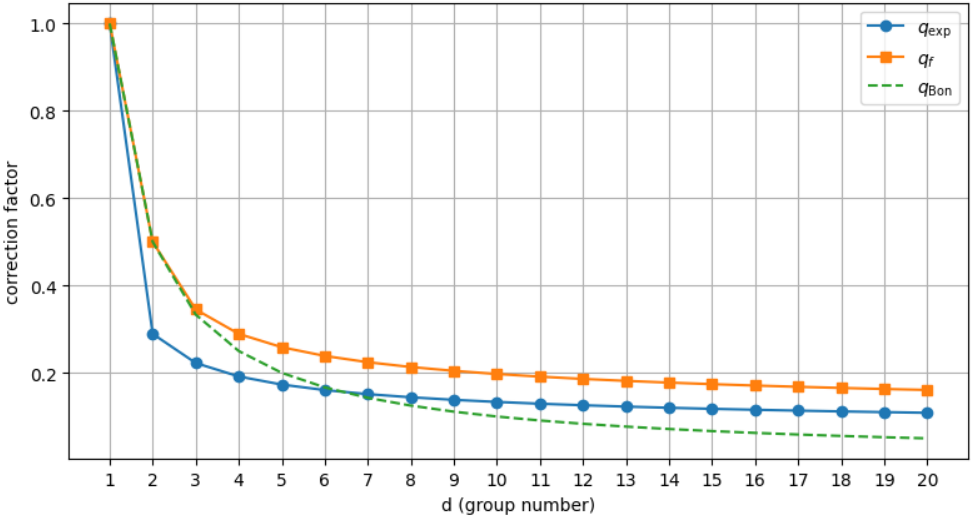}
    \caption{Comparison of different correction factors.}
    \label{fig:correction_factor}
\end{figure}

We define $q_{\rm{Bon}}(d,r)=1/d$ as the Bonferroni correction factor,
\[
q_{\exp}(d,r)=\sup_{\lambda\in(0,-\log\frac{r}{1+r})}\frac{\lambda}{\log d-\log(1+r)-\log\left(1-\ee^{\lambda}\frac{r}{1+r}\right)}
\]
as the correction factor given in Corollary~\ref{coro:GCT}, and
\[
q_{f}(d,r)=\left(\sum_{k=0}^{m-1}\frac{dr^k}{(1+r)^{k+1}}k^{a_d}+d\left(\frac{r}{1+r}\right)^mm^{a_d}\right)^{-\frac{1}{a_d}}
\]
as the simplified correction factor we just mentioned where $a_d = \log d$. This correction factor is slightly better than the one in Corollary~\ref{coro:GCT}, but its order is still $\mathcal{O}(\log d)$. Both are substantially better than the Bonferroni correction factor, as illustrated in Figure~\ref{fig:correction_factor}.

\section{FDR Control of GCT under Basic Independence}
\label{ssec:agct}

\subsection{Proofs of Theorems}

\begin{proof}[Proof of Theorem~\ref{thm:ft_gct}]

If the group competition statistics have basic independence and a symmetry parameter vector whose components are all equal, then conditioned on $\sum_{i=1}^dV^i_+(t_i) + \sum_{i=1}^dV^i_-(t_i)$, $\sum_{i=1}^dV^i_+(t_i)$ follows a binomial distribution $\mathrm{Bin}(\sum_{i=1}^dV^i_+(t_i) + \sum_{i=1}^dV^i_-(t_i),r/(1+r))$. Let $S_k\sim \mathrm{Bin}(k,r/(1+r))$, then we have
\begin{align*}
&\EE\left[\frac{\sum_{i=1}^dV^i_+(T_i)}{1\vee\sum_{i=1}^dR^i_+(T_i)}\right]\leq\frac{\alpha}{c(r)}\EE\left[\sup_{\boldsymbol{t}\in\mathbb{R}_*^d}\frac{\sum_{i=1}^dV^i_+(t_i)}{1+\sum_{i=1}^dV^i_-(t_i)}\right]\\
\leq&\frac{\alpha}{c(r)}\EE\left[\sup_{k\leq m_0}\frac{S_k}{1+k-S_k}\right]
\end{align*}
where $m_0=\sum_{i=1}^d|\mathcal{H}_0^i|$. Notice that we only consider the regime where $\sup_{k\leq m_0}\frac{S_k}{k-S_k+1}>1$,
\[
\EE\left[\sup_{k\leq m_0}\frac{S_k}{k-S_k+1}\right]\leq1+\int_1^{+\infty}\PP\left\{\sup_{k\leq m_0}\frac{S_k}{k-S_k+1}\geq x\right\}{\rm d}x.
\]
Let $D_k(x;r)=\exp\left\{\theta\left(S_k-xk/(x+1)-x/(x+1)\right)\right\}$, where $\theta=\theta(x,r)$, for $r<x$, is the positive root of the following equation
\[
1=\frac{r}{r+1}\exp\left\{\frac{\theta}{x+1}\right\}+\frac{1}{r+1}\exp\left\{\frac{-x\theta}{x+1}\right\},
\]
{\sloppy if $x\leq r$, $\theta =0$. Then $D_k(x;r)$ is a martingale with respect to the natural filtration $\left\{\sigma(D_i(x;r);i\leq k)\right\}$ because\par}
\begin{align*}
    \EE\left[\frac{D_{k+1}(x;r)}{D_k(x;r)}\Big|D_k(x;r)\right]=&\EE\left[\exp\left\{\theta\left(X_{k+1}-\frac{x}{x+1}\right)\right\}\right]\\
    =&\frac{r}{r+1}\exp\left\{\frac{\theta}{x+1}\right\}+\frac{1}{r+1}\exp\left\{\frac{-x\theta}{x+1}\right\}=1.
\end{align*}

With Doob's inequality,
\begin{align*}
    &\PP\left\{\sup_{k\leq m_0}\frac{S_k}{k-S_k+1}\geq x\right\}=\PP\left\{\sup_{k\leq m_0}D_k(x;r)\geq1\right\}\\
    \leq&\EE[D_{m_0}(x;r)]=\EE[D_0(x;r)]=D_0(x;r).
\end{align*}
To estimate an upper bound on the value of this integral, we first consider the equation in the case $r<x$. Denote $\exp\left\{\theta(x,r)/(x+1)\right\}$ by $\Theta(x,r)$, then
\[
1=\frac{r}{r+1}\Theta(x,r)+\frac{1}{r+1}\left[\Theta(x,r)\right]^{-x},
\]
and $\Theta(x,r)>1$,
\[
0=\frac{r}{r+1}\frac{\partial\Theta(x,r)}{\partial x}-\frac{1}{r+1}\left[\Theta(x,r)\right]^{-x}\log\Theta(x,r)-\frac{x}{r+1}\left[\Theta(x,r)\right]^{-x-1}\frac{\partial\Theta(x,r)}{\partial x}.
\]
which implies $\partial\Theta(x,r)/\partial x>0$ if and only if $x$ satisfies $\Theta(x,r)>\left(x/r\right)^{\frac{1}{x+1}}$.
For any $x_1>x_2>r\vee1$, we have
\[
\frac{r}{r+1}\Theta(x_2,r)+\frac{1}{r+1}\left[\Theta(x_2,r)\right]^{-x_1}<\frac{r}{r+1}\Theta(x_2,r)+\frac{1}{r+1}\left[\Theta(x_2,r)\right]^{-x_2}=1
\]
and
\[
\frac{r}{r+1}\frac{r+1}{r}+\frac{1}{r+1}\left[\frac{r+1}{r}\right]^{-x_1}>1.
\]
Define $\psi(u;x,r)=ru/(1+r)+u^{-x}/(1+r)$. This function has a unique critical point and satisfies $\psi(1;x_1,r)=\psi(\Theta(x_1,r);x_1,r)=1$; hence, by the implicit function theorem and the mean value theorem, $\Theta(x_1,r)>\Theta(x_2,r)$. With the above analysis, we have
\[
\Theta(x,r)>\left(\frac{x}{r}\right)^{\frac{1}{x+1}},\frac{\partial\Theta(x,r)}{\partial x}>0,\forall x>r\vee1.
\]
In summary,
\begin{align*}
    \EE\left[\sup_{k\leq m_0}\frac{S_k}{k-S_k+1}\right]\leq&\inf_{\varepsilon>0}(1+\varepsilon)r\vee1+\int_{(1+\varepsilon)r\vee1}^{+\infty}\left[\Theta(x,r)\right]^{-x}{\rm d}x\\
    \leq&\inf_{\varepsilon>0}(1+\varepsilon)r\vee1+\int_{(1+\varepsilon)r\vee1}^{+\infty}\left[\Theta\left((1+\varepsilon)r\vee1,r\right)\right]^{-x}{\rm d}x\\
    \leq&\inf_{\varepsilon>0}(1+\varepsilon)r\vee1+\int_{(1+\varepsilon)r\vee1}^{+\infty}\left((1+\varepsilon)\vee\frac{1}{r}\right)^{\frac{-x}{(1+\varepsilon)r\vee1+1}}{\rm d}x\\
    =&\inf_{\varepsilon>0}(1+\varepsilon)r\vee1+\frac{r}{\log\left((1+\varepsilon)\vee\frac{1}{r}\right)}.
\end{align*}

If $r=1$, let
\[
U_{k_0}(\boldsymbol{l})=\sup_{0\leq k\leq k_0}\frac{\sum_{i=1}^kl_i}{1+k-\sum_{i=1}^kl_i},\quad P_{k_0}(\boldsymbol{l})=\sum_{i=1}^{k_0}l_i,
\]
and $X_k=S_k-S_{k-1}$. Then we have
\begin{align*}
    &\PP\left\{\sup_{k\leq m_0}\frac{S_k}{1+k-S_k}\geq x\right\}\\
    =&\EE\left[\PP\left\{\sup_{k\leq m_0}\frac{S_k}{1+k-S_k}\geq x\mid X_1,X_2,\cdots,X_{k_0}\right\}\right]\\
    =&2^{-k_0}\sum_{\boldsymbol{l}\in\{0,1\}^{k_0}}\PP\left\{\sup_{k\leq m_0}\frac{S_k}{1+k-S_k}\geq x\mid X_1=l_1,X_2=l_2,\cdots,X_{k_0}=l_{k_0}\right\}\\
    =&2^{-k_0}\sum_{\boldsymbol{l}\in\{0,1\}^{k_0}}\left(\mathbf{1}_{\{U_{k_0}(\boldsymbol{l})\geq x\}}+\mathbf{1}_{\{U_{k_0}(\boldsymbol{l})< x\}}\PP\left\{\sup_{k\leq m_0}\frac{P_{k_0}(\boldsymbol{l})+\bar{S}_{k}}{1+k+k_0-\bar{S}_k-P_{k_0}(\boldsymbol{l})}\geq x\right\}\right)
\end{align*}
where
\[
\bar{S}_k=\sum_{i=k_0+1}^{k_0+k}X_i.
\]
From the previous analysis and the properties of the binomial distribution, it can be concluded that
\begin{align*}
&\PP\left\{\sup_{k\leq m_0}\frac{P_{k_0}(\boldsymbol{l})+\bar{S}_{k}}{1+k+k_0-\bar{S}_k-P_{k_0}(\boldsymbol{l})}\geq x\right\}\\
\leq& \exp\left\{\theta(x,1)\left(P_{k_0}(\boldsymbol{l})-\frac{xk_0}{1+x}\right)\right\}D_0(x;1)
\end{align*}
and then
\begin{align*}
    &\PP\left\{\sup_{k\leq m_0}\frac{S_k}{1+k-S_k}\geq x\right\}\leq2^{-k_0}\sum_{\boldsymbol{l}\in\{0,1\}^{k_0}}\Bigg(\mathbf{1}_{\{U_{k_0}(\boldsymbol{l})\geq x\}}\\
    &\quad+\mathbf{1}_{\{U_{k_0}(\boldsymbol{l})< x\}}\exp\left\{\theta(x,1)\left(P_{k_0}(\boldsymbol{l})-\frac{x(1+k_0)}{1+x}\right)\right\}\Bigg)
\end{align*}
and
\begin{align*}
    &\EE\left[\sup_{k\leq m_0}\frac{S_k}{1+k-S_k}\right]\\
    \leq&2^{-k_0}\sum_{\boldsymbol{l}\in\{0,1\}^{k_0}}\Bigg(\max\{U_{k_0}(\boldsymbol{l}),1\}
    +\int^{+\infty}_{\max\{U_{k_0}(\boldsymbol{l}),1\}}\\
    &\qquad\exp\left\{\theta(x,1)\left(P_{k_0}(\boldsymbol{l})-\frac{x(1+k_0)}{1+x}\right)\right\}\dd x\Bigg).
\end{align*}
Let $u_x=\theta(x,1)/(1+x)$, then
\[
x=\frac{-\log (2-\exp\{u_x\})}{u_x},\quad \dd x=\frac{\frac{\exp\{u_x\}}{2-\exp\{u_x\}}u_x+\log(2-\exp\{u_x\})}{u_x^2}\dd u_x,
\]
Substituting this into the integral, with
\[
g(u_x)=\dfrac{\dfrac{e^{u_x}}{2-e^{u_x}}u_x+\log(2-e^{u_x})}{u_x^2},
\]
we have
\begin{align*}
    &\EE\left[\sup_{k\leq m_0}\frac{S_k}{1+k-S_k}\right]\\
    \leq&2^{-k_0}\sum_{\boldsymbol{l}\in\{0,1\}^{k_0}}\Bigg(\max\{U_{k_0}(\boldsymbol{l}),1\}
    +\int^{\log2}_{u_{\max\{U_{k_0}(\boldsymbol{l}),1\}}}\\
    &\qquad\exp\left\{(1+k_0-P_{k_0}(\boldsymbol{l}))\log(2-\exp\{u_x\})+u_xP_{k_0}(\boldsymbol{l})\right\}
      g(u_x)\,\dd u_x\Bigg).
\end{align*}
This integral can be approximately evaluated numerically. We have
\[
\EE\left[\sup_{k\leq m_0}\frac{S_k}{1+k-S_k}\right]\leq1.922.
\]
\end{proof}

\begin{proof}[Proof of Lemma~\ref{lem:pt}]

If $\sum_j\mathbf{1}\{W_j\geq T^{a,b}(\beta;\mathbf{W},\mathbf{L}),L_j=1\}=0$, because $b>\beta$, $T^{a,b}(\beta)=+\infty$ and for any $t>0$
\[
\frac{a\sum_j\mathbf{1}\{W_j\geq t,L_j=0\}+b}{\left(\sum_j\mathbf{1}\{W_j\geq t,L_j=1\}\right)\vee1}>\beta,
\]
then we have
\[
\frac{a\sum_j\mathbf{1}\{W_j\geq t,L_j=0\}+b-c}{1\vee\sum_j\mathbf{1}\{W_j\geq t,L_j=1\}}>\beta-c
\]
which implies $T^{a,b-c}(\beta-c;\mathbf{W},\mathbf{L})=+\infty$.

So we focus on the case where $\sum_j\mathbf{1}\{W_j\geq T^{a,b}(\beta;\mathbf{W},\mathbf{L}),L_j=1\}>0$. For any $t<T^{a,b}(\beta;\mathbf{W},\mathbf{L})$,
\begin{align*}
\frac{a\sum_j\mathbf{1}\{W_j\geq t,L_j=0\}+b-c}{\sum_j\mathbf{1}\{W_j\geq t,L_j=1\}}>&\beta-\frac{c}{\sum_j\mathbf{1}\{W_j\geq t,L_j=1\}}\\
\geq&\beta-\frac{c}{\sum_j\mathbf{1}\{W_j\geq T^{a,b}(\beta;\mathbf{W},\mathbf{L}),L_j=1\}}
\end{align*}
which implies that
\[
T^{a,b-c}\left(\beta-\frac{c}{\sum_j\mathbf{1}\{W_j\geq T^{a,b}(\beta;\mathbf{W},\mathbf{L}),L_j=1\}};\mathbf{W},\mathbf{L}\right)\geq T^{a,b}(\beta;\mathbf{W},\mathbf{L}).
\]
And with the definition of competition threshold, we have
\begin{align*}
&\frac{a\sum_j\mathbf{1}\{W_j\geq T^{a,b}(\beta;\mathbf{W},\mathbf{L}),L_j=0\}+b-c}{\sum_j\mathbf{1}\{W_j\geq T^{a,b}(\beta;\mathbf{W},\mathbf{L}),L_j=1\}}\\
\leq&\beta-\frac{c}{\sum_j\mathbf{1}\{W_j\geq T^{a,b}(\beta;\mathbf{W},\mathbf{L}),L_j=1\}}
\end{align*}
which implies that
\[
T^{a,b-c}\left(\beta-\frac{c}{\sum_j\mathbf{1}\{W_j\geq T^{a,b}(\beta;\mathbf{W},\mathbf{L}),L_j=1\}};\mathbf{W},\mathbf{L}\right)\leq T^{a,b}(\beta;\mathbf{W},\mathbf{L}).
\]
Similarly, we have if
\[
\sum_j\mathbf{1}\left\{W_j\geq T^{a,b-c}(\beta;\mathbf{W},\mathbf{L}),L_j=1\right\}=0,
\]
then we clearly have $T^{a,b-c}(\beta;\mathbf{W},\mathbf{L})=+\infty$ and $T^{a,b}(\beta+c;\mathbf{W},\mathbf{L})=+\infty$. So we only consider the case where
\[
\sum_j\mathbf{1}\left\{W_j\geq T^{a,b-c}(\beta;\mathbf{W},\mathbf{L}),L_j=1\right\}>0,
\]
\begin{align*}
&\frac{a\sum_j\mathbf{1}\left\{W_j\geq T^{a,b-c}(\beta;\mathbf{W},\mathbf{L}),L_j=0\right\}+b}{\sum_j\mathbf{1}\left\{W_j\geq T^{a,b-c}(\beta;\mathbf{W},\mathbf{L}),L_j=1\right\}}\\
\leq&\beta+\frac{c}{\sum_j\mathbf{1}\left\{W_j\geq T^{a,b-c}(\beta;\mathbf{W},\mathbf{L}),L_j=1\right\}},
\end{align*}
which implies that
\[
T^{a,b-c}(\beta;\mathbf{W},\mathbf{L})\geq T^{a,b}\left(\beta+\frac{c}{\sum_j\mathbf{1}\left\{W_j\geq T^{a,b-c}(\beta;\mathbf{W},\mathbf{L}),L_j=1\right\}};\mathbf{W},\mathbf{L}\right).
\]

If the label is positive, the flip-one (FO) operation has no effect, so this case is trivial. We therefore focus on the case where the label is negative.

For any $t$,
\begin{align*}
&\frac{a\sum_j\mathbf{1}\{W_j'\geq t,L_j'=0\}+b}{1\vee\sum_j\mathbf{1}\{W_j'\geq t,L_j'=1\}}\\
=&\frac{a\sum_{j\neq i}\mathbf{1}\{W_j\geq t,L_j=0\}+b}{1\vee(\sum_{j\neq i}\mathbf{1}\{W_j\geq t,L_j=1\}+\mathbf{1}\{W_i\geq t\})}\\
=&\frac{a\sum_{j}\mathbf{1}\{W_j\geq t,L_j=0\}+b-a\mathbf{1}\{W_i\geq t,L_i=0\}}{1\vee(\sum_{j}\mathbf{1}\{W_j\geq t,L_j=1\}+\mathbf{1}\{W_i\geq t,L_i=0\})}
\end{align*}
Let $T_i=T_i^{a,b}(\beta;\mathbf{W},\mathbf{L})$, if $\mathbf{1}\{W_i\geq T_i,L_i=0\}=1$, we have
\begin{align*}
\frac{a\sum_j\mathbf{1}\{W_j'\geq T_i,L_j'=0\}+b}{1\vee\sum_j\mathbf{1}\{W_j'\geq T_i,L_j'=1\}}=\frac{a\sum_{j}\mathbf{1}\{W_j\geq T_i,L_j=0\}+b-a}{1+\sum_{j}\mathbf{1}\{W_j\geq T_i,L_j=1\}}\leq\beta
\end{align*}
and $\sum_{j}\mathbf{1}\{W_j\geq T_i,L_j=1\}+1>2$, so that
\begin{align*}
&\frac{a\sum_{j}\mathbf{1}\{W_j\geq T_i,L_j=0\}+b-a}{\sum_{j}\mathbf{1}\{W_j\geq T_i,L_j=1\}+1}-\frac{a\sum_{j}\mathbf{1}\{W_j\geq T_i,L_j=0\}+b-a-\beta}{\sum_{j}\mathbf{1}\{W_j\geq T_i,L_j=1\}}\\
=&-\frac{1}{\sum_{j}\mathbf{1}\{W_j\geq T_i,L_j=1\}}\frac{a\sum_{j}\mathbf{1}\{W_j\geq T_i,L_j=0\}+b-a}{\sum_{j}\mathbf{1}\{W_j\geq T_i,L_j=1\}+1}\\
&\quad+\frac{\beta}{\sum_{j}\mathbf{1}\{W_j\geq T_i,L_j=1\}}\\
\geq&0
\end{align*}
which implies that
\[
T^{a,b-a-\beta}(\beta;\mathbf{W},\mathbf{L})\leq T_i=T_i^{a,b}(\beta;\mathbf{W},\mathbf{L}).
\]
For any $t<T_i$, $\mathbf{1}\{W_i\geq t,L_i=0\}\geq\mathbf{1}\{W_i\geq T_i,L_i=0\}=1$,
\begin{align*}
\frac{a\sum_j\mathbf{1}\{W_j'\geq t,L_j'=0\}+b}{1\vee\sum_j\mathbf{1}\{W_j'\geq t,L_j'=1\}}=\frac{a\sum_{j}\mathbf{1}\{W_j\geq t,L_j=0\}+b-a}{\sum_{j}\mathbf{1}\{W_j\geq t,L_j=1\}+1}>\beta
\end{align*}
and so $\sum_{j}\mathbf{1}\{W_j\geq t,L_j=1\}+1\geq\sum_{j}\mathbf{1}\{W_j\geq T_i,L_j=1\}+1>2$
\begin{align*}
&\frac{a\sum_{j}\mathbf{1}\{W_j\geq t,L_j=0\}+b-a}{\sum_{j}\mathbf{1}\{W_j\geq t,L_j=1\}+1}-\frac{a\sum_{j}\mathbf{1}\{W_j\geq t,L_j=0\}+b-a-\beta}{\sum_{j}\mathbf{1}\{W_j\geq t,L_j=1\}}\\
=&-\frac{1}{\sum_{j}\mathbf{1}\{W_j\geq t,L_j=1\}}\frac{a\sum_{j}\mathbf{1}\{W_j\geq t,L_j=0\}+b-a}{\sum_{j}\mathbf{1}\{W_j\geq t,L_j=1\}+1}\\
&\quad+\frac{\beta}{\sum_{j}\mathbf{1}\{W_j\geq t,L_j=1\}}\\
<&0
\end{align*}
which implies that
\[
T^{a,b-a-\beta}(\beta;\mathbf{W},\mathbf{L})\geq T_i=T_i^{a,b}(\beta;\mathbf{W},\mathbf{L}).
\]
This completes the proof of the lemma.

\end{proof}

\begin{proof}[Proof of Theorem~\ref{thm:aGCT}]

Let $\tilde{r}_i=r_i\min\{1,R_+(T^{r_i,r_i}(\alpha);\mathbf{W}_i,\mathbf{L}_i)\}$ and let $\varepsilon>0$ be sufficiently small. Then we have
\begin{align*}
    \mathrm{FDR}=\EE\Bigg[&\frac{\max_{i}\tilde{r}_i+\sum_{i=1}^dr_iR_-(T^{r_i,r_i}(\alpha);\mathbf{W}_i,\mathbf{L}_i)}{1\vee\sum_{i=1}^dR_+(T^{r_i,r_i}(\alpha);\mathbf{W}_i,\mathbf{L}_i)}\cdot\\
    &\quad\frac{\sum_{i=1}^dV_+(T^{r_i,r_i}(\alpha);\mathbf{W}_i,\mathbf{L}_i,\mathcal{H}_0^i)}{\varepsilon\vee(\max_{i}\tilde{r}_i+\sum_{i=1}^dr_iR_-(T^{r_i,r_i}(\alpha);\mathbf{W}_i,\mathbf{L}_i))}\Bigg].
\end{align*}

Because
\begin{align*}
&\frac{\max_{i}\tilde{r}_i+\sum_{i=1}^dr_iR_-(T^{r_i,r_i}(\alpha);\mathbf{W}_i,\mathbf{L}_i)}{1\vee\sum_{i=1}^dR_+(T^{r_i,r_i}(\alpha);\mathbf{W}_i,\mathbf{L}_i)}\\
\leq&\max_{i\in[d]}\frac{r_i+r_iR_-(T^{r_i,r_i}(\alpha);\mathbf{W}_i,\mathbf{L}_i)}{1\vee R_+(T^{r_i,r_i}(\alpha);\mathbf{W}_i,\mathbf{L}_i)}\mathbf{1}_{R_+(T^{r_i,r_i}(\alpha);\mathbf{W}_i,\mathbf{L}_i)>0}\leq\alpha,
\end{align*}
then
\begin{align*}
    &\mathrm{FDR}\leq\alpha\EE\left[\frac{\sum_{i=1}^dV_+(T^{r_i,r_i}(\alpha);\mathbf{W}_i,\mathbf{L}_i,\mathcal{H}_0^i)}{\varepsilon\vee(\max_{k\in[d]}\tilde{r}_k+\sum_{k=1}^dr_kR_-(T^{r_k,r_k}(\alpha);\mathbf{W}_k,\mathbf{L}_k))}\right]\\
    =&\alpha\sum_{i=1}^d\sum_{j\in\mathcal{H}_0^i}\EE\left[\frac{\mathbf{1}\{W_{ij}\geq T^{r_i,r_i}(\alpha;\mathbf{W}_i,\mathbf{L}_i),L_{ij}=1\}}{\max_{k\in[d]}\tilde{r}_k+ \sum_{k=1}^dr_kR_-(T^{r_k,r_k}(\alpha);\mathbf{W}_k,\mathbf{L}_k)}\right]
\end{align*}
with Lemma~\ref{lem:pt}, we have
\begin{align*}
    &\EE\left[\frac{\mathbf{1}\{W_{ij}\geq T^{r_i,r_i}(\alpha;\mathbf{W}_i,\mathbf{L}_i),L_{ij}=1\}}{\max_{k\in[d]}\tilde{r}_k+ \sum_{k=1}^dr_kR_-(T^{r_k,r_k}(\alpha);\mathbf{W}_k,\mathbf{L}_k)}\right]\\
    \leq&\EE\left[\frac{\mathbf{1}\{W_{ij}\geq T^{r_i,r_i}(\alpha;\mathbf{W}_i,\mathbf{L}_i),L_{ij}=1\}}{r_i+ \sum_{k=1}^dr_kR_-(T^{r_k,r_k}(\alpha);\mathbf{W}_k,\mathbf{L}_k)}\right]\\
    =&\EE\left[\EE\left[\frac{\mathbf{1}\{W_{ij}\geq T_j^{r_i,r_i}(\alpha;\mathbf{W}_i,\mathbf{L}_i),L_{ij}=1\}}{r_i+ \sum_{k=1}^dr_kR_-(T_j^{r_k,r_k}(\alpha);\mathbf{W}_k,\mathbf{L}_k)}\mid\mathbf{W},\mathbf{L}_{-i},\mathbf{L}_{i,-j}\right]\right]\\
    =&\EE\left[\frac{\mathbf{1}\{W_{ij}\geq T_j^{r_i,r_i}(\alpha;\mathbf{W}_i,\mathbf{L}_i)\}\PP\{L_{ij}=1\mid\mathbf{W},\mathbf{L}_{-i},\mathbf{L}_{i,-j}\}}{r_i+ \sum_{k=1}^dr_kR_-(T_j^{r_k,r_k}(\alpha);\mathbf{W}_k,\mathbf{L}_k)}\right]\\
    =&\frac{r_i}{1+r_i}\left(\EE\left[\frac{\mathbf{1}\{W_{ij}\geq T_j^{r_i,r_i}(\alpha;\mathbf{W}_i,\mathbf{L}_i),L_{ij}=1\}}{r_i+ \sum_{k=1}^dR_-(T_j^{r_k,r_k}(\alpha);\mathbf{W}_k,\mathbf{L}_k)}\right]\right.\\
    &~~\left.+\EE\left[\frac{\mathbf{1}\{W_{ij}\geq T_j^{r_i,r_i}(\alpha;\mathbf{W}_i,\mathbf{L}_i),L_{ij}=0\}}{r_i+ \sum_{k=1}^dr_kR_-(T_j^{r_k,r_k}(\alpha);\mathbf{W}_k,\mathbf{L}_k)}\right]\right)\\
    =&\frac{r_i}{1+r_i}\EE\left[\frac{\mathbf{1}\{W_{ij}\geq T^{r_i,r_i}(\alpha;\mathbf{W}_i,\mathbf{L}_i),L_{ij}=1\}}{r_i+ \sum_{k=1}^dr_kR_-(T^{r_k,r_k}(\alpha);\mathbf{W}_k,\mathbf{L}_k)}\right]\\
    &~~+\frac{1}{1+r_i}\EE\left[\frac{r_i\mathbf{1}\{W_{ij}\geq T^{r_i,-\alpha}(\alpha;\mathbf{W}_i,\mathbf{L}_i),L_{ij}=0\}}{\varepsilon\vee \sum_{k=1}^dr_kR_-(T^{r_k,r_k-(r_k+\alpha)\mathbf{1}\{k=i\}}(\alpha);\mathbf{W}_k,\mathbf{L}_k)}\right]\\
\end{align*}
Therefore, the inequality is proved.

\end{proof}

\begin{proof}[Proof of Theorem~\ref{thm:medcondi}]

By Slutsky's theorem, we have for any $i$, when $m_i\to+\infty$
\[
\sup_{t\in(t^L,t^U)}\left|\frac{\sum_{j=1}^{m_i}\mathbf{1}\left\{W_{ij}\geq t,L_{ij}=0\right\}}{\sum_{j=1}^{m_i}\mathbf{1}\left\{W_{ij}\geq t,L_{ij}=1\right\}}-G^i(t)\right|\overset{p}{\longrightarrow}0,
\]
First, we use a lemma to show that for any $t\in(t^L,t^U)$, the size of the rejection set $R^i_+(t)$ is large enough.

\begin{Lemma}
    \label{slem:largeenoughrejection}
    If the positive-integer-valued random variables $X_n(t),Y_n(t)$, $t\in(t^L,t^U)$, $n\in\mathbb{Z}_+$, satisfy that $\left\{\left(X_n(t),Y_n(t)\right)_t\right\}$ is independent of $n$ and
    \[
    X_n(t_1)\leq X_n(t_2),Y_n(t_1)\leq Y_n(t_2),\forall t_1>t_2,
    \]
    for any $\varepsilon>0$
    \[
    \lim_{n\to+\infty}\PP\left\{\sup_{t\in(t^L,t^U)}\left|\frac{X_n(t)}{Y_n(t)}-G(t)\right|>\varepsilon\right\}=0,
    \]
    where $d^U,d^L$ are constants such that
    \[
    -d^U\leq\inf_{t\in(t^L,t^U)}\frac{\partial G}{\partial t}\leq\sup_{t\in(t^L,t^U)}\frac{\partial G}{\partial t}\leq -d^L<0,
    \]
    we have for any $M>0$ and $\delta_t\in\left(0,\frac{t^U-t^L}{2}\right)$,
    \[
    \lim_{n\to+\infty}\PP\left\{\sup_{t\in(t^L+\delta_t,t^U-\delta_t)}\left|\frac{1}{Y_n(t)}\right|>\frac{1}{M}\right\}=0.
    \]
\end{Lemma}

The condition of this lemma is satisfied, so the number of rejections is sufficiently large. Following the preceding analysis, we state the following lemma: bounded perturbations of the adjustment parameters of the parameterized threshold are negligible in this case.

\begin{Lemma}
    \label{slem:negligibleadjpara}
    With the same assumptions of Lemma~\ref{slem:largeenoughrejection}, for any given $c$ and any $\varepsilon>0,\delta_t\in\left(0,\frac{t^U-t^L}{2}\right)$,
    \[
    \lim_{n\to+\infty}\PP\left\{\sup_{t\in(t^L+\delta_t,t^U-\delta_t)}\left|\frac{X_n(t)+c}{Y_n(t)}-G(t)\right|>\varepsilon\right\}=0.
    \]
\end{Lemma}

With the assumption and Lemma~\ref{slem:negligibleadjpara}, it follows that all thresholds with finite parameters are close to the unique limit.

\begin{Lemma}
    \label{slem:uniquelimit}
    With the same assumptions of Lemma~\ref{slem:largeenoughrejection} and given $\alpha\in(0,1)$, if there is $t_{\alpha}\in(t^L,t^U)$ such that $G(t_{\alpha})=\alpha$, let
    \[
    T_n^c=\inf_{t\in(t^L,t^U)}\left\{t:\frac{X_n(t)+c}{Y_n(t)}\leq\alpha\right\},
    \]
    where $c$ is a constant and $\sup_{t\in(t^L,t^U)}\emptyset=t^U$, we have for any given $c>0$ and any $0<\varepsilon<\min\{t^U-t_{\alpha},t_{\alpha}-t^L\}$,
    \[
    \lim_{n\to+\infty}\PP\left\{|T_n^c-t_{\alpha}|>\varepsilon\right\}=0.
    \]
\end{Lemma}

Then, using techniques such as the triangle inequality, we obtain the following result: the sizes of the rejection sets obtained with different threshold parameters are asymptotically equal.

\begin{Lemma}
    \label{slem:asyfraction}
    With the same assumptions and definitions of Lemma~\ref{slem:uniquelimit} and there is $\delta>0$ such that
    \[
    \lim_{n\to+\infty}\PP\left\{\inf_{t\leq t^L}\frac{X_n(t)}{Y_n(t)}-\alpha>\delta\right\}=1,
    \]
    and for any $\varepsilon>0$, there exists $\delta_{\varepsilon}$
    \[
    \lim_{n\to+\infty}\PP\left\{\sup_{\substack{t_1,t_2\in(t^L,t^U)\\|t_1-t_2|\leq\delta_{\varepsilon}}}\left|\frac{Y_n(t_1)}{Y_n(t_2)}-1\right|>\varepsilon\right\}=0.
    \]
    then for any $c_1,c_2$ and $\varepsilon>0$,
    \[
    \lim_{n\to+\infty}\PP\left\{\left|\frac{Y_n(T_n^{c_1})}{Y_n(T_n^{c_2})}-1\right|>\varepsilon\right\}=0.
    \]
\end{Lemma}

Finally, with Lemma~\ref{slem:asyfraction} and the continuity of $G^i_-(t)$, we have that when $m_i\to+\infty$,
\[
\left|\frac{R_-\left(T^{r_i,r_i}\left(\alpha;\mathbf{W}_i,\mathbf{L}_i\right)\right)}{1\vee R_-\left(T^{r_i,-\alpha}\left(\alpha;\mathbf{W}_i,\mathbf{L}_i\right)\right)}-1\right|\overset{p}{\longrightarrow}0
\]
and so
\[
\left|\frac{V_-\left(T^{r_i,-\alpha}\left(\alpha;\mathbf{W}_i,\mathbf{L}_i\right)\right)-V_-\left(T^{r_i,r_i}\left(\alpha;\mathbf{W}_i,\mathbf{L}_i\right)\right)}{1\vee R_-\left(T^{r_i,-\alpha}\left(\alpha;\mathbf{W}_i,\mathbf{L}_i\right)\right)}\right|\overset{p}{\longrightarrow}0
\]
which implies that the second condition in Lemma~\ref{pro:ac} is satisfied, and the second condition is a sufficient condition for the first one.

\end{proof}

\begin{proof}[Proof of Theorem~\ref{thm:valid_ggt}, Theorem~\ref{thm:ggt} and Corollary~\ref{coro:ggt}]

First, we verify that the consistent threshold is well-defined. Suppose $(t_1, \dots, t_d)$ satisfies
\[
\sup_{\{i\in[d]:R^i_+(t_i)>0\}}\frac{r_i + r_i R^i_-(t_i)}{1 \vee \sum_{i=1}^d R^i_+(t_i)} \bigg/
\frac{m_i}{\sum_{k=1}^d m_k} \leq \alpha.
\]
Then for any $j$, if $t'_j \leq t_j$ satisfies
\[
\frac{r_j + r_j R^j_-(t'_j)}{m_j} \leq \sup_{\{i\in[d]:R^i_+(t_i)>0\}}\frac{r_i + r_i R^i_-(t_i)}{m_i},
\]
with the monotonicity of $R^j_+$ which implies $R^j_+(t'_j) \geq R^j_+(t_j)$, we have
\[
\sum_{i\neq j}^d R^i_+(t_i) + R^j_+(t_j)
\geq \sum_{i\neq j}^d R^i_+(t_i) + R^j_+(t'_j),
\]
and $(t_1,t_2,\cdots,t_{j-1},t'_j,t_{j+1},\cdots,t_d)$ satisfies the inequality constraint.

For given $\mathbf{W},\mathbf{L}$, let
\[
f(\boldsymbol{t})=f(\boldsymbol{t};\mathbf{W},\mathbf{L})=\sup_{\{i \in [d]:R^i_+(t_i) > 0\}}\frac{r_i + r_i R^i_-(t_i)}{1 \vee \sum_{i=1}^d R^i_+(t_i)} \bigg/
\frac{m_i}{\sum_{k=1}^d m_k}.
\]
From the above analysis, we obtain that for any vector $\boldsymbol{t}$, there exists a vector $\boldsymbol{t}'$ satisfying
\[
f(\boldsymbol{t})=f(\boldsymbol{t}'),\quad \boldsymbol{t}'\leq\boldsymbol{t},\quad \sum_{i=1}^d\mathbf{1}\{t_i=t_j\}\geq1.
\]
Moreover, for any $\boldsymbol{t}''$ such that
\[
f(\boldsymbol{t}'')=f(\boldsymbol{t}')
\]
we always have $\boldsymbol{t}''\geq\boldsymbol{t}'$. Hence we can define the equivalence class
\[
\mathcal{E}(\boldsymbol{t})=\{\boldsymbol{t}''\in\mathcal{A}_{\mathbf{W},\mathbf{L}} : f(\boldsymbol{t})=f(\boldsymbol{t}')\}.
\]
Although this equivalence class is not totally ordered, it possesses a unique minimal element, namely $\boldsymbol{t}'=\inf\mathcal{E}(\boldsymbol{t})$, which implies $\boldsymbol{t}'\leq\boldsymbol{t}'',\forall\boldsymbol{t}''\in\mathcal{E}(\boldsymbol{t})$. In the following, it suffices to show that $\{\mathcal{E}(\boldsymbol{t})\}$ forms a disjoint cover of the space $\mathcal{A}_{\mathbf{W},\mathbf{L}}$, which is also evident. At this point, the set $\{\inf\mathcal{E}(\boldsymbol{t}):\boldsymbol{t}\in\mathcal{A}_{\mathbf{W},\mathbf{L}}\}$ is totally ordered, and the final consistent threshold vector must take a value in this set. In summary, Theorem~\ref{thm:valid_ggt} is proved.

Write $A=\{j\in[d]:R^j_+(T^{\boldsymbol{r}}_j)>0\}$ for the set of groups with at least one rejection. Given Theorem~\ref{thm:valid_ggt}, if the group competition statistics have basic independence, then
\begin{align*}
    \mathrm{FDR}&= \EE\left[
    \frac{\sum_{i=1}^d V^i_+(T_i^{\boldsymbol{r}}) }
    { 1 \vee \sum_{i=1}^d R^i_+(T_i^{\boldsymbol{r}}) }
    \right] \\
    &= \EE\left[ \sup_{j\in A}\frac{\frac{r_j+r_jR_-^j(T^{\boldsymbol{r}}_j)}{1\vee\sum_{k=1}^dR^k_+(T^{\boldsymbol{r}}_k)}}{\frac{m_j}{\sum_{k=1}^dm_k}}\cdot
    \frac{\sum_{i=1}^d V^i_+(T_i^{\boldsymbol{r}}) }
    {\sup_{j\in A}\frac{r_j + r_j R^j_-(T_j^{\boldsymbol{r}})}{m_j}\cdot \sum_{k=1}^d m_k} \right] \\
    &\leq \alpha \sum_{i=1}^d \EE\left[
    \frac{V^i_+(T_i^{\boldsymbol{r}}) }
    {\sup_{j\in A}\frac{r_j + r_j R^j_-(T_j^{\boldsymbol{r}})}{m_j} \cdot \sum_{k=1}^d m_k}
    \right] \\
    &\leq \alpha \sum_{i=1}^d \frac{m_i}{\sum_{k=1}^d m_k} \,
    \EE\left[ \frac{V^i_+(T_i^{\boldsymbol{r}}) }{r_i + r_i R^i_-(T_i^{\boldsymbol{r}})} \right].
\end{align*}
Under the basic independence of the group competition statistics, we have
\[
\EE\left[ \frac{V^i_+(T_i^{\boldsymbol{r}}) }{r_i + r_i R^i_-(T_i^{\boldsymbol{r}})} \right]\leq1.
\]
Indeed, under basic independence the expectation can be evaluated as an iterated expectation. Conditioning first on the data of the other groups,
\[
\EE\left[ \frac{V^i_+(T_i^{\boldsymbol{r}}) }{r_i + r_i R^i_-(T_i^{\boldsymbol{r}})} \right]=\EE\left[\EE\left[ \frac{V^i_+(T_i^{\boldsymbol{r}}) }{r_i + r_i R^i_-(T_i^{\boldsymbol{r}})}\bigg|\mathbf W_{-i},\mathbf L_{-i}\right]\right],
\]
the threshold $T^{\boldsymbol{r}}_i$ is a stopping time with respect to the reversed filtration generated by the $i$-th group, and the conditional probability equality of Definition~\ref{def:gcs} continues to hold on that filtration. Therefore the argument in the proof of Theorem~\ref{thm:CT}, which is stated for a general stopping time, applies verbatim and bounds the conditional expectation by $1$; taking expectations over $(\mathbf W_{-i},\mathbf L_{-i})$ gives the claim.

If the group competition statistics do not have basic independence, then
\begin{align*}
    \mathrm{FDR}&= \EE\left[
    \frac{\sum_{i=1}^d V^i_+(T_i^{\boldsymbol{r}}) }
    { 1 \vee \sum_{i=1}^d R^i_+(T_i^{\boldsymbol{r}}) }
    \right] \\
    &= \EE\left[ \sup_{j\in A}\frac{\frac{c(r_j)r_j+c(r_j)r_jR_-^j(T^{\boldsymbol{r}}_j)}{1\vee\sum_{k=1}^dR^k_+(T^{\boldsymbol{r}}_k)}}{\frac{m_j}{\sum_{k=1}^dm_k}}\cdot
    \frac{\sum_{i=1}^d V^i_+(T_i^{\boldsymbol{r}}) }
    {\sup_{j\in A}\frac{c(r_j)r_j + c(r_j)r_j R^j_-(T_j^{\boldsymbol{r}})}{m_j}\cdot \sum_{k=1}^d m_k} \right] \\
    &\leq \alpha \sum_{i=1}^d \EE\left[
    \frac{V^i_+(T_i^{\boldsymbol{r}}) }
    {\sup_{j\in A}\frac{c(r_j)r_j + c(r_j)r_j R^j_-(T_j^{\boldsymbol{r}})}{m_j} \cdot \sum_{k=1}^d m_k}
    \right] \\
    &\leq \alpha \sum_{i=1}^d \frac{m_i}{\sum_{k=1}^d m_k} \,
    \EE\left[ \frac{V^i_+(T_i^{\boldsymbol{r}}) }{c(r_i)r_i + c(r_i)r_i R^i_-(T_i^{\boldsymbol{r}})} \right].
\end{align*}
and, with Theorem~\ref{thm:ft_gct},
\[
\EE\left[ \frac{V^i_+(T_i^{\boldsymbol{r}}) }{c(r_i)r_i + c(r_i)r_i R^i_-(T_i^{\boldsymbol{r}})} \right]\leq\frac{1}{c(r_i)}\EE\left[\sup_t\frac{V^i_+(t) }{r_i + r_i R^i_-(t)} \right]\leq 1.
\]
Here the argument follows the proof of Theorem~\ref{thm:GCT}: under the marginal condition, the statistics of the $i$-th group form a local competition statistic, so the conditional probability equality of Definition~\ref{def:gcs} continues to hold within the group; and the local threshold $T^{\boldsymbol r}_i$, whose dependence on the other groups would otherwise be difficult to handle, is eliminated by the supremum over $t$ taken as in the proof of Theorem~\ref{thm:ggt}, which requires no independence across groups.

In summary, Theorem~\ref{thm:ggt} and Corollary~\ref{coro:ggt} are proved.

\end{proof}

\subsection{Group competition filter with function-parameterized threshold}

We discuss further applications of the parameterized threshold in the competition test. The main idea for FDR control in the competition test is that the ratio of the probabilities for the competition label being $0$ to $1$ is constant and independent of the competition score. Therefore, the key to constructing a practical competition test lies in constructing labels that satisfy this condition. Methods such as knockoff and the Target-Decoy (TD) approach achieve this through a ``max-taking'' operation, which renders the competition labels and competition scores (conditionally) independent. Using the parameterized formulation, we show that this condition can potentially be relaxed, allowing the probability ratio to be a function of the score. We begin with the following definition.

\begin{Def}
    \label{def:fcs}
    The statistics $(\mathbf{W},\mathbf{L})\in\mathbb{R}^m\times\{0,1\}^m$ are functional competition statistics with symmetry function $r(\cdot)$ with respect to $\mathcal{N}\subseteq[m]$ if for any $j\in\mathcal{N}$
    \[
    \PP\{L_j=1 \mid \mathbf{W},\mathbf{L}_{-j}\}=r(W_j)\,\PP\{L_j=0 \mid \mathbf{W},\mathbf{L}_{-j}\}.
    \]
\end{Def}

In this setting, we generalize the constant symmetry parameter $r$ in the competition statistic to a function $r(\cdot)$. Following the definition of the functional competition statistic, we define the corresponding threshold and filter.

\begin{Def}
    \label{def:fpt}
    For the statistics $(\mathbf{W},\mathbf{L})$, the function-parameterized threshold is the solution to the following optimization problem with the functional parameter $a(\cdot)$ and the level parameter $q$:
    \begin{align*}
        &T^{a(\cdot)}(q)= \arg\max_{t} R_+(t)\\
        \mathrm{s.t.}&\quad\frac{\sum_{j=1}^ma(W_j)\mathbf{1}\{W_j\geq t,L_j=0\}+\max_{W_i\geq t}a(W_i)}{1\vee\sum_{j=1}^m\mathbf{1}\{W_j\geq t,L_j=1\}}\leq q\quad\text{or}\quad t=+\infty,\\
        &\quad t\in\{W_{i}\}\cup\{+\infty\}.
    \end{align*}
    Construct the rejection set
    \[
    \mathcal{R}_{q,a(\cdot)}(\mathbf{W},\mathbf{L})=\{j:W_j\geq T^{a(\cdot)}(q),L_j=1\},
    \]
    then $\mathcal{R}_{q,a(\cdot)}$ is a competition-like filter with functional parameter $a(\cdot)$ and level parameter $q$.
\end{Def}

As noted after Definition~\ref{def:ct}, we take the smallest admissible maximizer, and in the computation the candidates are restricted to the scores of the hypotheses with $L_i=1$.

Note that the function-parameterized threshold makes both the scale and adjustment of the threshold dependent on the functional parameter. When the functional parameter $r(\cdot)$ of the functional competition statistic satisfies certain regularity conditions, the proposed filter controls the FDR at the target level. We present the following theorem.

\begin{Thm}
    \label{thm:fcs}
    Suppose the competition statistics $(\mathbf{W},\mathbf{L})\in\RR^m\times\{0,1\}^m$ with respect to $\mathcal{H}_0$ have symmetry function $r(\cdot)$. Then the competition-like filter $\mathcal{R}_{\alpha,r(\cdot)}(\mathbf{W},\mathbf{L})$ controls the FDR such that
    \[
    \mathrm{FDR}=\EE\left[\frac{|\mathcal{H}_0\cap\mathcal{R}_{\alpha,r(\cdot)}(\mathbf{W},\mathbf{L})|}{1\vee|\mathcal{R}_{\alpha,r(\cdot)}(\mathbf{W},\mathbf{L})|}\right]\leq\alpha.
    \]
\end{Thm}

\begin{proof}[Proof of Theorem~\ref{thm:fcs}]

We begin with two lemmas similar to Lemma~\ref{lem:pt}. For the function-parameterized threshold, we similarly define its FO form $T^{a(\cdot)}_i(\beta;\mathbf{W},\mathbf{L})$.

\begin{Lemma}
    \label{slem:fpt1}
    Let $\mathcal{R}_-=\left\{j\in[m]:W_j\geq T^{r(\cdot)},L_j=0\right\}$. Then we have
    \begin{enumerate}
        \item $T^{r(\cdot)}_{(0:\mathcal{R}_-;\mathbf{W})}\geq T^{r(\cdot)}_{(1:\mathcal{R}_-;\mathbf{W})}\geq\cdots\geq T^{r(\cdot)}_{(|\mathcal{R}_-|:\mathcal{R}_-;\mathbf{W})}$, where $T^{r(\cdot)}_{(j:\mathcal{R}_-;\mathbf{W})}$ is the threshold obtained by flipping the label of the hypothesis with the $j$-th smallest score in the set $\mathcal{R}_-$ and $T^{r(\cdot)}_{(0:\mathcal{R}_-;\mathbf{W})}=T^{r(\cdot)}$.
        \item $\mathbf{1}\left\{W_k\geq T^{r(\cdot)}_{(|\mathcal{R}_-|:\mathcal{R}_-;\mathbf{W})},L_k=0\right\}=1\Rightarrow T^{r(\cdot)}_k\geq T^{r(\cdot)}_{(|\mathcal{R}_-|:\mathcal{R}_-;\mathbf{W})}$.
        \item $\mathbf{1}\left\{W_k\geq T^{r(\cdot)}_{(1:\mathcal{R}_-;\mathbf{W})},L_k=0\right\}=0\Rightarrow T^{r(\cdot)}_k=T^{r(\cdot)}$.
        \item $T^{r(\cdot)}_k<T^{r(\cdot)}\Rightarrow\mathbf{1}\left\{W_k\geq T^{r(\cdot)}_{(1:\mathcal{R}_-;\mathbf{W})},L_k=0\right\}=1$
    \end{enumerate}
\end{Lemma}

\begin{Lemma}
    \label{slem:fpt2}
    Let $\mathcal{R}^i_-=\left\{j\in[m]:W_j\geq T^{r(\cdot)}_i,L_j=0\right\}$. Then we have
    \begin{enumerate}
        \item $T^{r(\cdot)}_{(0:\mathcal{R}^i_-;\mathbf{W})}\geq T^{r(\cdot)}_{(1:\mathcal{R}^i_-;\mathbf{W})}\geq\cdots\geq T^{r(\cdot)}_{(|\mathcal{R}^i_-|:\mathcal{R}^i_-;\mathbf{W})}$, where $T^{r(\cdot)}_{(j:\mathcal{R}_-^i;\mathbf{W})}$ is the threshold obtained by flipping the label of the hypothesis with the $j$-th smallest score in the set $\mathcal{R}_-^i$ and $T^{r(\cdot)}_{(0:\mathcal{R}^i_-;\mathbf{W})}=T^{r(\cdot)}$.
        \item $\mathbf{1}\left\{W_k\geq T^{r(\cdot)}_{(1:\mathcal{R}^i_-;\mathbf{W})},L_k=0\right\}=0\Rightarrow T^{r(\cdot)}_k=T^{r(\cdot)}$.
        \item $T^{r(\cdot)}_k<T^{r(\cdot)}\Rightarrow \mathbf{1}\left\{W_k\geq T^{r(\cdot)}_{(1:\mathcal{R}^i_-;\mathbf{W})},L_k=0\right\}=1$.
    \end{enumerate}
\end{Lemma}

With such lemmas and the same analysis, we have
\begin{align*}
    &\mathrm{FDR} = \sum_{j\in\mathcal{H}_0}\EE\left[\frac{\mathbf{1}\left\{W_j\geq T^{r(\cdot)},L_j=1\right\}}{1\vee R_+\left(T^{r(\cdot)}\right)}\right]\\
    \leq& \alpha\sum_{j\in\mathcal{H}_0}\EE\left[\frac{\mathbf{1}\left\{W_j\geq T^{r(\cdot)},L_j=1\right\}}{\sum_ir(W_i)\mathbf{1}\left\{W_i\geq T^{r(\cdot)},L_i=0\right\}+\min_k\{r(W_k):W_k\geq T^{r(\cdot)}\}}\right]\\
    =& \alpha\sum_{j\in\mathcal{H}_0}\EE\left[\frac{\mathbf{1}\left\{W_j\geq T_j^{r(\cdot)},L_j=1\right\}}{\left(\sum_ir(W_i)\mathbf{1}\left\{W_i\geq T_j^{r(\cdot)},L_i=0\right\}\right)+\min_k\{r(W_k):W_k\geq T_j^{r(\cdot)}\}}\right]\\
    \leq& \alpha\sum_{j\in\mathcal{H}_0}\EE\left[\frac{r(W_j)\mathbf{1}\left\{W_j\geq T_j^{r(\cdot)},L_j=0\right\}}{\left(\sum_ir(W_i)\mathbf{1}\left\{W_i\geq T_j^{r(\cdot)},L_i=0\right\}\right)\vee \min_k{r(W_k)}}\right]\\
    \leq& \alpha\sum_{j\in\mathcal{H}_0}\EE\left[\frac{r(W_j)\mathbf{1}\left\{j\in \mathcal{R}_-^j\right\}}{\left(\sum_ir(W_i)\mathbf{1}\left\{i\in \mathcal{R}_-^j\right\}\right)\vee \min_k{r(W_k)}}\right]\\
    \leq& \alpha
\end{align*}
In summary, Theorem~\ref{thm:fcs} is proved.

\end{proof}

\section{Proofs of Auxiliary Lemmas}
\label{ssec:lem}

\begin{proof}[Proof of Lemma~\ref{slem:deltacontrol}]

First, for a fixed threshold $t_i$, we have the conditional probability
\begin{align*}
    &\PP\left\{\delta_i(t_i)=k_3,V^i_+(t_i)=k_1,V_-^i(t_i)=k_2\mid \mathbf{W}_i\right\}\\
    =&\sum_{\boldsymbol{l}\in\Omega_{k_1,k_2,k_3}}\sum_{j_1,\dots,j_{k_1+k_2}\in\mathcal{H}_0^i}\PP\left\{L_{ij_s}=l_s;s=1,\dots,k_1+k_2\mid\mathbf{W}_i\right\}\mathbf{1}_{B(t_i,k_1+k_2;\mathbf{W}_i)}
\end{align*}
where
\[
\Omega_{k_1,k_2,k_3}=\left\{l\in\{0,1\}^{k_1+k_2}:\sum_{s=1}^{k_1+k_2}l_s=k_1,\sum_{s=1}^{k_3}l_s=k_3,l_{k_3+1}=0\right\},
\]
and
\[
B(t_i,k_1+k_2;\mathbf{W}_i)=\left\{\sum_{j\in\mathcal{H}_0^i}\mathbf{1}\{W_{ij}\geq t_i\}=k_1+k_2\right\}.
\]
This equality holds because, under the condition of group competition statistics, the statistics of each local group are marginal competition statistics. Similarly, we have
\begin{align*}
    &\PP\left\{V^i_+(t_i)=k_1,V_-^i(t_i)=k_2\mid \mathbf{W}_i\right\}\\
    =&\sum_{k_3=0}^{k_1}\sum_{\boldsymbol{l}\in\Omega_{k_1,k_2,k_3}}\sum_{j_1,\dots,j_{k_1+k_2}\in\mathcal{H}_0^i}\PP\left\{L_{ij_s}=l_s;s=1,\dots,k_1+k_2\mid\mathbf{W}_i\right\}\mathbf{1}_{B(t_i,k_1+k_2;\mathbf{W}_i)}
\end{align*}
This equality holds because the sets $\Omega_{k_1,k_2,k_3}$ are disjoint with respect to $k_3$ and this conditional probability on the label is independent of $k_3$, in the sense that for all $\boldsymbol{l}\in\Omega_{k_1,k_2,k_3}$ and $j_1,\dots,j_{k_1+k_2}\in\mathcal{H}^i_0$,
\[
\PP\left\{L_{ij_s}=l_s;s=1,\dots,k_1+k_2\mid\mathbf{W}_i\right\}=\left(\frac{r_i}{1+r_i}\right)^{k_1}\left(\frac{1}{1+r_i}\right)^{k_2}
\]
So we have, with $S_{k_1,k_2,k_3}=\sum_{\boldsymbol{l}\in\Omega_{k_1,k_2,k_3}}\sum_{j_1,\dots,j_{k_1+k_2}\in\mathcal{H}_0^i}\mathbf{1}_{B(t_i,k_1+k_2;\mathbf{W}_i)}$,
\begin{align*}
    &\PP\left\{\delta_i(t_i)=k_3\mid V_+^i(t_i)=k_1,V_-^i(t_i)=k_2,\mathbf{W}_i\right\}\\
    =&\frac{\PP\left\{\delta_i(t_i)=k_3,V_+^i(t_i)=k_1,V_-^i(t_i)=k_2\mid \mathbf{W}_i\right\}}{\PP\left\{V_+^i(t_i)=k_1,V_-^i(t_i)=k_2\mid \mathbf{W}_i\right\}}\\
    =&\frac{S_{k_1,k_2,k_3}}{\sum_{k_3=0}^{k_1}S_{k_1,k_2,k_3}}
      =\frac{\binom{k_1-k_3+k_2-1}{k_2-1}}{\binom{k_1+k_2}{k_2}}\mathbf{1}_{B(t_i,k_1+k_2;\mathbf{W}_i)}
\end{align*}
which is independent of the value $t_i$. We note in passing that if $\mathbf{1}_{B(t_i,k_1+k_2;\mathbf{W}_i)}=0$, both the fraction and the conditional probability equal $0$.
Then we show that the result is the same when choosing a data-driven threshold.
\begin{align*}
    &\PP\left\{\delta_i(T_i)=k_3\mid V_+^i(T_i)=k_1,V_-^i(T_i)=k_2,\mathbf{W}_i\right\}\\
    =&\sum_{t}\PP\left\{\delta_i(T_i)=k_3,T_i=t\mid V_+^i(T_i)=k_1,V_-^i(T_i)=k_2,\mathbf{W}_i\right\}\\
    =&\sum_{t}\PP\left\{\delta_i(t)=k_3\mid T_i=t,V_+^i(t)=k_1,V_-^i(t)=k_2,\mathbf{W}_i\right\}\\
    &\quad\PP\left\{T_i=t\mid V_+^i(T_i)=k_1,V_-^i(T_i)=k_2,\mathbf{W}_i\right\}\\
    =&\frac{\binom{k_1-k_3+k_2-1}{k_2-1}}{\binom{k_1+k_2}{k_2}}\sum_t\PP\left\{T_i=t\mid V_+^i(T_i)=k_1,V_-^i(T_i)=k_2,\mathbf{W}_i\right\}\mathbf{1}_{B(t,k_1+k_2;\mathbf{W}_i)}\\
    =&\frac{\binom{k_1-k_3+k_2-1}{k_2-1}}{\binom{k_1+k_2}{k_2}}\sum_t\PP\left\{T_i=t\mid V_+^i(T_i)=k_1,V_-^i(T_i)=k_2,\mathbf{W}_i\right\}\\
    =&\frac{\binom{k_1-k_3+k_2-1}{k_2-1}}{\binom{k_1+k_2}{k_2}}=f(k_3;k_1,k_2)
\end{align*}
The second-to-last equality holds because when $\mathbf{1}_{B(t,k_1+k_2;\mathbf{W}_i)}=0$, the conditional probability must also be $0$. Here we omit the trivial case where the values of $k_1,k_2$ themselves are impossible.

\end{proof}

\begin{proof}[Proof of Lemma~\ref{slem:cher}]
With the convexity of the function $f$,
\[
f\left(\EE\left[\lambda \max_{1\leq i\leq D}X_i\right]\right)\leq\EE\left[f\left(\lambda \max_{1\leq i\leq D}X_i\right)\right]
\]
Then because $\PP\left(f(\lambda X_i)<0\right)=0,\forall \lambda>0$,
\[
\EE\left[f\left(\lambda \max_{1\leq i\leq D}X_i\right)\right]\leq\EE\left[\sum_{i=1}^Df\left(\lambda X_i\right)\right]
\]
Then, since the function $f$ is monotonically increasing,
\[
\EE\left[\lambda \max_{1\leq i\leq D}X_i\right]\leq f^{-1}\left(\EE\left[f\left(\lambda \max_{1\leq i\leq D}X_i\right)\right]\right)\leq f^{-1}\left(\EE\left[\sum_{i=1}^Df\left(\lambda X_i\right)\right]\right)
\]

\end{proof}

\begin{proof}[Proof of Lemma~\ref{slem:largeenoughrejection}]

Assume, for contradiction, that there exist $M>0$ and $\delta_t\in\left(0,\frac{t^U-t^L}{2}\right)$ for which there is a constant $\delta_1>0$ such that for any $N>0$ there exists $n>N$
\[
\PP\left\{\sup_{t\in(t^L+\delta_t,t^U-\delta_t)}\left|\frac{1}{Y_n(t)}\right|>\frac{1}{M}\right\}>\delta_1>0,
\]
because $Y_n(t)$ is integer-valued and monotone, we have
\[
\PP\left\{\left|Y_n(t)\right|<M,\forall t\in[t^U-\delta_t,t^U)\right\}>\delta_1>0,
\]
and
\[
\inf_{t\in[t^U-\delta_t,t^U)}\PP\left\{\left|Y_n(t)\right|<M\right\}\geq \PP\left\{\left|Y_n(t)\right|<M,\forall t\in[t^U-\delta_t,t^U)\right\},
\]
so for any $t\in[t^U-\delta_t,t^U)$, for any $N>0$ there exists $n>N$
\[
\PP\left\{\left|Y_n(t)\right|<M\right\}>\delta_1>0.
\]
If there is $t_1\in(t^L,t^U)$ such that $Y_n(t_1)<M<+\infty$ for given $n$, because $Y_n(t),X_n(t)$ are both positive integers for any $t$, we have for any $t_2\in(t_1,t^U)$,
\[
\left|\frac{X_n(t_1)}{Y_n(t_1)}-\frac{X_n(t_2)}{Y_n(t_2)}\right|=\left|\frac{Y_n(t_2)X_n(t_1)-Y_n(t_1)X_n(t_2)}{Y_n(t_1)Y_n(t_2)}\right|\in\{0\}\cup\left[\frac{1}{M^2},+\infty\right),
\]
which implies that
\[
\left\{\left|\frac{Y_n(t_2)X_n(t_1)-Y_n(t_1)X_n(t_2)}{Y_n(t_1)Y_n(t_2)}\right|\in\{0\}\cup\left[\frac{1}{M^2},+\infty\right)\right\}\subseteq\left\{Y_n(t_1)<M\right\}.
\]
Then we discuss two cases. First, if $(d^L+d^U)\delta_t>M^{-2}$, let $t_{\delta}=t^U-\delta_t+\frac{1}{M^2(d^L+d^U)}\in(t^U-\delta_t,t^U)$, which satisfies
\[
\frac{d^L+d^U}{2}(t_{\delta}-t^U+\delta_t)=\frac{1}{2M^2}.
\]
With the triangle inequality, we have
\begin{align*}
&\left|\frac{X_n(t_{\delta})}{Y_n(t_{\delta})}-G(t_{\delta})\right|\\
=&\left|\left(\frac{X_n(t_{\delta})}{Y_n(t_{\delta})}-\frac{X_n(t^U-\delta_t)}{Y_n(t^U-\delta_t)}\right)+\left(\frac{X_n(t^U-\delta_t)}{Y_n(t^U-\delta_t)}-G(t^U-\delta_t)\right)+\left(G(t^U-\delta_t)-G(t_{\delta})\right)\right| \\
\geq&\left|\left|\frac{X_n(t_{\delta})}{Y_n(t_{\delta})}-\frac{X_n(t^U-\delta_t)}{Y_n(t^U-\delta_t)}\right|-\left|G(t^U-\delta_t)-G(t_{\delta})\right|\right|-\left|\frac{X_n(t^U-\delta_t)}{Y_n(t^U-\delta_t)}-G(t^U-\delta_t)\right|
\end{align*}
where with the condition,
\[
0<\frac{1}{M^2}\frac{d^L}{d^L+d^U}\leq G(t^U-\delta_t)-G(t_{\delta})\leq\frac{1}{M^2}\frac{d^U}{d^L+d^U}.
\]
Let $\varepsilon_1=\frac{1}{M^2}\frac{d^L}{3(d^L+d^U)}$, we have
\begin{align*}
&\left\{\left|\left|\frac{X_n(t_{\delta})}{Y_n(t_{\delta})}-\frac{X_n(t^U-\delta_t)}{Y_n(t^U-\delta_t)}\right|-\left|G(t^U-\delta_t)-G(t_{\delta})\right|\right|>2\varepsilon_1\right\}\\
\supseteq& \left\{\left|\frac{X_n(t_{\delta})}{Y_n(t_{\delta})}-\frac{X_n(t^U-\delta_t)}{Y_n(t^U-\delta_t)}\right|<\frac{d^L}{3(d^L+d^U)M^2}\right\}\\
&\quad\cup\left\{\left|\frac{X_n(t_{\delta})}{Y_n(t_{\delta})}-\frac{X_n(t^U-\delta_t)}{Y_n(t^U-\delta_t)}\right|>\frac{\frac{2}{3}d^L+d^U}{(d^L+d^U)M^2}\right\}\\
\supseteq &\left\{\left|\frac{X_n(t_{\delta})}{Y_n(t_{\delta})}-\frac{X_n(t^U-\delta_t)}{Y_n(t^U-\delta_t)}\right|\in\{0\}\cup\left[\frac{1}{M^2},+\infty\right)\right\}\supseteq \left\{Y_n(t^U-\delta_t)<M\right\}
\end{align*}
so for any $n$, with $A_n=X_n(t_\delta)/Y_n(t_\delta)$, $B_n=X_n(t^U-\delta_t)/Y_n(t^U-\delta_t)$, $G_1=G(t_\delta)$, $G_2=G(t^U-\delta_t)$,
\begin{align*}
&\PP\left\{\left|A_n-G_1\right|>\varepsilon_1\right\}\\
\geq& \PP\left\{\bigl|\,\bigl|A_n-B_n\bigr|-\left|G_2-G_1\right|\,\bigr|-\left|B_n-G_2\right|>\varepsilon_1\right\}\\
\geq& \PP\left\{\left|B_n-G_2\right|\leq\varepsilon_1<\frac{1}{2}\bigl|\,\bigl|A_n-B_n\bigr|-\left|G_2-G_1\right|\,\bigr|\right\}\\
\geq& \PP\left\{\bigl|\,\bigl|A_n-B_n\bigr|-\left|G_2-G_1\right|\,\bigr|>2\varepsilon_1\right\}
      -\PP\left\{\left|B_n-G_2\right|>\varepsilon_1\right\}\\
\geq& \PP\left\{Y_n(t^U-\delta_t)<M\right\}-\PP\left\{\left|B_n-G_2\right|>\varepsilon_1\right\}
\end{align*}
which implies for any $n$
\begin{align*}
\PP\left\{Y_n(t^U-\delta_t)<M\right\}\leq \PP\left\{\left|A_n-G_1\right|>\varepsilon_1\right\}+\PP\left\{\left|B_n-G_2\right|>\varepsilon_1\right\},
\end{align*}
then let $\delta_2=\delta_1/3$. There exists $N_1$ such that for any $n>N_1$,
\[
\PP\left\{\left|\frac{X_n(t^U-\delta_t)}{Y_n(t^U-\delta_t)}-G(t^U-\delta_t)\right|\leq\varepsilon_1\right\}<\delta_2,
\]
\[
\PP\left\{\left|\frac{X_n(t_\delta)}{Y_n(t_\delta)}-G(t_\delta)\right|\leq\varepsilon_1\right\}<\delta_2,
\]
which contradicts our initial assumption. Second, if $(d^L+d^U)\delta_t\leq M^{-2}$, let $t'_{\delta}=t^U-\delta_t/3$, which implies $t'_{\delta}-t^U+\delta_t=2\delta_t/3$ and, under the condition,
\[
0<\frac{2d^L\delta_t}{3}\leq G(t^U-\delta_t)-G(t'_{\delta})\leq\frac{2d^U\delta_t}{3}<\frac{1}{M^2}-\frac{2d^L\delta_t}{3}.
\]
Let $\varepsilon_2=2d^L\delta_t/9$, we similarly have
\begin{align*}
\left|\frac{X_n(t'_{\delta})}{Y_n(t'_{\delta})}-G(t'_{\delta})\right|\geq&\left|\left|\frac{X_n(t'_{\delta})}{Y_n(t'_{\delta})}-\frac{X_n(t^U-\delta_t)}{Y_n(t^U-\delta_t)}\right|-\left|G(t^U-\delta_t)-G(t'_{\delta})\right|\right|\\
&\quad-\left|\frac{X_n(t^U-\delta_t)}{Y_n(t^U-\delta_t)}-G(t^U-\delta_t)\right|,
\end{align*}
and
\begin{align*}
&\left\{\left|\left|\frac{X_n(t'_{\delta})}{Y_n(t'_{\delta})}-\frac{X_n(t^U-\delta_t)}{Y_n(t^U-\delta_t)}\right|-\left|G(t^U-\delta_t)-G(t'_{\delta})\right|\right|>2\varepsilon_2\right\}\\
&\supseteq\left\{Y_n(t^U-\delta_t)<M\right\}.
\end{align*}
Therefore, for any $n$,
\begin{align*}
&\PP\left\{Y_n(t^U-\delta_t)<M\right\}\\
\leq& \PP\left\{\left|\frac{X_n(t^U-\delta_t)}{Y_n(t^U-\delta_t)}-G(t^U-\delta_t)\right|>\varepsilon_2\right\}+\PP\left\{\left|\frac{X_n(t'_{\delta})}{Y_n(t'_{\delta})}-G(t'_{\delta})\right|>\varepsilon_2\right\}.
\end{align*}
Then, under the condition, there exists $N_2$ such that for any $n>N_2$,
\[
\PP\left\{\left|\frac{X_n(t^U-\delta_t)}{Y_n(t^U-\delta_t)}-G(t^U-\delta_t)\right|\leq\varepsilon_2\right\}<\delta_2,
\]
\[
\PP\left\{\left|\frac{X_n(t'_\delta)}{Y_n(t'_\delta)}-G(t'_\delta)\right|\leq\varepsilon_2\right\}<\delta_2,
\]
so
\[
\PP\left\{Y_n(t^U-\delta_t)<M\right\}\leq2\delta_2<\frac{2\delta_1}{3},
\]
which contradicts our initial assumption. Thus, the proof is complete.

\end{proof}

\begin{proof}[Proof of Lemma~\ref{slem:negligibleadjpara}]

Because of the triangle inequality, for any $t\in(t^L+\delta_t,t^U-\delta_t)$,
\begin{align*}
    \left|\frac{X_n(t)+c}{Y_n(t)}-G(t)\right|=&\left|\frac{X_n(t)}{Y_n(t)}+\frac{c}{Y_n(t)}-G(t)\right|\\
		\leq&\left|\frac{X_n(t)}{Y_n(t)}-G(t)\right|+\left|\frac{c}{Y_n(t)}\right|,
\end{align*}
so for any $n$ and $\varepsilon>0$,
\begin{align*}
    &\PP\left\{\sup_{t\in(t^L+\delta_t,t^U-\delta_t)}\left|\frac{X_n(t)+c}{Y_n(t)}-G(t)\right|>\varepsilon\right\}\\
    \leq& \PP\left\{\sup_{t\in(t^L+\delta_t,t^U-\delta_t)}\left\{\left|\frac{X_n(t)}{Y_n(t)}-G(t)\right|+\left|\frac{c}{Y_n(t)}\right|\right\}>\varepsilon\right\}\\
    \leq& \PP\left\{\sup_{t\in(t^L+\delta_t,t^U-\delta_t)}\left|\frac{X_n(t)}{Y_n(t)}-G(t)\right|>\frac{\varepsilon}{2}\right\}+\PP\left\{\sup_{t\in(t^L+\delta_t,t^U-\delta_t)}\left|\frac{c}{Y_n(t)}\right|>\frac{\varepsilon}{2}\right\}.
\end{align*}
By definition and Lemma~\ref{slem:largeenoughrejection},
\[
\lim_{n\to+\infty}\PP\left\{\sup_{t\in(t^L+\delta_t,t^U-\delta_t)}\left|\frac{X_n(t)+c}{Y_n(t)}-G(t)\right|>\varepsilon\right\}\leq0+0=0.
\]

\end{proof}

\begin{proof}[Proof of Lemma~\ref{slem:uniquelimit}]

According to Lemma~\ref{slem:negligibleadjpara}, it is sufficient for us to consider the case where $c=0$ and
\[
\PP\left\{|T_n^c-t_{\alpha}|>\varepsilon\right\}=\PP\left\{T_n^c>t_{\alpha}+\varepsilon\right\}+\PP\left\{T_n^c<t_{\alpha}-\varepsilon\right\}.
\]
For any $n$ and $\varepsilon\in\left(0,\min\left\{t^U-t_{\alpha},t_{\alpha}-t^L\right\}\right)$,
\begin{align*}
&\left\{T_n^0<t_{\alpha}-\varepsilon\right\}
\subseteq\left\{\exists t\in(t^L,t_{\alpha}-\varepsilon),\frac{X_n(t)}{Y_n(t)}\leq\alpha\right\}\\
=& \left\{\exists t\in(t^L,t_{\alpha}-\varepsilon),\frac{X_n(t)}{Y_n(t)}\leq G(t_{\alpha})\right\}\\
=& \left\{\exists t\in(t^L,t_{\alpha}-\varepsilon),\frac{X_n(t)}{Y_n(t)}-G(t)\leq G(t_{\alpha})-G(t)<0\right\}\\
\subseteq&\left\{\exists t\in(t^L,t_{\alpha}-\varepsilon),\left|\frac{X_n(t)}{Y_n(t)}-G(t)\right|\geq |G(t)-G(t_{\alpha})|\right\}\\
\subseteq&\left\{\exists t\in(t^L,t_{\alpha}-\varepsilon),\left|\frac{X_n(t)}{Y_n(t)}-G(t)\right|\geq d^L\varepsilon\right\}
\end{align*}
so we have
\[
\PP\left\{T_n^0<t_{\alpha}-\varepsilon\right\}\leq \PP\left\{\sup_{t\in(t^L,t_{\alpha}-\varepsilon)}\left|\frac{X_n(t)}{Y_n(t)}-G(t)\right|\geq d^L\varepsilon\right\},
\]
and similarly, we have
\begin{align*}
&\left\{T_n^0>t_{\alpha}+\varepsilon\right\}
\subseteq\left\{\forall t>t_{\alpha}+\varepsilon,\frac{X_n(t)}{Y_n(t)}>\alpha\right\}\\
=&\left\{\forall t>t_{\alpha}+\varepsilon,\frac{X_n(t)}{Y_n(t)}> G(t_{\alpha})\right\}\\
=&\left\{\forall t>t_{\alpha}+\varepsilon,\frac{X_n(t)}{Y_n(t)}-G(t)> G(t_{\alpha})-G(t)\right\}\\
\subseteq&\left\{\left|\frac{X_n\left(t_{\alpha}+\frac{\varepsilon}{2}\right)}{Y_n\left(t_{\alpha}+\frac{\varepsilon}{2}\right)}-G\left(t_{\alpha}+\frac{\varepsilon}{2}\right)\right|\geq \left|G(t_{\alpha})-G\left(t_{\alpha}+\frac{\varepsilon}{2}\right)\right|\right\}\\
\subseteq&\left\{\left|\frac{X_n\left(t_{\alpha}+\frac{\varepsilon}{2}\right)}{Y_n\left(t_{\alpha}+\frac{\varepsilon}{2}\right)}-G\left(t_{\alpha}+\frac{\varepsilon}{2}\right)\right|\geq \frac{d^L\varepsilon}{2}\right\}
\end{align*}
and then
\[
\PP\left\{T_n^0>t_{\alpha}+\varepsilon\right\}\leq \PP\left\{\left|\frac{X_n\left(t_{\alpha}+\frac{\varepsilon}{2}\right)}{Y_n\left(t_{\alpha}+\frac{\varepsilon}{2}\right)}-G\left(t_{\alpha}+\frac{\varepsilon}{2}\right)\right|\geq \frac{d^L\varepsilon}{2}\right\}.
\]
In summary, we can obtain the following equation
\[
\lim_{n\to+\infty}\PP\left\{|T_n^0-t_{\alpha}|>\varepsilon\right\}=0,
\]
and this result can be easily generalized to the case of any given $c$. Thus, the proof is complete.

\end{proof}

\begin{proof}[Proof of Lemma~\ref{slem:asyfraction}]

For any $\varepsilon>0$, there is $\delta_{\varepsilon}>0$ such that for any $\delta_1>0$ there exists $N_{\delta_1}$ such that for any $n>N_{\delta_1}$,
\[
\PP\left\{\sup_{\substack{t_1,t_2\in(t^L,t^U)\\ |t_1-t_2|\leq\delta_{\varepsilon}}}\left|\frac{Y_n(t_1)}{Y_n(t_2)}-1\right|>\varepsilon\right\}\leq\delta_1
\]
then with Lemma~\ref{slem:uniquelimit}, for any $\delta_2$ we have $N_{\delta_2}$ which satisfies that for any $n>N_{\delta_2}$,
\[
\PP\left\{\left|T^{c_1}_n-t_{\alpha}\right|>\frac{\delta_{\varepsilon}}{2}\right\}\leq1-\sqrt{1-\delta_2},\quad \PP\left\{\left|T^{c_2}_n-t_{\alpha}\right|>\frac{\delta_{\varepsilon}}{2}\right\}\leq1-\sqrt{1-\delta_2}
\]
and so
\[
\PP\left(\left\{\left|T^{c_1}_n-T^{c_2}_n\right|>\delta_{\varepsilon}\right\}\cup\left\{T^{c_1}_n\notin (t^L,t^U)\right\}\cup\left\{T^{c_2}_n\notin (t^L,t^U)\right\}\right)\leq\delta_2.
\]
Then because
\begin{align*}
&\left\{\left|\frac{Y_n(T^{c_1}_n)}{Y_n(T^{c_2}_n)}-1\right|>\varepsilon\right\}\\
\subseteq&\left\{\sup_{\substack{t_1,t_2\in(t^L,t^U)\\ |t_1-t_2|\leq\delta_{\varepsilon}}}\left|\frac{Y_n(t_1)}{Y_n(t_2)}-1\right|>\varepsilon\right\}\cup\left\{T^{c_1}_n,T^{c_2}_n\in(t^L,t^U),|T^{c_1}_n-T^{c_2}_n|\leq\delta_{\varepsilon}\right\}^c,
\end{align*}
we have for any $n>\max\{N_{\delta_1},N_{\delta_2}\}$,
\[
\PP\left\{\left|\frac{Y_n(T^{c_1}_n)}{Y_n(T^{c_2}_n)}-1\right|>\varepsilon\right\}\leq\delta_1+\delta_2.
\]
Since $\delta_1$ and $\delta_2$ are arbitrary, we finally have
\[
\lim_{n\to+\infty}\PP\left\{\left|\frac{Y_n(T^{c_1}_n)}{Y_n(T^{c_2}_n)}-1\right|>\varepsilon\right\}=0.
\]

\end{proof}

\begin{proof}[Proof of Lemma~\ref{slem:fpt1}]

For any $t>0$, if $\mathbf{1}\left\{W_i\geq W_j\geq t,L_i=L_j=0\right\}=1$, we have
\begin{align*}
&\frac{\sum_{k\neq i}r(W_k)\mathbf{1}\left\{W_k\geq t,L_k=0\right\}+\max_k\left\{r(W_k):W_k\geq t\right\}}{\sum_{k\neq i}\mathbf{1}\left\{W_k\geq t,L_k=1\right\}+\mathbf{1}\left\{W_k\geq t\right\}}\\
=&\frac{\sum_{k}r(W_k)\mathbf{1}\left\{W_k\geq t,L_k=0\right\}+\max_k\left\{r(W_k):W_k\geq t\right\}-r(W_i)}{\sum_{k}\mathbf{1}\left\{W_k\geq t,L_k=1\right\}+1}\\
\leq&\frac{\sum_{k}r(W_k)\mathbf{1}\left\{W_k\geq t,L_k=0\right\}+\max_k\left\{r(W_k):W_k\geq t\right\}-r(W_i)}{\sum_{k}\mathbf{1}\left\{W_k\geq t,L_k=1\right\}+1}\\
=&\frac{\sum_{k\neq j}r(W_k)\mathbf{1}\left\{W_k\geq t,L_k=0\right\}+\max_k\left\{r(W_k):W_k\geq t\right\}}{\sum_{k\neq j}\mathbf{1}\left\{W_k\geq t,L_k=1\right\}+\mathbf{1}\left\{W_j\geq t\right\}}
\end{align*}
so $T_i^{r(\cdot)}\leq T_j^{r(\cdot)}$, which implies $T^{r(\cdot)}_{(0:\mathcal{R}_-;\mathbf{W})}\geq T^{r(\cdot)}_{(1:\mathcal{R}_-;\mathbf{W})}\geq\cdots\geq T^{r(\cdot)}_{(|\mathcal{R}_-|:\mathcal{R}_-;\mathbf{W})}$.

For the case that $\mathbf{1}\left\{W_k\geq T^{r(\cdot)}_{(|\mathcal{R}_-|:\mathcal{R}_-;\mathbf{W})}\right\}=1$, if $L_k=1$, obviously $T^{r(\cdot)}_k=T^{r(\cdot)}\geq T^{r(\cdot)}_{(|\mathcal{R}_-|:\mathcal{R}_-;\mathbf{W})}$. If $L_k=0$ and $W_k>T^{r(\cdot)}$, we have $T^{r(\cdot)}_k\geq T^{r(\cdot)}_{(|\mathcal{R}_-|:\mathcal{R}_-;\mathbf{W})}$, if $L_k=0$ and $T^{r(\cdot)}\geq W_k\geq T^{r(\cdot)}_{(|\mathcal{R}_-|:\mathcal{R}_-;\mathbf{W})}$, we have $W_k\leq\min_{i\in A(T)}\{W_i\}$ which implies that $r(W_k)\leq\min_{i\in A(T)}\{r(W_i)\}$, so $T^{r(\cdot)}_k\geq T^{r(\cdot)}_{(|\mathcal{R}_-|:\mathcal{R}_-;\mathbf{W})}$.

For the case that $\mathbf{1}\left\{W_k\geq T^{r(\cdot)}_{(1:\mathcal{R}_-;\mathbf{W})},L_k=0\right\}=0$, if $L_k=1$, obviously, $T^{r(\cdot)}_k=T^{r(\cdot)}$, if $W_k< T^{r(\cdot)}_{(1:\mathcal{R}_-;\mathbf{W})}$ and $L_k=0$, for any $t<T^{r(\cdot)}$,
\begin{align*}
&\frac{\sum_{i\neq k}r(W_i)\mathbf{1}\left\{W_i\geq t,L_i=0\right\}+\max_i\left\{r(W_i):W_i\geq t\right\}}{\sum_{i\neq k}\mathbf{1}\left\{W_i\geq t,L_i=1\right\}+\mathbf{1}\left\{W_k\geq t\right\}} \\
=&\frac{\sum_{i}r(W_i)\mathbf{1}\left\{W_i\geq t,L_i=0\right\}+\max_i\left\{r(W_i):W_i\geq t\right\}-r(W_k)\mathbf{1}\left\{W_k\geq t\right\}}{\sum_{k}\mathbf{1}\left\{W_k\geq t,L_k=1\right\}+\mathbf{1}\left\{W_k\geq t\right\}} \\
\geq&\frac{\sum_{i}r(W_i)\mathbf{1}\left\{W_i\geq t,L_i=0\right\}+\max_i\left\{r(W_i):W_i\geq t\right\}-r(T)\mathbf{1}\left\{W_k\geq t\right\}}{\sum_{k}\mathbf{1}\left\{W_k\geq t,L_k=1\right\}+\mathbf{1}\left\{W_k\geq t\right\}}
\end{align*}
and this inequality holds with equality if and only if $r(T^{r(\cdot)})=r(W_k)$. If $\max_{i\in \mathcal{R}_-}\left\{r(W_i)\right\}=r(T^{r(\cdot)})$, there is $T^{r(\cdot)}_{(1:\mathcal{R}_-;\mathbf{W})}=T^{r(\cdot)}_{(|\mathcal{R}_-|:\mathcal{R}_-;\mathbf{W})}$, which implies that
\begin{align*}
&\frac{\sum_{i\neq k}r(W_i)\mathbf{1}\left\{W_i\geq t,L_i=0\right\}+\max_i\left\{r(W_i):W_i\geq t\right\}}{\sum_{i\neq k}\mathbf{1}\left\{W_i\geq t,L_i=1\right\}+\mathbf{1}\left\{W_k\geq t\right\}} \\
=&\frac{\sum_{i\neq j}r(W_i)\mathbf{1}\left\{W_i\geq t,L_i=0\right\}+\max_i\left\{r(W_i):W_i\geq t\right\}}{\sum_{i\neq j}\mathbf{1}\left\{W_i\geq t,L_i=1\right\}+\mathbf{1}\left\{W_j\geq t\right\}},\quad\forall j\in \mathcal{R}_-
\end{align*}
if and only if $r(T^{r(\cdot)})=r(W_k)$ and $t\leq W_k< T^{r(\cdot)}_{(1:\mathcal{R}_-;\mathbf{W})}=T^{r(\cdot)}_{(|\mathcal{R}_-|:\mathcal{R}_-;\mathbf{W})}$. In that case,
\[
\frac{\sum_{i\neq j}r(W_i)\mathbf{1}\left\{W_i\geq t,L_i=0\right\}+\max_i\left\{r(W_i):W_i\geq t\right\}}{\sum_{i\neq j}\mathbf{1}\left\{W_i\geq t,L_i=1\right\}+\mathbf{1}\left\{W_j\geq t\right\}}>\alpha,\quad\forall j\in \mathcal{R}_-
\]
and if the condition is not satisfied, the inequalities cannot be taken to be equal but will be strictly greater, so $T^{r(\cdot)}_k>W_k$ and then $T^{r(\cdot)}_k=T^{r(\cdot)}$.

\end{proof}

\begin{proof}[Proof of Lemma~\ref{slem:fpt2}]

It is obvious that $T^{r(\cdot)}_i\leq T^{r(\cdot)}$ for any $i$ because of the definition of the FO threshold. With the same analysis as the above lemma, if $i,j\in \mathcal{R}_-$ and $W_i\geq W_j$, we have $T^{r(\cdot)}_i\leq T^{r(\cdot)}_j\leq T^{r(\cdot)}$. If $L_k=0$ but $k\notin \mathcal{R}^i_-$, $W_k<T^{r(\cdot)}_k\leq T^{r(\cdot)}$ which also implies that $T^{r(\cdot)}_k=T^{r(\cdot)}$. Furthermore, for the case that $\mathbf{1}\left\{W_k\geq T^{r(\cdot)}_{(1:\mathcal{R}^i_-;\mathbf{W})},L_k=0\right\}=0$, $T^{r(\cdot)}_k=T^{r(\cdot)}$ obviously.

\end{proof}

\section{Details of the Simulation Experiments and PSM}
\label{ssec:sim}

\subsection{Groups of multiple hypothesis testing}

For the single-model experiment with different parameters across groups, we select each parameter with equal probability from the following list:

\begin{itemize}
    \item $m_i$: $300, 400, 500, 750, 1000$;
    \item $\pi_{i0}$: $0.6, 0.7, 0.8, 0.9$;
    \item $c^{\text{sca}}_i$: $1.0, 2.0, 3.0, 4.0$;
    \item $c^{\text{loc}}_i$: $0.0, 1.0, 2.0$.
\end{itemize}

For the multi-model experiment, we select each parameter with unequal probabilities from Table~\ref{tab:hypotheses_config}.
\begin{table}[t]
\centering
\begin{tabular}{c p{4cm} p{4cm}}
\hline
\textbf{Distribution} & \textbf{$H_0$ parameters} & \textbf{$H_1$ parameters} \\
\hline
normal & $\mu = 0.0$, $\sigma = 1.0$ & $\mu = 2.7$, $\sigma = 1.0$ \\
uniform & $a = 0.0$, $b = 1.0$ & $a = 0.2$, $b = 2.0$ \\
exponential & $\lambda = 1.0$ & $\lambda = 1/6.0$ \\
lognormal & $\mu = 0.0$, $\sigma = 1.0$ & $\mu = 2.5$, $\sigma = 1.0$ \\
gamma & $\alpha = 2.0$, $\beta = 1.0$ & $\alpha = 2.0$, $\beta = 4.0$ \\
weibull & $k = 2.0$, $\lambda = 1.0$ & $k = 2.0$, $\lambda = 2.8$ \\
beta & $\alpha = 2.0$, $\beta = 4.0$ & $\alpha = 8.0$, $\beta = 2.0$ \\
chi-square & $k = 2$ & $k = 10$ \\
F & $d_1 = 2$, $d_2 = 10$ & $d_1 = 20$, $d_2 = 1$ \\
binomial & $n = 40$, $p = 0.5$ & $n = 40$, $p = 0.7$ \\
poisson & $\lambda = 2$ & $\lambda = 6$ \\
geometric & $p = 0.5$ & $p = 0.1$ \\
negative binomial & $r = 3$, $p = 0.5$ & $r = 6$, $p = 0.3$ \\
\hline
\end{tabular}
\caption{Parameter configuration for hypothesis testing.}
\label{tab:hypotheses_config}
\end{table}

\subsection{Edge inference in Gaussian graphical models}

\begin{figure}[t]
    \centering
    \includegraphics[width=0.8\linewidth]{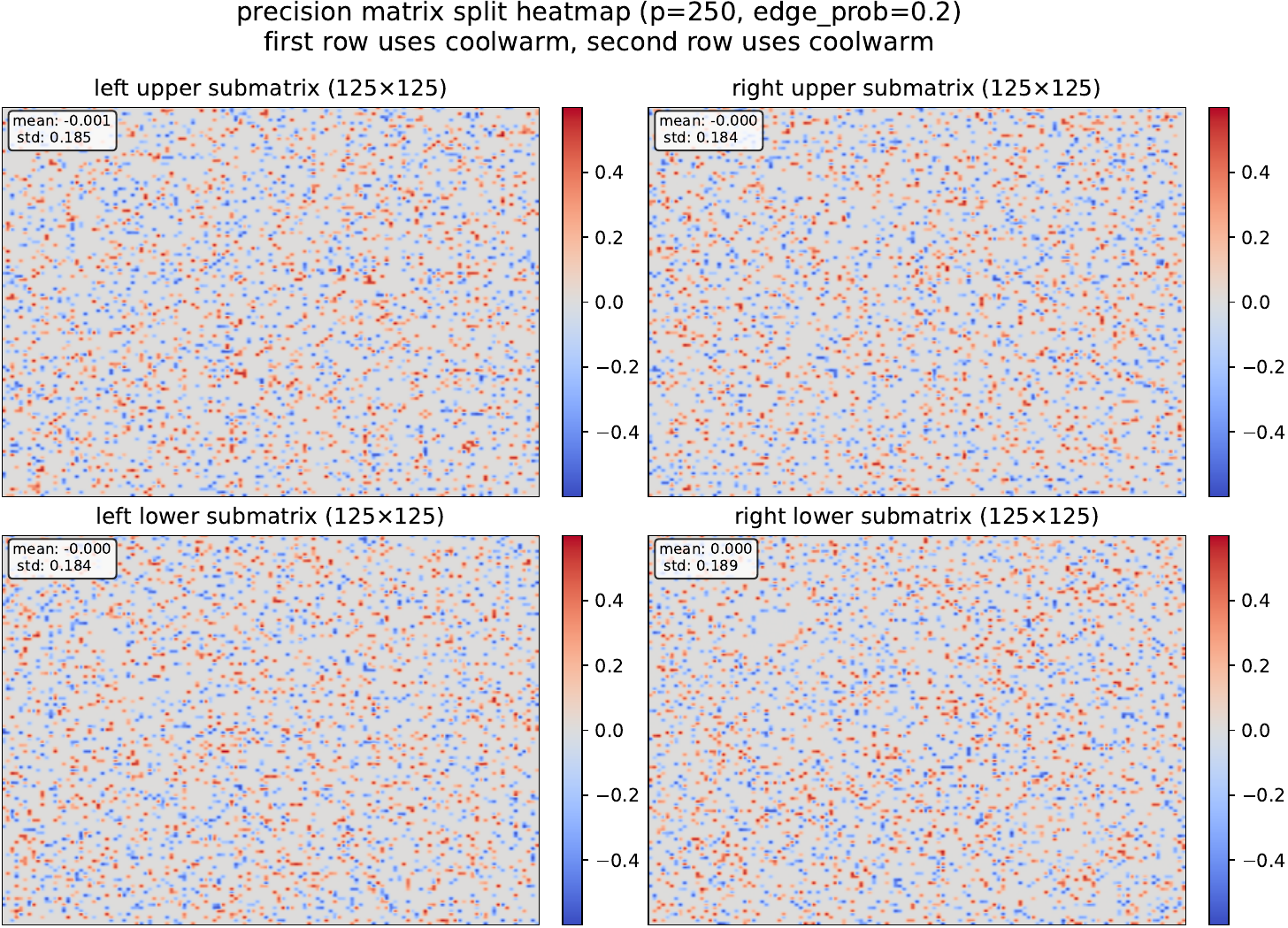}
    \caption{Heatmaps of the precision matrix.}
    \label{fig:oneheatmap}
\end{figure}

\begin{figure}[t]
    \centering
    \includegraphics[width=0.8\linewidth]{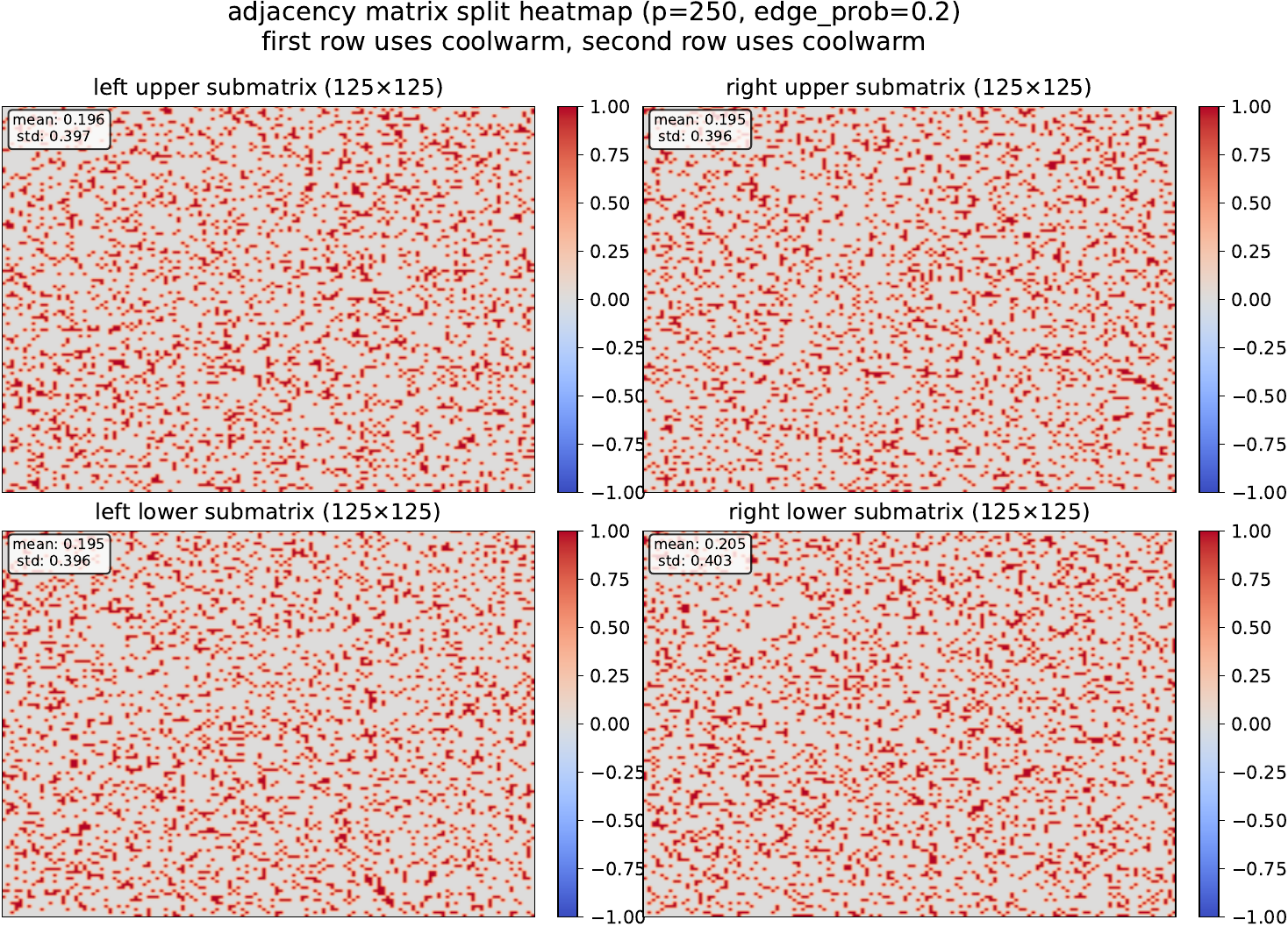}
    \caption{Heatmaps of the adjacency matrix.}
    \label{fig:oneheatmap_adj}
\end{figure}

We display the heatmaps (Figure~\ref{fig:oneheatmap} and Figure~\ref{fig:oneheatmap_adj}) of the precision matrix and the adjacency matrix generated by the method described in Section~\ref{sec:ggm}. The diagonal elements are set to $0$.

\begin{figure}[t]
    \centering
    \includegraphics[width=1\linewidth]{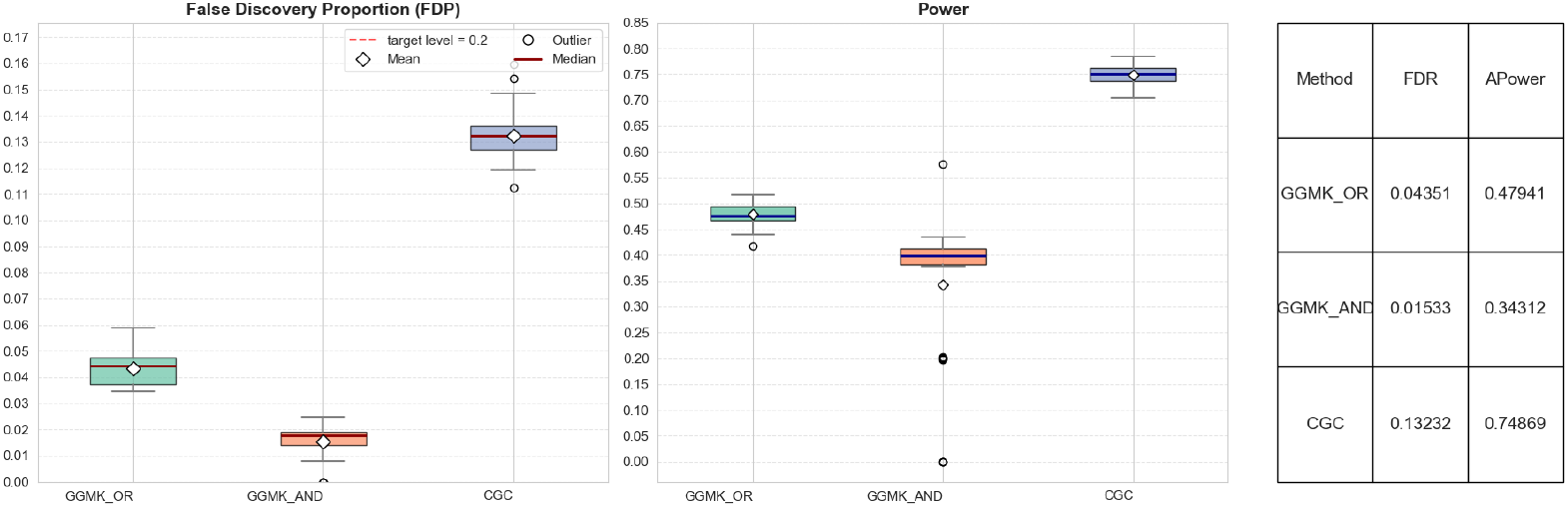}
    \caption{Results of edge inference in GGMs with a target FDR control level of $0.2$ and a sample size of $3000$.}
    \label{fig:exp_GGM}
\end{figure}

Furthermore, we supplement the results with a box-plot (Figure~\ref{fig:exp_GGM}) based on $300$ replications, a target FDR control level of $0.2$ and a sample size of $3000$.

\subsection{PSM analysis}

Peptide Post-Translational Modifications (PTMs): Table~\ref{tab:peptidedata} provides an overview of spectra with various modifications. The \textbf{PTM} column lists the type of post-translational modification, along with the specific amino acid residue affected. The \textbf{Subset} column indicates the subset category to which each PTM belongs and the subscript of $S$ describes the specific modification and residue. The \textbf{Mass} column presents the exact mass shift caused by each PTM, measured in Daltons (Da). The \textbf{Number} column presents the number of valid spectra without missing values.

\begin{table}[t]
    \centering
    \begin{tabular}{cccc}
        \hline
        \textbf{PTM} & \textbf{Subset} & \textbf{Mass} & \textbf{Number} \\
        \hline
        Acetyl[ProteinN-term] & $S_{acetyl,N-term}$ & 42.010565 & 5000\\
        Deamidated[N] & $S_{deamidated,N}$ & 0.984016 & 7506\\
        Dimethyl[K] & $S_{dimethyl,K}$ & 28.031300 & 500\\
        Methyl[K] & $S_{methyl,K}$ & 14.015650 & 1000\\
        Oxidation[M] & $S_{oxidation,M}$ & 15.994915 & 6771\\
        Oxidation[W] & $S_{oxidation,W}$ & 15.994915 & 10000\\
        Dioxidation[M] & $S_{dioxidation,M}$ & 31.989829 & 100\\
        Phospho[S] & $S_{phospho,S}$ & 79.966331 & 25\\
        Phospho[T] & $S_{phospho,T}$ & 79.966331 & 25\\
        Trimethyl[K] & $S_{trimethyl,K}$ & 42.046950 & 5000\\
        Unmodified & $S_{unmodified}$ & N/A & 60000\\
        \hline
        \multicolumn{3}{c}{\textbf{Total}} & 95927 \\
        \hline
    \end{tabular}
    \caption{Modification types involved in the experiments and the corresponding number of spectra.}
    \label{tab:peptidedata}
\end{table}

\begin{figure}[t]
    \centering
    \includegraphics[width=0.8\linewidth]{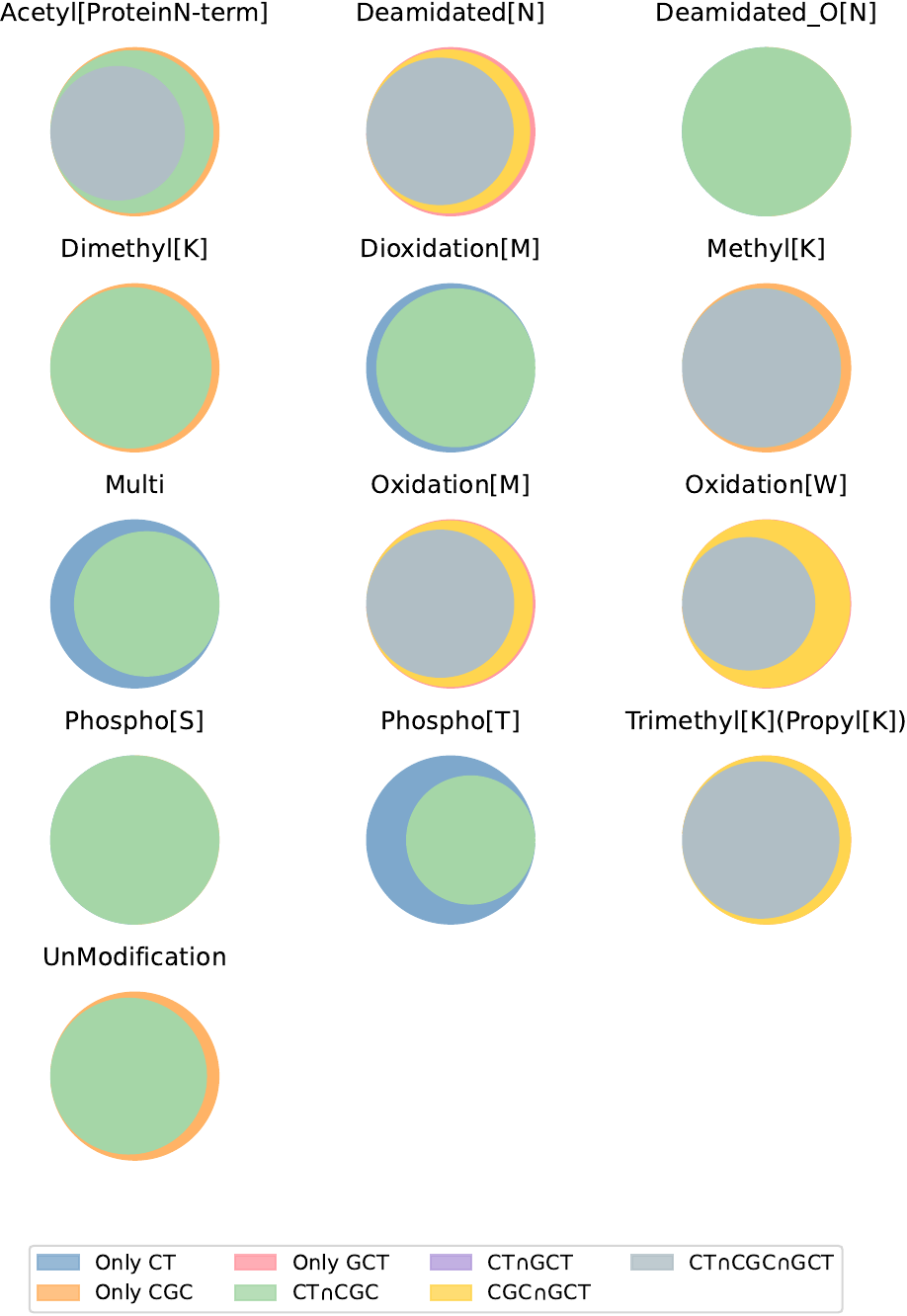}
    \caption{Venn diagram of the raw scores under various modifications, with the overlap counts among the methods.}
    \label{fig:venn_raw}
\end{figure}

\begin{figure}[t]
    \centering
    \includegraphics[width=0.8\linewidth]{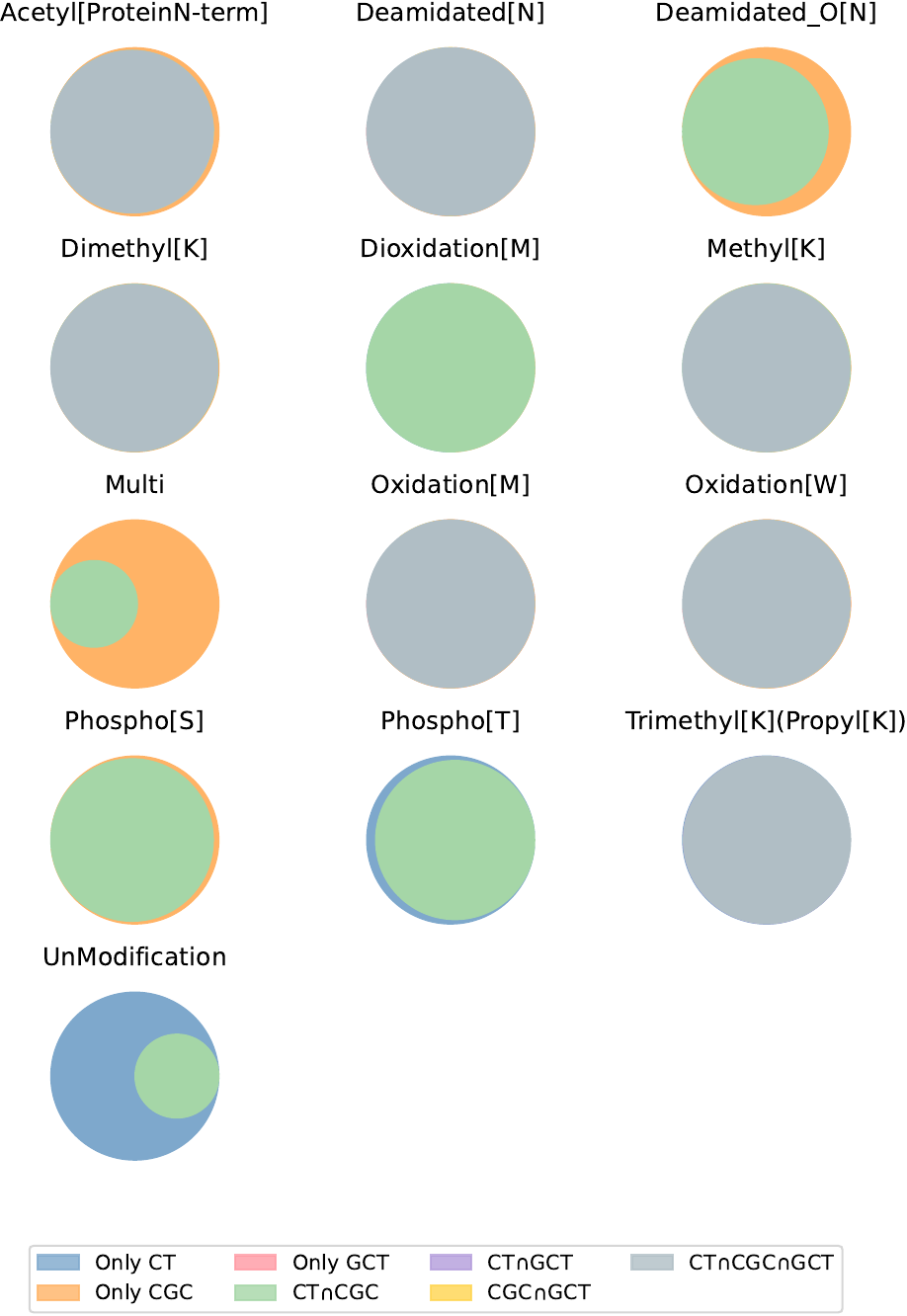}
    \caption{Venn diagram of the final scores under various modifications, with the overlap counts among the methods.}
    \label{fig:venn_final}
\end{figure}

We next present Venn diagrams (Figure~\ref{fig:venn_raw} and Figure~\ref{fig:venn_final}) and tables (Table~\ref{tab:modification_summary_raw} and Table~\ref{tab:modification_summary_final}) to display the matching results and modification identifications under the two scoring schemes (raw score and final score) for the three methods.

\begin{table}[t]
  \centering
  \begin{tabular}{lcccccccc}
    \hline
    Modification
    & CT
    & CGC
    & GCT
    & C$\cap$G
    & C$\cap$T
    & G$\cap$T
    & ALL \\
    \hline
    Acetyl[ProteinN-term]  & 285  & 307  & 193  & 285  & 193  & 193  & 193  \\
    Deamidated[N]           & 4255 & 5279 & 5622 & 4255 & 4255 & 5279 & 4255 \\
    Deamidated\_O[N]        & 2    & 2    & 0    & 2    & 0    & 0    & 0    \\
    Dimethyl[K]             & 257  & 283  & 0    & 257  & 0    & 0    & 0    \\
    Dioxidation[M]          & 59   & 52   & 0    & 52   & 0    & 0    & 0    \\
    Methyl[K]               & 494  & 560  & 492  & 494  & 492  & 492  & 492  \\
    Multi                   & 121  & 89   & 0    & 89   & 0    & 0    & 0    \\
    Oxidation[M]            & 3981 & 5057 & 5216 & 3981 & 3981 & 5057 & 3981 \\
    Oxidation[W]            & 5740 & 9192 & 9303 & 5740 & 5740 & 9192 & 5740 \\
    Phospho[S]              & 21   & 21   & 0    & 21   & 0    & 0    & 0    \\
    Phospho[T]              & 19   & 11   & 0    & 11   & 0    & 0    & 0    \\
    Trimethyl[K](Propyl[K]) & 3148 & 3645 & 3631 & 3148 & 3148 & 3631 & 3148 \\
    Unmodified          & 18   & 21   & 0    & 18   & 0    & 0    & 0    \\
    \hline
  \end{tabular}
  \caption{Matching results with raw scores under various modifications, and their overlap counts. Abbreviations: C = CT, G = CGC, T = GCT, ALL = $\text{C}\cap \text{G}\cap \text{T}$.}
  \label{tab:modification_summary_raw}
\end{table}

\begin{table}[t]
  \centering
  \begin{tabular}{lcccccccc}
    \hline
    Modification
    & CT
    & CGC
    & GCT
    & C$\cap$G
    & C$\cap$T
    & G$\cap$T
    & ALL \\
    \hline
    Acetyl[ProteinN-term]  & 492  & 526  & 492  & 492  & 492  & 492  & 492  \\
    Deamidated[N]           & 7057 & 7063 & 7074 & 7057 & 7057 & 7063 & 7057 \\
    Deamidated\_O[N]        & 3    & 4    & 0    & 3    & 0    & 0    & 0    \\
    Dimethyl[K]             & 455  & 460  & 453  & 455  & 453  & 453  & 453  \\
    Dioxidation[M]          & 99   & 99    & 0    & 99    & 0    & 0    & 0    \\
    Methyl[K]               & 907  & 910  & 900  & 907  & 900  & 900  & 900  \\
    Multi                    & 34   & 129   & 0    & 34   & 0    & 0    & 0    \\
    Oxidation[M]             & 6281 & 6290 & 6304 & 6281 & 6281 & 6290 & 6281 \\
    Oxidation[W]            & 9291 & 9326 & 9350 & 9291 & 9291 & 9326 & 9291 \\
    Phospho[S]              & 29   & 31   & 0    & 29   & 0    & 0    & 0    \\
    Phospho[T]              & 29   & 26   & 0    & 26   & 0    & 0    & 0    \\
    Trimethyl[K](Propyl[K]) & 5539 & 5482 & 5490 & 5482 & 5490 & 5482 & 5482 \\
    Unmodified          & 288  & 71   & 0    & 71   & 0    & 0    & 0    \\
    \hline
  \end{tabular}
  \caption{Matching results with final scores under various modifications, and their overlap counts. Abbreviations: C = CT, G = CGC, T = GCT, ALL = $\text{C}\cap \text{G}\cap \text{T}$.}
  \label{tab:modification_summary_final}
\end{table}

\clearpage
\subsection{Numerical study of the generic e-value transformation}
\label{ssec:evalue}

\begin{figure}[ht]
    \centering
    \includegraphics[width=1\linewidth]{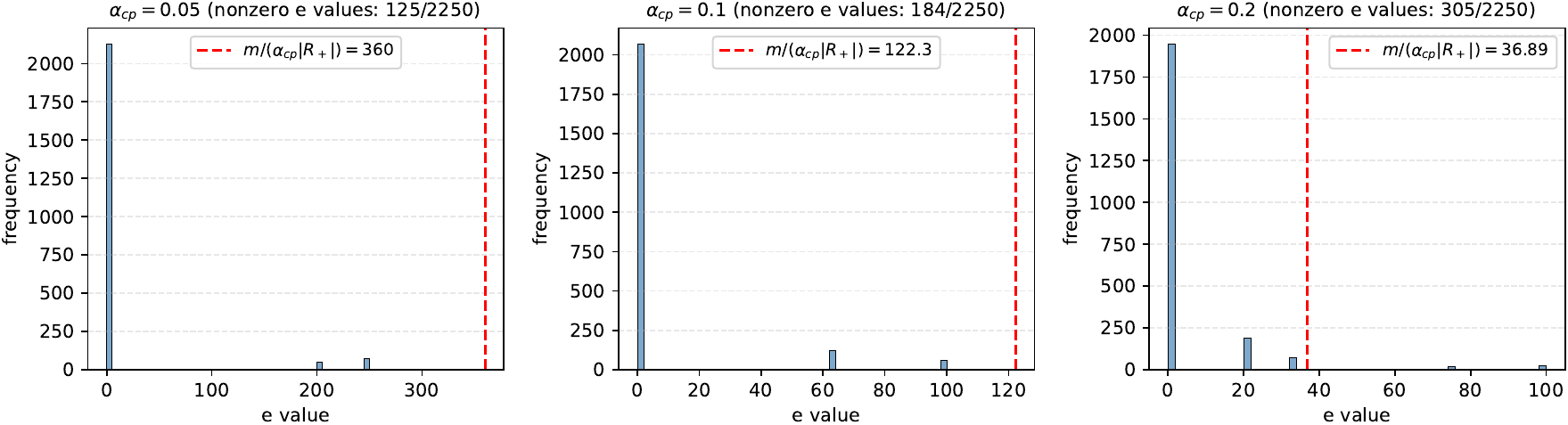}
    \caption{Histograms of the e-values induced by the generic transformation, for one data set with $d=5$ groups. The three panels use $\alpha_{\text{cp}}=0.05$, $0.1$ and $0.2$. The dashed line marks $m/(\alpha_{\text{cp}}|R_+|)$, which is the level of the competition filter written on the e-value scale.}
    \label{fig:evalue_hist}
\end{figure}

We first show that the e-values produced by the generic transformation carry almost no information beyond the threshold of the filter itself. We start from a procedure that controls the FDR at level $\alpha_{\text{cp}}$, and we convert it into e-values by assigning $E_j=m_i/(1+|R^i_-|)$ to every $j\in R^i_+$ and $E_j=0$ to every other hypothesis. Here $m_i$ is the size of group $i$. The sets $R^i_+$ and $R^i_-$ collect the hypotheses of group $i$ whose score is at least the threshold of the filter. Their label is $1$ in $R^i_+$ and $0$ in $R^i_-$. The set $R^i_+$ is exactly the set of rejections of group $i$. The hypotheses of $R^i_-$ lie above the threshold, but the filter keeps them because their label is $0$. Figure~\ref{fig:evalue_hist} shows the histogram of these e-values for one data set with $d=5$ groups, at $\alpha_{\text{cp}}=0.05$, $0.1$ and $0.2$. The e-values concentrate on a pile at $0$ together with one spike per group that has at least one rejection, because all hypotheses of such a group receive the same value. As $\alpha_{\text{cp}}$ grows, fewer hypotheses receive a nonzero e-value, from $125$ of $2250$ at $\alpha_{\text{cp}}=0.05$ to $305$ of $2250$ at $\alpha_{\text{cp}}=0.2$, and the spikes become smaller, from the range $[200,250]$ to the range $[20.8,100]$. The dashed line marks $m/(\alpha_{\text{cp}}|R_+|)$, which is the level of the competition filter written on the e-value scale, where $m=\sum_{i=1}^d m_i$ and $|R_+|=\sum_{i=1}^d|R^i_+|$. Every spike lies within a factor of three of this line. The transformed e-values therefore add nothing to what the threshold of the filter already contains, and this is the limitation that Section~\ref{sec:cad} describes.

\begin{figure}[ht]
    \centering
    \includegraphics[width=1\linewidth]{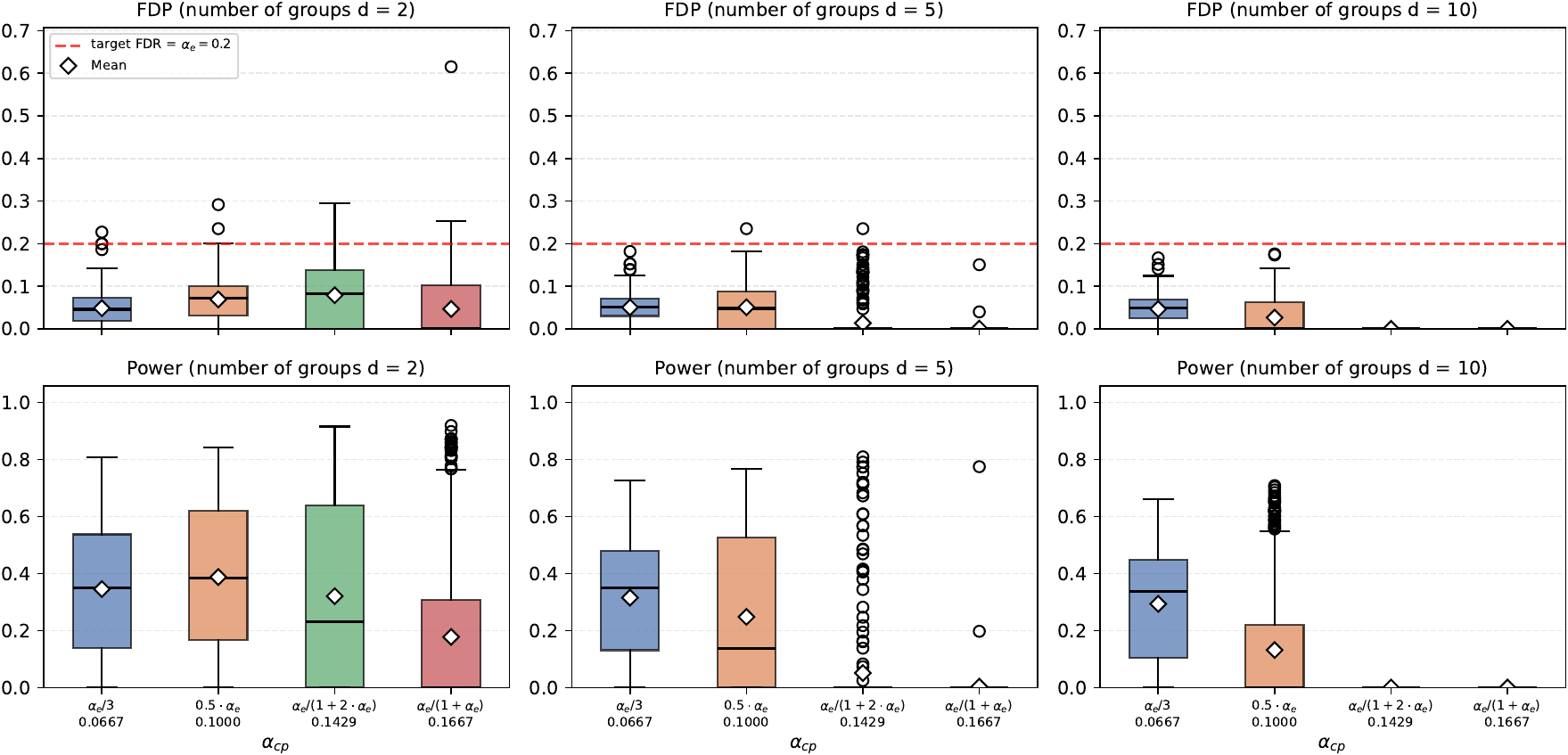}
    \caption{FDP and power of e-BH at level $\alpha_{\text{e}}=0.2$ under four rules for the level $\alpha_{\text{cp}}$ of the competition filter, over $300$ replications. Each column uses a different number of groups $d$. The dashed line marks the target FDR level.}
    \label{fig:evalue_boxplot}
\end{figure}

We next ask which level $\alpha_{\text{cp}}$ the transformation should start from. Figure~\ref{fig:evalue_boxplot} reports the FDP and the power of e-BH at level $\alpha_{\text{e}}=0.2$ for four choices of $\alpha_{\text{cp}}$ and for $d=2$, $5$ and $10$ groups, over $300$ replications. In each replication we draw the group size $m_i$ and the null proportion $\pi_{i0}$ as in Section~\ref{ssec:sim}, fix the scale and the location at $1$ and $0$, and draw the mean of $H_1$ in each group from $\{1.5,2.0,2.5,3.0,2.7\}$, so that the groups differ in size, in signal proportion and in signal strength. The four rules are $\alpha_{\text{cp}}=\alpha_{\text{e}}/(1+\alpha_{\text{e}})=0.1667$, $\alpha_{\text{e}}/(1+2\alpha_{\text{e}})=0.1429$, $0.5\alpha_{\text{e}}=0.1$ and $\alpha_{\text{e}}/3=0.0667$. The FDP of the e-value route stays below the target in every panel, and the largest average FDP is $0.0791$. The power behaves quite differently. It depends strongly on $\alpha_{\text{cp}}$ and on $d$. At $d=2$ the largest power is $0.3875$, attained at $0.5\alpha_{\text{e}}$, while the rule with the largest level, $\alpha_{\text{e}}/(1+\alpha_{\text{e}})$, gives only $0.1776$. At $d=5$ and $d=10$ the power decreases as $\alpha_{\text{cp}}$ grows. It is $0.3152$ and $0.2936$ at $\alpha_{\text{e}}/3$, but only $0.0509$ and $0.0000$ at $\alpha_{\text{e}}/(1+2\alpha_{\text{e}})$, and $0.0032$ and $0.0000$ at $\alpha_{\text{e}}/(1+\alpha_{\text{e}})$. Once $\alpha_{\text{cp}}$ is too large, e-BH rejects almost nothing and the power collapses to $0$. The reason is that all nonzero e-values of a group are equal. Raising $\alpha_{\text{cp}}$ adds more nonzero e-values, but it makes each of them smaller, and the e-BH boundary $m/(\alpha_{\text{e}}k)$ also moves down with $k$. Past a point the boundary wins and the whole block is lost. The best choice of $\alpha_{\text{cp}}$ is therefore $0.5\alpha_{\text{e}}$ at $d=2$ and $\alpha_{\text{e}}/3$ at $d=5$ and $d=10$. No single transformation level is optimal for every number of groups.

\begin{figure}[ht]
    \centering
    \includegraphics[width=1\linewidth]{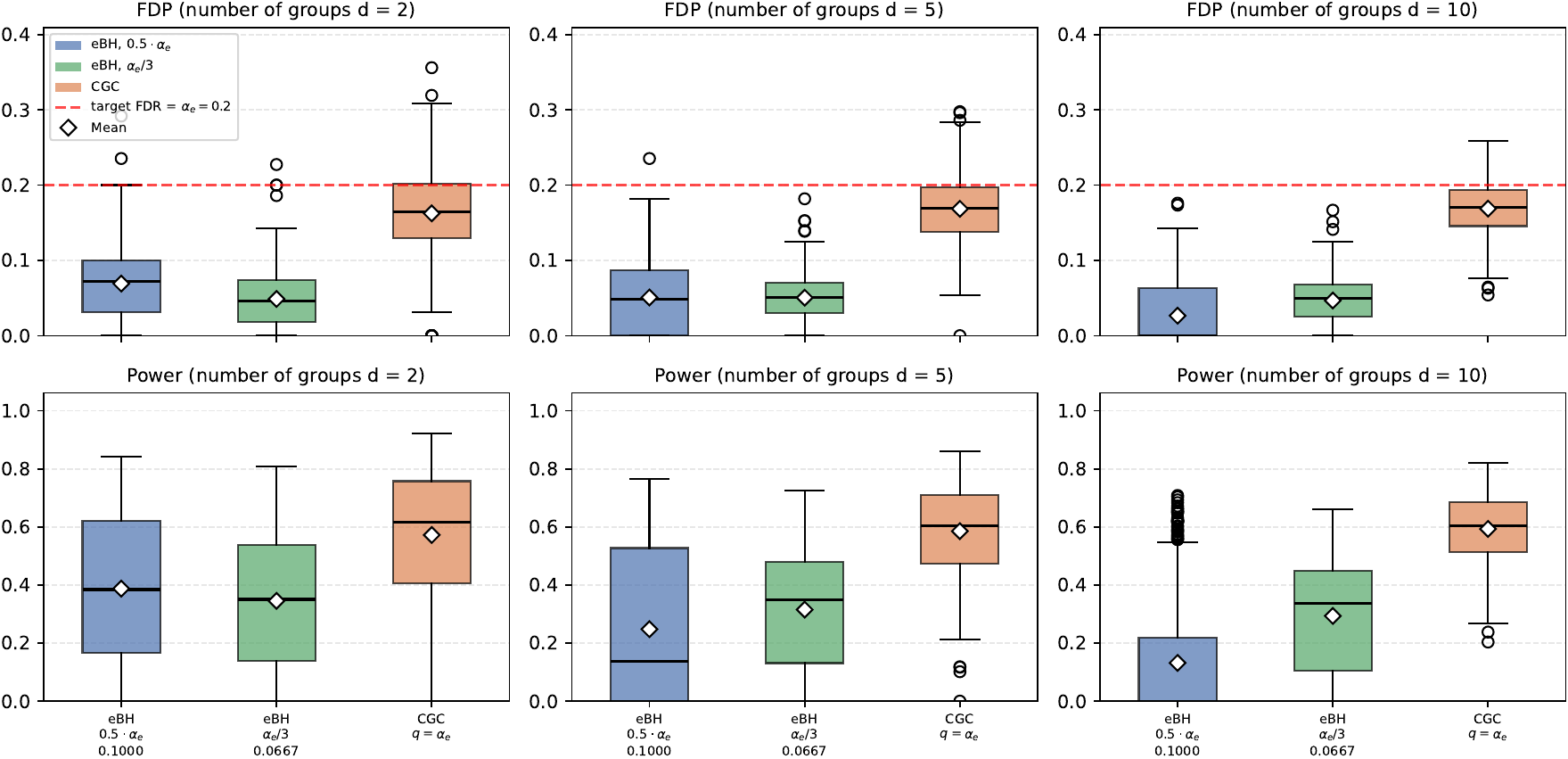}
    \caption{FDP and power of the two e-value routes and of the CGC filter on the same data, over $300$ replications. The CGC filter runs at level $q=\alpha_{\text{e}}=0.2$ and e-BH at the same level. Each column uses a different number of groups $d$. The dashed line marks the target FDR level.}
    \label{fig:evalue_vs_cgc}
\end{figure}

We finally compare the e-value route with the CGC filter. Figure~\ref{fig:evalue_vs_cgc} reports both routes on the same data, over $300$ replications. The CGC filter runs at level $q=\alpha_{\text{e}}=0.2$ and e-BH at the same level, and the e-value route uses the two transformation levels that did best above, $0.5\alpha_{\text{e}}$ and $\alpha_{\text{e}}/3$. The FDP of the CGC filter is centered below the target in every panel, with averages $0.1625$, $0.1684$ and $0.1687$ for $d=2$, $5$ and $10$ groups. The power of the CGC filter is $0.5724$, $0.5855$ and $0.5933$, against $0.3875$ and $0.3454$, $0.2479$ and $0.3152$, and $0.1316$ and $0.2936$ for the two e-value routes. The CGC filter therefore keeps a larger power than either e-value route in every panel, by a factor of about $1.5$ at $d=2$ and about $2$ at $d=10$. The reason is that the CGC filter requires no additional tuning level. It aggregates over the groups in the denominator of the threshold and uses the nominal level directly, so no power is spent on choosing a transformation level. The e-value route requires such a level, and its best choice still costs a large part of the power. For heterogeneous groups the CGC filter is thus clearly the better route, and the generic transformation cannot replace it.

